\documentclass[fleqn,usenatbib]{mnras}

\usepackage{newtxtext,newtxmath}
\usepackage[T1]{fontenc}

\DeclareRobustCommand{\VAN}[3]{#2}
\let\VANthebibliography\thebibliography
\def\thebibliography{\DeclareRobustCommand{\VAN}[3]{##3}\VANthebibliography}

\usepackage{graphicx}	% Including figure files
\usepackage{amsmath}	% Advanced maths commands

\usepackage{dsfont}     % for \mathds{1} for indicator functions
\usepackage{pdflscape}  % Landscape pages (for wide TOI-544 model tables)
\usepackage{bm}         % Use \bm for bold-italic maths (for vectors)
\usepackage{subcaption} % Used for subfigures

\newcommand{\ms}{m\,s$^{-1}$}  % metres per second
\newcommand{\kms}{km\,s$^{-1}$} % kilometres per second
\newcommand{\msd}{{\ms}\,d$^{-1}$} % metres per second per day
\newcommand{\msdd}{{\ms}\,d$^{-2}$} % metres per second per day^2

\newcommand{\wstar}{\omega_\star}  % argument of periastron
\newcommand{\secosw}{\sqrt{e}\cos{\wstar}}  % reparameterisations for argument of periastron
\newcommand{\sesinw}{\sqrt{e}\sin{\wstar}}

\newcommand{\sjit}{\sigma_\mathrm{jit}}  % sigma jitter
\newcommand{\srv}{\sigma_\mathrm{RV}}  % sigma RV (RV errorbars)

\newcommand{\Z}{\mathcal{Z}} % bayesian evidence
\newcommand{\dlnZ}{\Delta\ln{\Z}}  % delta ln Z (Bayes Factor)
\newcommand{\Like}{\mathcal{L}} % L (likelihood)
\newcommand{\lnL}{\ln{\Like}} % ln L

\newcommand{\Mp}{M_\mathrm{p}}  % planetary mass
\newcommand{\Mpsini}{\Mp \sin{i}}  % minimum mass M_p*sin(i)

\newcommand{\eb}{e_{\mathrm{b}}}  % eccentricity b
\newcommand{\ec}{e_{\mathrm{c}}}  % eccentricity c

\title[Improving exoplanet masses using the LHME]{Improving exoplanet mass characterisation with Bayesian model selection using the Learned Harmonic Mean Estimator}

\author[R. S. Dobson et al.]{
Ross S. Dobson,$^{1}$\thanks{E-mail: \href{mailto:ross.dobson@ucl.ac.uk}{ross.dobson@ucl.ac.uk} (RSD)}
Vincent Van~Eylen,$^{1}$
Ioanna Manolopoulou$^{2}$ and
Kendall Sullivan$^{1}$
\\
$^{1}$Mullard Space Science Laboratory, Department of Space \& Climate Physics, University College London, Holmbury St Mary, Dorking, Surrey RH5 6NT, UK\\
$^{2}$Department of Statistical Science, University College London, Gower Street, London WC1E 6BT, UK
}

\date{Accepted XXX. Received YYY; in original form ZZZ}

\pubyear{\the\year{}}

\begin{document}
\label{firstpage}
\pagerange{\pageref{firstpage}--\pageref{lastpage}}
\maketitle

\begin{abstract}
Radial velocity (RV) analyses require modelling choices (such as eccentricity treatment, noise model, velocity trends, and number of planets) that can significantly affect derived planetary masses. Current practice often relies on information criteria to compare and select models, but these have known limitations: they lack the built-in Occam's razor of Bayesian model comparison, and they do not incorporate prior information. Computing the Bayesian evidence needed for Bayes factor model comparison has traditionally required dedicated algorithms such as nested sampling. The learned harmonic mean estimator (LHME) offers an alternative, estimating the Bayesian evidence directly from MCMC posterior samples, with less computational cost and with no modification to the fitting procedure. We present the first application of the LHME to RV model selection, fitting 18 model variants -- comparing circular and eccentric orbits, white noise and Gaussian Process noise models, and long-term velocity trends -- to six single-planet systems, and 72 variants to a seventh system for an $N$ versus $N+1$ planet model comparison. We find that no single model is universally preferred, reinforcing the need for model comparison to select the most appropriate model for a system, thereby ensuring robust mass characterisation. The LHME, implemented in the open-source \textsc{harmonic} package, makes rigorous Bayesian model comparison accessible to existing MCMC-based RV workflows, and we encourage its wider use for other model comparisons in astrophysics.
\end{abstract}


%%%%%%%%%%%%%%%%%%%%%%%%%%%%%%%%%%%%%%%%%%%%%%%%%%
%%%%%%%%%%%%%%%%% BODY OF PAPER %%%%%%%%%%%%%%%%%%
%%%%%%%%%%%%%%%%%%%%%%%%%%%%%%%%%%%%%%%%%%%%%%%%%%

%%%%%%%%%%%% SECTION 1: INTRODUCTION %%%%%%%%%%%%%
%
\section{Introduction}\label{sec:introduction}

Mass characterisation is key to our understanding of exoplanets. Precise masses are essential for constraining atmospheric properties \citep{Batalha.etal2019_PrecisionMassMeasurements} and for understanding bulk densities and interior structures in order to better understand the compositions and formation pathways of exoplanets \citep[e.g.][]{Seager.etal2007_MassRadiusRelationshipsSolid, Dorn.etal2015_CanWeConstrain}. As of 24th February 2026, there are 2343 exoplanets on the NASA Exoplanet Archive\footnote{\url{https://exoplanetarchive.ipac.caltech.edu/}} \citep{Christiansen.etal2025_NASAExoplanetArchive} with masses derived using the Radial Velocity (RV) method.

When fitting RV data, modelling choices must be made: should orbits be assumed circular, or allowed to be eccentric? What priors should be placed on eccentricity? Should a Gaussian Process (GP) \citep{Rasmussen.Williams2006_GaussianProcessesMachine} be used to mitigate stellar activity? Should a long-term linear or quadratic trend be included in the model? 

Recent large-scale homogeneous RV studies have highlighted that modelling choices can significantly impact the accuracy and precision of derived planetary masses \citep{Mignon.etal2024_RadialVelocityHomogeneous, Polanski.etal2024_TESSKeckSurveyXX, Osborne.etal2025_HomogeneousPlanetMasses} and eccentricities \citep{Morgan.etal2025_ExploringWarmJupiter}, with downstream consequences for inferred bulk densities, compositions, formation pathways and habitability. 

One of the most challenging model selection problems in RV analysis is identifying additional bodies beyond the known planets in a system. Non-transiting companions detected solely through RV must be distinguished from stellar activity or instrumental systematics -- without a transit signal, it can be difficult to confirm an additional Keplerian signal over other potential sources of RV variation \citep{Hara.Ford2023_StatisticalMethodsExoplanet}. A related challenge arises from long-term linear or quadratic trends in the RV, which can have several origins: instrumental drift, stellar activity cycles, or a Keplerian contribution from a long-period companion. This last case of an unseen long-period companion -- whether a planetary or stellar object -- is perhaps the most interesting. Strong evidence for a radial acceleration therefore motivates further investigation into its origin; a confirmed companion, especially an additional planet, would have implications for system architecture, derived masses, and formation history.

Another common choice is whether to model orbits as circular or eccentric. Eccentricity is a difficult parameter to constrain, requiring both good signal-to-noise ratio and dense RV phase coverage across the entire orbital period \citep[e.g.][]{Hara.etal2019_BiasRobustnessEccentricity, VanEylen.etal2019_OrbitalEccentricitySmall}. For short and ultra-short period planets, circular orbits are sometimes assumed on physical grounds based on tidal circularisation timescales \citep[e.g.][]{VanEylen.etal2019_OrbitalEccentricitySmall}. However, even a small departure from $e=0$ will have some effect on the RV semi-amplitude $K$, and therefore on the derived planetary mass.

The choice of prior on eccentricity can also have an impact, with uninformative priors sometimes leading to spurious high $e$ fits when data is sparse \citep{Hara.etal2019_BiasRobustnessEccentricity}. In general, the influence of the prior upon a parameter is reduced as the number of data points increase, but RV analyses are often limited in the number of observations available. Therefore, the choice of prior function on eccentricity should also be considered carefully as part of the modelling choices \citep{Osborne.etal2025_HomogeneousPlanetMasses}.

Especially when searching for small Earth-like planets, the amplitude of stellar activity can match or exceed the planetary signal, and quasi-periodic behaviour in stellar activity can be hard to distinguish from the periodic Keplerian signals of planets, making it harder to detect and mass-characterise exoplanets. Gaussian process noise models have been widely adopted to mitigate stellar activity contamination \citep[e.g.][]{Aigrain.etal2012_SimpleMethodEstimate,  Haywood.etal2014_PlanetsStellarActivity, Rajpaul.etal2015_GaussianProcessFramework, Ahrer.etal2021_HARPSSearchSouthern, Barragan.etal2023_RevisitingK2233Spectroscopic}, but the flexibility they add to a model carries a risk of over-fitting \citep{Blunt.etal2023_OverfittingAffectsReliability}, especially with sparse data or if using a one-dimensional GP on the RV time series only without additional activity indicator time series \citep[see the recommendations given in][]{Osborne.etal2025_HomogeneousPlanetMasses}.

Because of these many possible modelling choices, model comparison is crucial for interpreting RV data -- but how should one choose the `best' model? In Bayesian statistics, comparing models via Bayes factors is desirable because it quantifies how much more probable one model is compared to another, given the observed data. However, calculating Bayes factors requires evaluating the marginal likelihood (Bayesian evidence), which involves a high-dimensional integral that in practice is computationally intractable. MCMC methods -- widely used in RV fitting and available in many packages, including \textsc{allesfitter} \citep{Gunther.Daylan2021_AllesfitterFlexibleStar}, \textsc{EXOFASTv2} \citep{Eastman.etal2019_EXOFASTv2PublicGeneralized}, \textsc{ExoFit} \citep{Balan.Lahav2009_EXOFITOrbitalParameters}, \textsc{exoplanet} \citep{Foreman-Mackey.etal2024_ExoplanetGradientbasedProbabilistic}, \textsc{Exo-Striker} \citep{Trifonov2019_ExoStrikerTransitRadial}, \textsc{emperor} \citep{PenaR..Jenkins2025_EMPERORExoplanetMCMC}, \textsc{pyaneti} \citep{Barragan.etal2019_PyanetiFastPowerful, Barragan.etal2022_PyanetiIIMultidimensional}, \textsc{PyORBIT} \citep{Malavolta.etal2016_PyORBIT1, Malavolta.etal2018_PyORBIT2} and \textsc{RadVel} \citep{Fulton.etal2018_RadVelRadialVelocity} -- sample the posterior distribution only up to a normalising constant (the Bayesian evidence), and so cannot provide the Bayesian evidence required for Bayes factor comparison \citep{Ford.Gregory2007_BayesianModelSelection, Hogg.Foreman-Mackey2018_DataAnalysisRecipes}.

Without direct access to the Bayesian evidence, alternative quantitative metrics are commonly used for judging competing models, such as $\chi^2$ and $\lnL$ statistics \citep[e.g.][]{Cale.etal2021_DivingSeaStellar} and information criteria such as the AIC and BIC \citep[e.g.][]{Cloutier.etal2021_MorePreciseMass, Morgan.etal2025_ExploringWarmJupiter, Osborne.etal2025_HomogeneousPlanetMasses}, as well as qualitative physical and domain-knowledge arguments \citep[e.g.][]{VanEylen.etal2019_OrbitalEccentricitySmall}. However, these approaches have known limitations for RV model comparison: $\chi^2$ and $\lnL$ statistics do not account for model complexity; AIC and BIC penalise more complex models but treat all free parameters the same regardless of how sensitive the model is to them, and none of these metrics incorporate prior knowledge on parameters -- for RV fitting in particular, none of them are a rigorous substitute for Bayesian evidence \citep{Ford.Gregory2007_BayesianModelSelection}.

While the Bayesian evidence cannot be feasibly computed analytically for RV model comparison, it can be estimated numerically. Perhaps the most commonly used method in astrophysics for estimating Bayesian evidence is nested sampling \citep{Skilling2004_NestedSampling}: it has been implemented in many RV fitting codes -- including \textsc{allesfitter} \citep{Gunther.Daylan2021_AllesfitterFlexibleStar}, \textsc{Exo-Striker} \citep{Trifonov2019_ExoStrikerTransitRadial}, \textsc{emperor} \citep{PenaR..Jenkins2025_EMPERORExoplanetMCMC}, \textsc{juliet} \citep{Espinoza.etal2019_JulietVersatileModelling}, \textsc{kima} \citep{Faria.etal2018_KimaExoplanetDetection, Faria.etal2023_KimaExoplanetDetection}, \textsc{PyORBIT} \citep{Malavolta.etal2016_PyORBIT1, Malavolta.etal2018_PyORBIT2} and \textsc{RadVel} \citep{Fulton.etal2018_RadVelRadialVelocity} -- and has been successfully applied to model comparison in RV exoplanet analyses \citep[e.g.][]{Baycroft.etal2023_NestedSamplingExample, Desidera.etal2023_TOI179YoungSystem, John.etal2023_NestedSamplingExample, Dreizler.etal2025_NestedSamplingExample, Stevenson.etal2025_RVexoplanetEccentricitiesGood, Lienhard.etal2026_NestedSamplingExample, Naponiello.etal2026_NestedSamplingExample}. However, nested sampling must be used in place of standard MCMC methods rather than alongside them, coupling evidence estimation to a specific sampling strategy (thereby restricting the range of methods that can be employed) and can be computationally expensive in higher-dimensional problems.

Recently, \citet{McEwen.etal2023_MachineLearningAssisted} and \citet{Polanska.etal2025_LearnedHarmonicMean} developed the learned harmonic mean estimator (LHME) which estimates Bayesian evidence from the posterior samples generated by any MCMC method. This decoupling of posterior sampling from evidence estimation allows the use of any MCMC algorithm, and enables post-hoc model comparison using existing saved-down MCMC chains. The LHME is implemented in the open-source \textsc{harmonic} Python package\footnote{\url{https://github.com/astro-informatics/harmonic}}. Although the LHME has been applied to evidence estimation in cosmology \citep[e.g.][]{Carrion.etal2025_HarmonicExample, Du.etal2025_HarmonicExample, Stiskalek.etal2025_HarmonicExample, Stiskalek.etal2026_HarmonicExample}, its application to exoplanet RV analysis is novel. In this work, we demonstrate the use of the LHME as a way to estimate Bayesian evidence for RV model selection within standard MCMC workflows. 

We select six single-planet systems from the homogeneous re-analysis of archival HARPS \citep[High Accuracy Radial Velocity Planet Searcher;][]{Mayor.etal2003_HARPS} data presented in \citet{Osborne.etal2025_HomogeneousPlanetMasses}. For each of these systems, we fit 18 model variants spanning different treatments of eccentricity (circular vs.\ eccentric with uniform or informative priors), stellar activity (white noise vs.\ Gaussian Process), and long-term velocity trends (none, linear, or quadratic). We additionally apply \textsc{harmonic} to TOI-544, a two-planet system observed with HARPS and HARPS-N, to demonstrate an $N$ vs.\ $N+1$ planet model comparison; for this system we fit 72 model variants (18 one-planet and 54 two-planet configurations).

For each model, we perform MCMC sampling with \textsc{ravest} (see Section~\ref{sec:models}), then apply the LHME using \textsc{harmonic} to the posterior samples in order to compute the Bayesian evidence for that model.

The structure of this paper is as follows. Section~\ref{sec:bayesian_comparison} provides an overview of Bayesian model comparison and the learned harmonic mean estimator. Section~\ref{sec:data} describes our target systems and data preparation. Section~\ref{sec:models} details the RV models and fitting procedure. Section~\ref{sec:results} presents the results of the fitting, the model comparisons and resulting planetary parameters. Section~\ref{sec:discussion} discusses the performance of Bayesian model comparison and the implications for RV analysis, and alternative model comparison methods used in previous studies and the limitations of our approach. We conclude in Section~\ref{sec:conclusions}.

%%%%%%%%%%%%%%%% SECTION 2 BAYES %%%%%%%%%%%%%%%%%

\section{Bayesian Model Comparison}\label{sec:bayesian_comparison}
Here we provide a brief overview of Bayesian parameter inference and model comparison, with emphasis on computing the Bayesian evidence. We introduce the learned harmonic mean estimator (implemented in the \textsc{harmonic} package) as an evidence estimator that can be applied to posterior samples produced by standard MCMC methods. For full technical details, we refer readers to \citet{McEwen.etal2023_MachineLearningAssisted} and \citet{Polanska.etal2025_LearnedHarmonicMean}.

%%%%%%%%%%%%%%%% SECTION 2.1 BAYESIAN PARAMETER INFERENCE %%%%%%%%%%%%%%%%%
%
\subsection{Bayesian Parameter Inference}
When fitting RV data, we aim to infer model parameters $\bm{\theta}$ (e.g.\ the semi-amplitude $K$, orbital period $P$, transit time / inferior conjunction time $T_C$, eccentricity $e$). Given observed RV data $\bm{y}$ with uncertainties $\bm{\sigma}$, for a model $M$ the posterior distribution $p(\bm{\theta}\,|\,\bm{y},M)$ over the parameters is given by Bayes' theorem \citep{Bayes1763_LIIEssaySolving, Gelman.etal2014_BayesianDataAnalysis}:
\begin{equation}\label{eq:bayes}
   p(\bm{\theta} \,|\, \bm{y},M) = \frac{p(\bm{y} \,|\, \bm{\theta},M)\,p(\bm{\theta}\,|\,M)}{p(\bm{y}\,|\,M)} \equiv \frac{\Like(\bm{\theta})\,\pi(\bm{\theta})}{\Z},
\end{equation}
where $\Like(\bm{\theta}) \equiv p(\bm{y}\,|\,\bm{\theta},M)$ is the likelihood, $\pi(\bm{\theta}) \equiv p(\bm{\theta}\,|\,M)$ is the prior distribution, and
\begin{equation}\label{eq:evidence}
    \Z \equiv p(\bm{y}\,|\,M) = \int \Like(\bm{\theta})\,\pi(\bm{\theta})\,\mathrm{d}\bm{\theta}
\end{equation}
is the Bayesian evidence (also called the marginal likelihood). We adopt a compact notation $\Like(\bm{\theta})$, $\pi(\bm{\theta})$ and $\Z$, dropping $\bm{y}$ and $M$ for brevity except where needed.

For parameter inference under a fixed model $M$, $\Z$ enters only as a normalisation constant, and in the MCMC acceptance step it cancels from the ratio of posterior densities at proposed and current parameter values -- therefore, it is never required (see Appendix~\ref{app:app1} for further details).

%%%%%%%%%%%%%%%% SECTION 2.2 MODEL COMPARISON AND THE BAYESIAN EVIDENCE %%%%%%%%%%%%%%%%%
%
\subsection{Model comparison and the Bayesian evidence}\label{subsec:model_comparison}
Whilst the evidence is unnecessary for parameter inference, it becomes central for Bayesian model comparison. The Bayesian evidence naturally implements Occam's razor by balancing goodness-of-fit against model complexity. Since $\Z$ is the prior-weighted average of the likelihood (equation~(\ref{eq:evidence})), introducing additional parameters dilutes the prior mass over a larger volume of parameter space. For a model with those additional parameters to be favoured, there must be a sufficiently large increase in likelihood. If the additional parameters are well constrained by the data and the fit improves, then the increase in likelihood can outweigh the dilution and increase $\Z$. If instead they are not supported by the data and contribute little to the fit, the likelihood does not improve over most of the prior support, so $\Z$ changes little or may even decrease. The evidence therefore penalises unnecessary model complexity automatically \citep{Trotta2008_BayesSkyBayesian}, without requiring an explicit penalty based solely on parameter count, such as in AIC or BIC.

Suppose we wish to compare two competing models $M_1$ and $M_2$ (e.g.\ a circular versus eccentric orbit, or two choices of prior function, or a 1-planet versus 2-planet model) by comparing their posterior probabilities, to determine which model is preferred in light of the observed data. Applying Bayes' theorem to obtain the posterior over models $p(M\,|\,\bm{y})$ yields the posterior odds ratio:
\begin{equation}\label{eq:posterior_odds}
    \frac{p(M_1\,|\,\bm{y})}{p(M_2\,|\,\bm{y})}
    = \frac{p(\bm{y}\,|\,M_1)}{p(\bm{y}\,|\,M_2)} \,
      \frac{p(M_1)}{p(M_2)}
    = \frac{\Z_1}{\Z_2} \,
      \frac{p(M_1)}{p(M_2)},
\end{equation}
where $p(\bm{y}\,|\,M_i) \equiv \Z_i$ is the Bayesian evidence for model $M_i$ (equation~(\ref{eq:evidence})), and $p(M_i)$ is the prior probability assigned to model $M_i$ before seeing the data. If we adopt equal prior model probabilities ($p(M_1)=p(M_2)$), which can be a reasonable decision in the absence of any prior belief preferring one model over the other \citep{Trotta2008_BayesSkyBayesian}, then the posterior odds ratio shown in equation~(\ref{eq:posterior_odds}) reduces to the Bayes factor $\Z_1/\Z_2$ alone.

Once we have the evidence estimate $\ln\hat{\Z}$ for each model, we can compare two models by calculating the log Bayes factor
\begin{equation}\label{eq:ln_bf}
    \ln(B_{12}) = \ln\!\left(\frac{\Z_1}{\Z_2}\right) = \ln(\Z_1) - \ln(\Z_2) \equiv \dlnZ_{12}.
\end{equation}
\citet{Trotta2008_BayesSkyBayesian} assigns a strength of evidence to the absolute log Bayes factor $|\dlnZ|$ when comparing two models: inconclusive for $\dlnZ < 1.0$, weak evidence for $1.0 \leq \dlnZ <2.5$, moderate evidence for $2.5 \leq \dlnZ < 5.0$, and strong evidence for $\dlnZ \geq 5.0$.

The Bayesian evidence is, however, rarely computationally tractable and therefore needs to be estimated. Because MCMC methods sample the posterior distribution only up to a normalising constant (the evidence), they do not provide or require calculation of $\Z$ for parameter inference within a single model \citep{Ford.Gregory2007_BayesianModelSelection, Hogg.Foreman-Mackey2018_DataAnalysisRecipes}. This is true of any MCMC-based parameter inference (not just RV fitting) and is the reason why Bayesian model comparison requires specialised evidence estimation techniques, e.g.\ nested sampling \citep{Ford.Gregory2007_BayesianModelSelection}. 

In practice, the computational difficulty of estimating $\Z$ has often led to the use of other model comparison metrics, such as traditional goodness-of-fit metrics $\chi^2$ and $\lnL_\mathrm{max}$, or information criteria. The Bayesian Information Criterion (BIC; \citealt{Schwarz1978_EstimatingDimensionModel}) -- which can be viewed as an approximation to the Bayes factor \citep{Trotta2008_BayesSkyBayesian} -- and the similar Akaike Information Criterion (AIC; \citealt{Akaike1974_NewLookStatistical}) are defined as
\begin{equation}\label{eq:BIC}
  \mathrm{BIC} = k\ln(n) - 2\ln(\Like_\mathrm{max}),
\end{equation}
and
\begin{equation}\label{eq:AIC}
  \mathrm{AIC} = 2k - 2\ln(\Like_\mathrm{max}),
\end{equation}
where $n$ is the number of data points, $k$ is the number of free parameters, and $\Like_\mathrm{max}$ is the maximum likelihood value. Both criteria penalise model complexity through the parameter count $k$, but depend only on the maximum likelihood and do not incorporate prior information -- yet the prior is as much a part of the model specification as the likelihood \citep{Thorngren.etal2026_BayesianModelComparison}. Their known limitations for RV model selection are discussed further in Section~\ref{subsec:discussion_bayesian}. In contrast, Bayesian evidence integrates both the likelihood and prior (see equation~(\ref{eq:evidence})), enabling a comparison that accounts for the full model specification.

%%%%%%%%%%%%%%%% SECTION 2.3 THE LEARNED HARMONIC MEAN ESTIMATOR AND HARMONIC %%%%%%%%%%%%%%%%%
%
\subsection{The learned harmonic mean estimator and \textsc{harmonic}}\label{subsec:harmonic}
The learned harmonic mean estimator (LHME; \citealt{McEwen.etal2023_MachineLearningAssisted, Polanska.etal2025_LearnedHarmonicMean}) provides an alternative to nested sampling or model selection criteria like BIC or AIC by estimating $\Z$ directly from posterior samples generated by any MCMC method. This means that any choice of MCMC-based methods can be used, without restriction to a subset of methods (such as nested methods). It also means that existing archived MCMC chains can be re-used for evidence estimation and model comparison, without having to re-perform the entire RV fitting process.

Evidence estimates from \textsc{harmonic} have been empirically validated against common Bayesian evidence benchmark problems including the Rosenbrock and Normal-Gamma models \citep{Polanska.etal2025_LearnedHarmonicMean}. Results have also been shown to be consistent with other evidence estimation methods, including nested sampling \citep[e.g.\ ][]{Piras.etal2024_FutureCosmologicalLikelihoodbased}.

The LHME in \textsc{harmonic} uses normalising flows, a machine learning method, to learn a normalised target distribution from MCMC posterior samples. This learned distribution approximates the posterior distribution, while addressing the issues of the original harmonic mean estimator \citep{Newton.Raftery1994_ApproximateBayesianInference, Neal1994_DiscussionPaperNewton, Clyde.etal2007_CurrentChallengesBayesian, Robert.Wraith2009_ComputationalMethodsBayesian}, discussed below.

%%%%%%%%%%%%%%%% SECTION 2.3.1 ORIGINAL HARMONIC MEAN ESTIMATOR %%%%%%%%%%%%%%%%%
%
\subsubsection{Original harmonic mean estimator}
\citet{Newton.Raftery1994_ApproximateBayesianInference} introduced the original harmonic mean estimator, showing that the harmonic mean of samples from the likelihood can be an estimator for the reciprocal of the evidence $\rho = \Z^{-1}$:
\begin{equation}\label{eq:rho_hat}
    \hat{\rho} = \frac{1}{N}\sum_{i=1}^N \frac{1}{\Like(\bm{\theta}_i)}, \quad \bm{\theta}_i\sim p(\bm{\theta}\,|\,\bm{y}).
\end{equation}
The motivation for using the harmonic mean can be seen from considering the expected value, with respect to the posterior, of the reciprocal of the likelihood \citep{McEwen.etal2023_MachineLearningAssisted}:
\begin{equation}\label{eq:rho}
\begin{split}
    \rho = \mathbb{E}_{p(\bm{\theta}\,|\,\bm{y})} \left[ \frac{1}{\Like(\bm{\theta})} \right]
    &= \int \frac{1}{\Like(\bm{\theta})}\,p(\bm{\theta}\,|\,\bm{y})\,\mathrm{d}\bm{\theta}\\
    &= \int \frac{1}{\Like(\bm{\theta})}\,\frac{\Like(\bm{\theta})\,\pi(\bm{\theta})}{\Z}\,\mathrm{d}\bm{\theta}\\
    &= \frac{1}{\Z}\int \pi(\bm{\theta})\,\mathrm{d}\bm{\theta}\\
    &= \frac{1}{\Z},
\end{split}
\end{equation}
using equation~(\ref{eq:bayes}) to substitute the posterior and noting that a proper prior distribution $\pi(\bm{\theta})$ is normalised and integrates to $1$. 

This appears to offer a computationally convenient route to estimating $\Z$, since the likelihood evaluations are already available from MCMC parameter inference. However, as was later noted, the harmonic mean estimator was unstable and could fail due to large, often non-finite variances \citep{Neal1994_DiscussionPaperNewton} causing very slow convergence \citep{Clyde.etal2007_CurrentChallengesBayesian, Ford.Gregory2007_BayesianModelSelection}. 

The reason for this can be seen by viewing the first integral of  equation~(\ref{eq:rho}) as importance sampling, with the posterior $p(\bm{\theta}\,|\,\bm{y})$ as the sampling density and the prior $\pi(\bm{\theta})$ as the target distribution \citep{McEwen.etal2023_MachineLearningAssisted}. Importance sampling requires the target distribution to have thinner tails than the sampling density -- in this case, the prior must have thinner tails than the posterior. However, this often is not the case: the posterior is concentrated around the region of high likelihood, typically leading to thinner tails than the prior \citep{McEwen.etal2023_MachineLearningAssisted} and therefore the instability of the original harmonic mean estimator.

%%%%%%%%%%%%%%%% SECTION 2.3.2 LEARNED HARMONIC MEAN ESTIMATOR %%%%%%%%%%%%%%%%%
%
\subsubsection{Learned harmonic mean estimator}\label{subsubsec:LHME}
\citet{Gelfand.Dey1994_BayesianModelChoice} attempted to remedy this when they introduced a normalised arbitrary density $\varphi(\bm{\theta})$, allowing equation~(\ref{eq:rho}) to be rewritten as
\begin{equation}
    \rho = \mathbb{E}_{p(\bm{\theta}\,|\,\bm{y})} \left[ \frac{\varphi(\bm{\theta})}{\Like(\bm{\theta}) \pi(\bm{\theta})} \right]
\end{equation}
motivating a new estimator, the re-targeted harmonic mean estimator
\begin{equation}\label{eq:rhme}
    \hat{\rho} = \frac{1}{N}\sum_{i=1}^N \frac{\varphi(\bm{\theta}_i)}{\Like(\bm{\theta}_i)\pi(\bm{\theta}_i)}, \quad \bm{\theta}_i\sim p(\bm{\theta}\,|\,\bm{y})
\end{equation}
where the target distribution is now the arbitrary density $\varphi(\bm{\theta})$. This means a target distribution can be chosen, provided it is normalised, so that its tails are narrower than the posterior tails, mitigating the problem.

However, selecting a suitable target distribution $\varphi(\bm{\theta})$ is difficult. \citet{McEwen.etal2023_MachineLearningAssisted} realised that the optimal choice would be to use the normalised posterior (equation~(\ref{eq:bayes})):
\begin{equation}
    \varphi_\mathrm{optimal}(\bm{\theta}) = \frac{\Like(\bm{\theta})\,\pi(\bm{\theta})}{\Z}
\end{equation}
which, by inspection of equation~(\ref{eq:rhme}), would yield an unbiased estimator with zero variance. However, normalising the posterior requires knowing $\Z$ -- but this is the very quantity being estimated in the first place.

\citet{McEwen.etal2023_MachineLearningAssisted} showed that machine learning techniques can be used to learn an approximate model of the normalised posterior density from the posterior samples, while maintaining the condition that the target distribution tails must be thinner than the sampling density tails (which is more important than having a close approximation of the posterior). \citet{Polanska.etal2025_LearnedHarmonicMean} builds upon this by using normalising flows in place of the original bespoke machine learning techniques.

This allows us to use our MCMC posterior samples from RV fitting and directly estimate the Bayesian evidence required to calculate Bayes factors, enabling robust model comparison.

%%%%%%%%%%%%%%%% SECTION 3 DATA %%%%%%%%%%%%%%%%%
%
\section{Data and Systems}\label{sec:data}
\begin{table*}
\centering
\small
\caption{Summary of systems and observations. $N_{\mathrm{obs}}$ is the total number of RV datapoints used after data preparation (see Section~\ref{sec:data_prep}). HARPS$_{03}$, HARPS$_{15}$, and HARPS$_{20}$ denote HARPS observations acquired before the 2015 June 2\textsuperscript{nd} fibre upgrade, between the upgrade and the 2020 March 23\textsuperscript{rd} COVID warm-up, and after the 2020 warm-up respectively.}
\label{tab:systems}

\begin{tabular}{lllll}
    \hline
    TIC ID        & Name     & $N_{\rm obs}$ & Instruments & Priors Reference \\
    \hline
    TIC 146364192 & K2-265   & 140 &
      {HARPS$_{15}$: 140} &
      \citet{Lam.etal2018_K2265TransitingRocky} \\
    TIC 150098860 & TOI-220  & 91  &
      {HARPS$_{15}$: 91} &
      \citet{Hoyer.etal2021_TOI220WarmSubNeptune} \\
    TIC 207141131 & HD 18599 & 106 &
      {HARPS$_{03}$: 7, HARPS$_{15}$: 99} &
      \citet{Desidera.etal2023_TOI179YoungSystem} \\
    TIC 260004324 & LHS 1815 & 71  &
      {HARPS$_{03}$: 10, HARPS$_{15}$: 61} &
      \citet{Gan.etal2020_LHS1815bFirst} \\
    TIC 320004517 & TOI-1055 & 71  &
      {HARPS$_{03}$: 36, HARPS$_{15}$: 20, HARPS$_{20}$: 15} &
      \citet{Bonfanti.etal2023_TOI1055NeptunianPlanet} \\
    TIC 467929202 & GJ 1214  & 164 &
      {HARPS$_{03}$: 89, HARPS$_{15}$: 75} &
      \citet{Cloutier.etal2021_MorePreciseMass} \\

    TIC 50618703  & TOI-544  & 119 &
      {HARPS$_{20}$: 105, HARPS-N: 14} &
      \citet{Osborne.etal2024_TOI544PotentialWaterworld} \\
    \hline
    \end{tabular}
  \end{table*}
We analyse radial velocity observations of seven systems. Six are drawn from the homogeneous re-analysis of archival HARPS \citep[High Accuracy Radial Velocity Planet Searcher;][]{Mayor.etal2003_HARPS} observations presented in \citet{Osborne.etal2025_HomogeneousPlanetMasses}, each hosting a single confirmed transiting planet. The data in that study are publicly available from the ESO archive \citep{Barbieri2023_ESOHARPSRadial} and were reduced from spectra via the online HARPS pipeline \citep{Lovis.Pepe2007_NewListThorium}. We additionally include TOI-544, a recently characterised system hosting two confirmed planets \citep{Osborne.etal2024_TOI544PotentialWaterworld} observed with both HARPS and HARPS-N, making it a suitable example to demonstrate an $N$ vs.\ $N+1$ planet model comparison with the LHME. Table~\ref{tab:systems} summarises the key properties of the systems and observations used in this work.

%%%%%%%%%%%%%%%% SECTION 3.1 DATA PREPARATION %%%%%%%%%%%%%%%%%
%
\subsection{Data Preparation}\label{sec:data_prep}
For the six single-planet systems that were previously studied in \citet{Osborne.etal2025_HomogeneousPlanetMasses}, we adopt the same data preparation procedure to ensure data quality. For some individual systems they performed further system-specific processing \citep[see appendix A.1 of][]{Osborne.etal2025_HomogeneousPlanetMasses} which we replicate: for LHS~1815 (TIC~260004324), heavily outlying data are removed by cutting data with velocities below $42.5$\,\kms and above $100$\,\kms. For K2-265 (TIC~146364192), we additionally exclude 10 observations identified by \citet{Lam.etal2018_K2265TransitingRocky} as Moon-contaminated on the basis of anomalous FWHM values. For TOI-220 (TIC~150098860), we additionally replicate the exclusions identified by \citet{Hoyer.etal2021_TOI220WarmSubNeptune}: 8 observations from 2018 November 25\textsuperscript{th}--27\textsuperscript{th}, one observation from 2019 January 19\textsuperscript{th}, and one from 2019 April 17\textsuperscript{th} are excluded.

For all seven systems, we then apply a two-iteration $3\sigma$ sigma-clip using \textsc{Astropy} \citep{astropy:2013, astropy:2018, astropy:2022} to remove outliers, discarding measurements more than $3\sigma$ from the median RV. The final number of observations $N_\mathrm{obs}$ for each system is listed in Table~\ref{tab:systems}.

HARPS underwent a fibre exchange in 2015 and also experienced a warmup due to the COVID-19 pandemic in 2020\footnote{\url{https://www.eso.org/sci/facilities/lasilla/instruments/harps/news.html}}. These events introduced shifts in the RV zero-point \citep{LoCurto.etal2015_HARPSGetsNew, Barbieri2023_ESOHARPSRadial, Mignon.etal2024_RadialVelocityHomogeneous}, therefore observations taken before and after both 2015 June 2\textsuperscript{nd} and 2020 March 23\textsuperscript{rd} need to be modelled as separate instruments (with separate RV offsets $\gamma$ and jitter $\sjit$ as free model parameters). In this work, we adopt and extend the nomenclature of \citet{Mignon.etal2024_RadialVelocityHomogeneous} and refer to the pre- and post-fibre upgrade series (sometimes referred to in previous studies as $\mathrm{HARPS}_\mathrm{pre}$ and $\mathrm{HARPS}_\mathrm{post}$) as $\mathrm{HARPS}_{03}$ and $\mathrm{HARPS}_{15}$, and the post-warmup series as $\mathrm{HARPS}_{20}$. The total numbers of observations within each HARPS era for each system can be seen in Table~\ref{tab:systems}. 

For readability, throughout this work, we convert observation times from the original UTC Barycentric Julian Date ($\mathrm{BJD}_\mathrm{UTC}$) to Barycentric TESS Julian Date (BTJD = BJD $- 2 457 000$).

Values and uncertainties for some planetary parameters (orbital period $P$ and mid-transit time $T_C$) and stellar parameters (rotation period $P_\mathrm{rot}$, effective temperature $T_\mathrm{eff}$, stellar mass $M_\star$) are obtained from the NASA Exoplanet Archive via automated Table Access Protocol queries. $P$ and $T_C$ are used directly as informative priors, while the stellar rotation period $P_\mathrm{rot}$ (or, where unavailable, the effective temperature $T_\mathrm{eff}$) is used to inform priors on the GP hyperparameters, if published hyperparameter values are not available (see Section~\ref{subsec:priors} and Table~\ref{tab:priors}). The stellar mass $M_\star$ is used to convert the RV semi-amplitude $K$ into a planetary minimum mass estimate $\Mpsini$.

%%%%%%%%%%%%%%%% SECTION 4 MODELS %%%%%%%%%%%%%%%%%
%
\section{Models and Fitting Procedure}\label{sec:models}
%
%%%%%%%%%%%%%%%% SECTION 4.1 THE RAVEST PACKAGE %%%%%%%%%%%%%%%%%
%
\subsection{The \textsc{ravest} package}
We perform all RV fitting using \textsc{ravest}, an open-source radial velocity fitting package written in Python\footnote{\url{https://github.com/ross-dobson/ravest}}. Full details are the subject of a separate paper \citetext{Dobson et al., in prep};  we summarise the key features here. The \textsc{ravest} package can model single- or multi-planet Keplerian orbits, with support for eccentric orbits in different parameterisations e.g.\  $(e, \wstar)$ or $(\secosw, \sesinw)$. It can use data from multiple instruments simultaneously, modelling them with separate RV offset $\gamma$ and jitter $\sjit$ terms, and it can fit for long-term linear and quadratic velocity trends. It has support for Gaussian Process noise models via the \textsc{tinygp} package \citep{foreman_mackey_2024_10463641}. MCMC sampling for parameter inference and uncertainty estimation is performed using the affine-invariant ensemble sampler \textsc{emcee} \citep{Foreman-Mackey.etal2013_EmceeMCMCHammer}, and it has been designed to be very easily used with \textsc{harmonic} for Bayesian evidence estimation. We clarify that when reporting values of argument of periastron $\wstar$, \textsc{ravest} operates in a coordinate system where the positive z-axis points away from the observer, such that a positive RV corresponds to a stellar redshift \citep{Householder.Weiss2022_InconsistentUseomega}, using the same convention as \textsc{EXOFAST} and \textsc{RadVel}.

%%%%%%%%%%%%%%%% SECTION 4.2 MODEL VARIANTS %%%%%%%%%%%%%%%%%
%
\subsection{Model Variants}
\begin{table}
	\centering
	\caption{Radial velocity models compared in this work. For the two-planet models for TOI-544, two eccentricity characters are used for planet b and c respectively.}
	\label{tab:model_params}
	\begin{tabular}{llll}
		\hline
		Model & Eccentricity & Noise Model & Trend \\
		\hline
		\texttt{CW0}     & Circular & White & None \\
		\texttt{CW1}     & Circular & White & Linear \\
		\texttt{CW2}     & Circular & White & Quadratic \\
        \texttt{CG0}     & Circular & GP    & None \\
		\texttt{CG1}     & Circular & GP    & Linear \\
		\texttt{CG2}     & Circular & GP    & Quadratic \\
        \texttt{UW0}     & Eccentric (uniform) & White & None \\
        \texttt{UW1}     & Eccentric (uniform) & White & Linear \\
        \texttt{UW2}     & Eccentric (uniform) & White & Quadratic \\
        \texttt{UG0}     & Eccentric (uniform) & GP    & None \\
        \texttt{UG1}     & Eccentric (uniform) & GP    & Linear \\
        \texttt{UG2}     & Eccentric (uniform) & GP    & Quadratic \\
        \texttt{HW0}     & Eccentric (half-normal) & White & None \\
        \texttt{HW1}     & Eccentric (half-normal) & White & Linear \\
        \texttt{HW2}     & Eccentric (half-normal) & White & Quadratic \\
        \texttt{HG0}     & Eccentric (half-normal) & GP    & None \\
        \texttt{HG1}     & Eccentric (half-normal) & GP    & Linear \\
        \texttt{HG2}     & Eccentric (half-normal) & GP    & Quadratic \\
		\hline
	\end{tabular}
\end{table}

For the six single-planet systems we fit 18 model variants, differing in eccentricity treatment, velocity trends, and noise modelling (listed in Table~\ref{tab:model_params}). The model naming scheme is \texttt{XYZ}, where:

\begin{itemize}
    \item \texttt{X} indicates eccentricity: \texttt{C} (circular, fixed $e=0$), \texttt{U} (eccentric with uniform prior on $\secosw$ and $\sesinw$), or \texttt{H} (eccentric with half-normal prior for single-planets on $e$, from \citealt{VanEylen.etal2019_OrbitalEccentricitySmall})
    \item \texttt{Y} indicates noise model: \texttt{W} (white noise with jitter $\sjit$) or \texttt{G} (Gaussian Process with quasi-periodic kernel)
    \item \texttt{Z} indicates trend: \texttt{0} (none), \texttt{1} (linear), or \texttt{2} (quadratic)
\end{itemize}

For TOI-544 where we also have two-planet models, we extend the scheme to have two eccentricity characters, \texttt{X}$_\mathrm{b}$ and \texttt{X}$_\mathrm{c}$, leading to 72 model variants in total (18 one-planet and 54 two-planet).

The RV model $\mu_i$ at observation time $t_i$ is
\begin{equation}\label{eq:mu}
  \mu_i = v(t_i) + \gamma_\mathrm{inst} + \dot{\gamma}(t_i - t_0) + \ddot{\gamma}(t_i - t_0)^2
\end{equation}
where $v(t_i)$ is the summed Keplerian RV contribution from all planets and $\gamma_\mathrm{inst}$ is the systemic velocity offset for the instrument used to take observation $i$. Long-term velocity trends are modelled with the $\dot{\gamma}$ (in \msd) and $\ddot{\gamma}$ (in \msdd) linear and quadratic trend coefficients (shared across all instruments), with reference time $t_0$ set as the median BTJD of the RV observations for that system. In \texttt{0} models both trend coefficients are fixed to zero; in \texttt{1} models $\dot{\gamma}$ is a free parameter and $\ddot{\gamma}$ is fixed to zero; in \texttt{2} models both trend coefficients are free.

Each instrument is modelled separately with an independent velocity offset $\gamma_\mathrm{inst}$. As described in Section~\ref{sec:data}, HARPS observations that span the 2015 HARPS fibre upgrade or 2020 COVID warm-up are modelled as separate instruments (HARPS$_{03}$, HARPS$_{15}$, HARPS$_{20}$), to handle the shifts in RV offsets introduced by those events. TOI-544 additionally includes HARPS-N observations, which are modelled as a fourth instrument. To account for additional uncorrelated astrophysical or instrumental noise, we also fit a white noise jitter term $\sigma_{\mathrm{jit,inst}}$ separately for each instrument, which is added to the formal observed RV uncertainties $\srv$ in quadrature \citep[e.g.][]{Ford2006_ImprovingEfficiencyMarkov}.

The log-likelihood for white noise \texttt{W} models is \citep{Ford.Gregory2007_BayesianModelSelection}
\begin{equation}\label{eq:lnL_W}
     \lnL_\mathrm{W} = -\frac{1}{2}\sum_{i=1}^{N_\mathrm{obs}}\left[\frac{(y_i - \mu(t_i))^2}{\srv^2(t_i) + \sigma_{\mathrm{jit,inst}}^2} + \ln\!\left(2\pi\!\left(\srv^2(t_i) + \sigma_{\mathrm{jit,inst}}^2\right)\right)\right]
\end{equation}
where $y_i$ and $\srv(t_i)$ are the observed radial velocity and uncertainty at time $t_i$, and $\mu(t_i)$ is the model prediction (equation~(\ref{eq:mu})), and $N_\mathrm{obs}$ is the number of observations (see Table~\ref{tab:systems}).

For the \texttt{G} models, the noise is instead modelled with a Gaussian Process (GP), to mitigate time-correlated structure -- primarily rotationally modulated stellar variability due to starspots, faculae and plages -- that is otherwise difficult to distinguish from Keplerian planet signals \citep{Rajpaul.etal2015_GaussianProcessFramework}. We assume previous knowledge of Gaussian Processes; for further details we recommend the textbook \citet{Rasmussen.Williams2006_GaussianProcessesMachine} and the review \citet{Aigrain.Foreman-Mackey2023_GaussianProcessRegression}.

A common choice of GP kernel for modelling stellar activity is a quasi-periodic (QP) kernel \citep[e.g.][]{Haywood.etal2014_PlanetsStellarActivity, Rajpaul.etal2015_GaussianProcessFramework, Barragan.etal2023_RevisitingK2233Spectroscopic, Stock.etal2023_GaussianProcessesRadial}. The QP kernel function as defined by \citet{Roberts.etal2013_GaussianProcessesTimeseries} is
\begin{equation}\label{eq:QP}
    k_\mathrm{QP}(t_i,t_j) = A_\mathrm{GP}^2\exp\left[ -\frac{\sin^2\left[\pi(t_i-t_j)/P_\mathrm{GP}\right]}{2\lambda^2_\mathrm{p}}  -\frac{(t_i-t_j)^2}{2\lambda^2_\mathrm{e}} \right]
\end{equation}
with hyperparameters $A_\mathrm{GP}$ (amplitude, in \ms), $P_\mathrm{GP}$ (characteristic period, in d), $\lambda_\mathrm{p}$ (inverse of the harmonic complexity), and $\lambda_\mathrm{e}$ (active region evolution timescale, in d). The kernel function describes how correlated the stellar activity noise should be between any two observation times.

The covariance matrix $\bm{K}$ is constructed by evaluating the QP kernel function for every pair of observations $(t_i,t_j)$, and adding the per-observation uncorrelated noise (observed RV uncertainty and jitter in quadrature, the same as in equation~(\ref{eq:lnL_W})) on the diagonal:
\begin{equation}\label{eq:K_matrix}
    K_{ij} = k_{\mathrm{QP}} (t_i, t_j) + \left(\srv^2 (t_i) + \sigma_{\mathrm{jit,inst}}^2 \right)\delta_{ij}
\end{equation}
where $\delta_{ij}$ is the Kronecker delta. Note that the white noise model of equation~(\ref{eq:lnL_W}) can be seen as the special case where a kernel function $k(t_i, t_j) = 0$: the covariance matrix $\bm{K}$ becomes purely diagonal with entries $\srv^2 (t_i) + \sigma_{\mathrm{jit,inst}}^2$ on the main diagonal, with the log-likelihood reducing to the white noise log-likelihood (equation~(\ref{eq:lnL_W})).

The log-likelihood function for the GP models is the multivariate Gaussian log-likelihood \citep{Rajpaul.etal2015_GaussianProcessFramework}
\begin{equation}\label{eq:lnL_GP}
   \lnL_\mathrm{GP} = -\frac{1}{2}\,\bm{r}^\top \bm{K}^{-1} \bm{r} - \frac{1}{2}\ln|\bm{K}| - \frac{N_\mathrm{obs}}{2}\ln 2\pi
\end{equation}
where $\bm{r} = \bm{y} - \bm{\mu}$ is the $N_\mathrm{obs}$-dimensional vector of residuals of the observed radial velocities $\bm{y}$ from the modelled RV $\bm{\mu}$ (equation~(\ref{eq:mu})). Both the RV model (Keplerian, trend, and per-instrument) parameters and the GP hyperparameters are sampled simultaneously in the MCMC, so the deterministic model and the correlated noise are fit jointly. We adopt the QP kernel for all \texttt{G} models.

%%%%%%%%%%%%%%%% SECTION 4.3 PRIORS %%%%%%%%%%%%%%%%%
%
\subsection{Priors}\label{subsec:priors}

\begin{table*}  
\centering
\caption{Prior functions used in all model families. 
$\mathcal{N}[\mu,\sigma]$ denotes a Normal distribution with mean $\mu$ and standard deviation $\sigma$. 
$\mathcal{N_H}[\sigma]$ denotes a Half-Normal distribution with standard deviation $\sigma$. 
$\mathcal{U}[a,b]$ denotes a Uniform distribution between lower bound $a$ and upper bound $b$. 
$\mathcal{F}[x]$ denotes a fixed value.
$\mu_{\mathrm{arch}}$ and $\sigma_{\mathrm{arch}}$ denote the published value and quoted uncertainty adopted from the literature (retrieved via the NASA Exoplanet Archive). 
For systems without published constraints on quasi-periodic GP hyperparameters, the prior on the GP period $P_\mathrm{GP}$ is informed by the stellar rotation period $P_{\mathrm{rot}}$ if available, and otherwise by the uniform prior of \citet{Osborne.etal2025_HomogeneousPlanetMasses}, in which $P_{\mathrm{GP}}$ is uniform between 0 and a maximum value $P_{\mathrm{GP,max}}$ that depends on stellar effective temperature $T_{\mathrm{eff}}$\!: $P_{\mathrm{GP,max}} = 60, 50, 40,$ and $20$~d for $T_{\mathrm{eff}} < 4000\,\mathrm{K}$, $4000$--$5000\,\mathrm{K}$, $5000$--$6000\,\mathrm{K}$, and $>6000\,\mathrm{K}$ respectively. The remaining three hyperparameters ($\lambda_{\mathrm{e}}$, $\lambda_{\mathrm{p}}$ and $A_{\mathrm{GP}}$) likewise use the uniform priors of \citet{Osborne.etal2025_HomogeneousPlanetMasses}. For TOI-544, broader uniform priors on $P_\mathrm{GP}$ and $\lambda_\mathrm{e}$ are used; see Section~\ref{subsec:priors}.}
\label{tab:priors}
  \begin{tabular}{llp{3.3in}}
  \hline
  Parameter & Prior & Notes \\
  \hline
  
  \multicolumn{3}{l}{\textbf{Orbital Parameters}} \\
  
  Period $P$ [d] 
    & $\mathcal{N}[\mu_{\mathrm{arch}},\sigma_{\mathrm{arch}}]$ 
    & Where $\mu_{\mathrm{arch}}$ and $\sigma_{\mathrm{arch}}$ are published values and uncertainties from the NASA Exoplanet Archive. \\
    
  Semi-amplitude $K$ [\ms] 
    & $\mathcal{U}[0,100]$ 
    &  \\
    
  Transit time $T_{C}$ [BTJD] 
    & $\mathcal{N}[\mu_{\mathrm{arch}},\sigma_{\mathrm{arch}}]$ 
    & Where $\mu_{\mathrm{arch}}$ and $\sigma_{\mathrm{arch}}$ are published values and uncertainties from the NASA Exoplanet Archive. \\
  
    \hline
    \multicolumn{3}{l}{\textbf{Eccentricity Parameters} -- Circular \texttt{C} models only} \\
    
    Eccentricity $e$
      & $\mathcal{F}[0]$
      & \\
    
    Argument of periastron $\wstar$ [rad]
      & $\mathcal{F}[\pi/2]$
      & \\
    
    \hline
    \multicolumn{3}{l}{\textbf{Eccentricity Parameters} -- Uniform \texttt{U} models only} \\
    
    $\secosw$
      & $\mathcal{U}[-1,1]$
      & Sampling parameter, transformed to $(e, \wstar)$ for reporting. \\
    
    $\sesinw$
      & $\mathcal{U}[-1,1]$
      & Sampling parameter; transformed to $(e,\wstar)$ for reporting. \\
    
    \hline
    \multicolumn{3}{l}{\textbf{Eccentricity Parameters} -- Half-Normal \texttt{H} models only} \\
    
    Eccentricity $e$
      & $\mathcal{N_H}[0.32]$
      & Single-planet eccentricity prior from \citet{VanEylen.etal2019_OrbitalEccentricitySmall}. Sampled in $(\secosw,\sesinw)$; prior defined in $(e,\wstar)$. \\
    
    Argument of periastron $\wstar$ [rad]
      & $\mathcal{U}[-\pi,\pi]$
      & Sampled in $(\secosw,\sesinw)$; prior defined in $(e,\wstar)$. \\

  \hline
  
  \multicolumn{3}{l}{\textbf{Per-instrument Parameters}} \\
  
  Systemic velocity $\gamma$ [\ms] 
    & $\mathcal{U}[\mathrm{RV}_{\min},\mathrm{RV}_{\max}]$ 
    & Where $\mathrm{RV}_{\min}$ and $\mathrm{RV}_{\max}$ are the minimum and maximum observed radial velocities for that system. \\
    
  Jitter $\sjit$ [\ms]  
    & $\mathcal{U}[0,200]$ 
    &  \\
  
  \hline
  
  \multicolumn{3}{l}{\textbf{Trend Parameters}} \\
  
  Linear trend $\dot{\gamma}$ [\msd] 
    & $\mathcal{U}[-0.1,0.1]$ 
    & Used in \texttt{1} and \texttt{2} models; otherwise $\mathcal{F}[0]$. \\
    
  Quadratic trend $\ddot{\gamma}$ [\msdd] 
    & $\mathcal{U}[-0.001,0.001]$ 
    & Used in \texttt{2} models only; otherwise $\mathcal{F}[0]$. \\
  
  \hline
  
  \multicolumn{3}{l}{\textbf{Gaussian Process Hyperparameters} -- Gaussian Process \texttt{G} models only} \\
  
  Amplitude $A_\mathrm{GP}$ [\ms] 
    & $\mathcal{N}[\mu_{\mathrm{arch}},\sigma_{\mathrm{arch}}]$ or $\mathcal{U}[0,100]$ 
    & Uniform prior used if no literature values available. \\
  
  GP period $P_{\mathrm{GP}}$ [d] 
    & $\mathcal{N}[\mu_{\mathrm{arch}},\sigma_{\mathrm{arch}}]$ or $\mathcal{U}[0,P_{\mathrm{GP,max}}]$ 
    & Uniform prior used if no literature values available. \\
  
  Evolution timescale $\lambda_\mathrm{e}$ [d] 
    & $\mathcal{N}[\mu_{\mathrm{arch}},\sigma_{\mathrm{arch}}]$ or $\mathcal{U}[1,2P_{\mathrm{GP,max}}]$ 
    & Uniform prior used if no literature values available. \\
  
  Inverse of the harmonic complexity $\lambda_\mathrm{p}$ 
    & $\mathcal{N}[\mu_{\mathrm{arch}},\sigma_{\mathrm{arch}}]$ or $\mathcal{U}[0.01,2]$ 
    & Uniform prior used if no literature values available. \\
  
  \hline
  \end{tabular}
  
\end{table*}

The priors for parameters (and hyperpriors on GP hyperparameters) are listed in Table~\ref{tab:priors}. For planetary orbital period $P$ and mid-transit time $T_C$, we use informative priors based on literature values obtained from the NASA Exoplanet Archive's Planetary Systems Composite Parameters table, queried via the Table Access Protocol. For RV semi-amplitude $K$ we adopt a broad uninformative uniform distribution prior between 0\,\ms and 100\,\ms, this being much wider than the range of the observed radial velocities for the target systems.

For circular \texttt{C} models, eccentricity $e$ is fixed at 0, and argument of periastron $\wstar$ is fixed at $\pi/2$ by convention \citep{Eastman.etal2013_EXOFASTFastExoplanetary}. For eccentric orbit \texttt{U} and \texttt{H} models, to mitigate the Lucy--Sweeney bias \citep{Lucy.Sweeney1971_SpectroscopicBinariesCircular} caused by the boundary condition at $e=0$, and to shorten convergence time, we sample in the commonly-used parameterisation $\secosw$ and $\sesinw$ \citep{Anderson.etal2010_WASP30b61MJup, Eastman.etal2013_EXOFASTFastExoplanetary} rather than in $e$ and $\wstar$ directly. For \texttt{U} models, the priors on both $\secosw$ and $\sesinw$ are uniform distributions between -1 and 1, while for \texttt{H} models we place a half-normal distribution prior on $e$ with a standard deviation $\sigma=0.32$ truncated to the physically allowed range $0 \leq e < 1$, and a uniform prior on $\wstar$ from $-\pi$ to $+\pi$. The choice of half-normal prior is motivated by previous studies investigating transiting planets around FGK stars and M-dwarfs, finding the single-planet and multi-planet eccentricity populations are best described by separate underlying distributions \citep{Xie.etal2016_ExoplanetOrbitalEccentricities, VanEylen.etal2019_OrbitalEccentricitySmall, Sagear.Ballard2023_OrbitalEccentricityDistribution, Stevenson.etal2025_RVexoplanetEccentricitiesGood}; we adopt the half-normal prior with $\sigma=0.32$ for single-transiting planet systems from \citet{VanEylen.etal2019_OrbitalEccentricitySmall}. We highlight that for Bayesian evidence estimation, care must be taken in the eccentric models to ensure the eccentricity priors are correctly normalised (see Appendix~\ref{app:app1}). 

For the per-instrument parameters (the systemic velocity offset $\gamma_\mathrm{inst}$ and the additional jitter noise term $\sigma_{\mathrm{jit,inst}}$) we adopt uniform priors. The minimum and maximum RV values are used as boundaries for $\gamma_\mathrm{inst}$, and $\sigma_{\mathrm{jit,inst}}$ has a large uninformative prior from 0 to 200\,\ms.

For the \texttt{1} and \texttt{2} models with linear and quadratic trends respectively, we adopt priors on the trend coefficients of $\mathcal{U}[-0.1,0.1]$ \msd for linear coefficient $\dot{\gamma}$ and $\mathcal{U}[-0.001,0.001]$ \msdd for quadratic coefficient $\ddot{\gamma}$.

For GP hyperparameters, if there are published values and uncertainties for the hyperparameters of a QP kernel (or, for stellar rotation period $P_\mathrm{rot}$ in place of GP period $P_\mathrm{GP}$), we adopt those; else we adopt the same broad uniform priors used in \citet{Osborne.etal2025_HomogeneousPlanetMasses} where GP period $P_\mathrm{GP}$ is uniform between 0 and a maximum value based on stellar effective temperature $T_\mathrm{eff}$: $P_\mathrm{GP,max} = 60, 50, 40,$ and $20$~d for $T_\mathrm{eff} < 4000\,\mathrm{K}$, $4000$--$5000\,\mathrm{K}$, $5000$--$6000\,\mathrm{K}$, and $>6000\,\mathrm{K}$ respectively. The prior on active region evolution timescale $\lambda_\mathrm{e}$ is then uniform between $1$ and $2P_\mathrm{GP,max}$~d, the prior on inverse harmonic complexity $\lambda_\mathrm{p}$ is uniform between $0.01$ and $2$, and the prior on GP amplitude $A_\mathrm{GP}$ is uniform between $0$ and $100$\,\ms.

For the two-planet models in TOI-544, the same prior structure applies independently to each planet: $P_{\mathrm{b}}$, $K_{\mathrm{b}}$, $T_{C,{\mathrm{b}}}$ and the eccentricity parameters for planet~b follow the priors above, with independent priors of the same form assigned to the corresponding parameters for planet~c. The HARPS-N instrument offset $\gamma_{\mathrm{HARPS-N}}$ and jitter $\sigma_{\mathrm{jit}, \mathrm{HARPS-N}}$ follow the same uniform prior scheme as the HARPS epochs. The GP amplitude $A_\mathrm{GP}$ and inverse harmonic complexity $\lambda_\mathrm{p}$ follow the standard priors listed above. However for $P_\mathrm{GP}$ we adopt $\mathcal{U}[15, 25]$\,d, following \citet{Osborne.etal2024_TOI544PotentialWaterworld}, which encompasses the stellar rotation period ($P_\mathrm{rot} \sim 19.4$\,d) while avoiding the first and second harmonics at ${\sim}9.8$ and ${\sim}6.5$\,d. For $\lambda_\mathrm{e}$ we adopt a wide uniform prior $\mathcal{U}[6, 196]$\,d, with the lower bound set to exclude unphysically short coherence timescales relative to the stellar rotation period. As \citet{Osborne.etal2024_TOI544PotentialWaterworld} report GP hyperparameter values only for their two-planet model, we adopt broad uniform priors on the $P_\mathrm{GP}$ and $\lambda_\mathrm{e}$ hyperparameters, to ensure that in one-planet \texttt{G} models the sampler can more fully explore the range of timescales when the signal from TOI-544~c is not being modelled by a second Keplerian.

%%%%%%%%%%%%%%%% SECTION 4.4 FITTING PROCEDURE %%%%%%%%%%%%%%%%%
%
\subsection{Fitting Procedure}\label{subsec:fitting}
We sample the posterior using MCMC. Chains are considered converged when their length exceeds 50 times the estimated autocorrelation length $\hat{\tau}$ for all parameters, once $\hat{\tau}$ has been stable ($\Delta\hat{\tau}<1\%$) compared to the estimate 1000 steps ago. We use 200 walkers, apply a thin factor of 10 and then retain the final 1000 iterations, yielding 200\,000 posterior samples for each free parameter.

We then compute the Bayesian evidence $\Z$ for each model by applying the LHME to the posterior samples. The posterior chains are divided into training and testing sets on a 25\%/75\% split (50 chains for training, 150 chains for testing), which \textsc{harmonic} then uses to learn the target distribution required to estimate the evidence (see Section~\ref{subsec:harmonic}).

To ensure reliable evidence estimates, we calculate several diagnostic metrics as recommended by \citet{McEwen.etal2023_MachineLearningAssisted}. As well as the reciprocal evidence estimate $\hat{\rho}$, we can also calculate $\hat{\sigma}^2$, the estimator for $\mathrm{var}(\hat{\rho})$, and $\hat{\nu}^4$, the estimator of $\mathrm{var}(\hat{\sigma}^2)$. We can also estimate the kurtosis $\hat{\kappa}$, which should be approximately $3$ for a well-behaved estimator with Gaussian-like properties. We therefore exclude models that fail any of three checks: firstly, we exclude models where $|\hat{\kappa} - 3| > 3$, indicating the normalising flow has failed to adequately learn a distribution that approximates (but is constrained within the tails of) the posterior distribution. Secondly, we verify the variance stability by computing the ratio $\hat{\nu}^4 / \hat{\sigma}^2$, which should match the expected value for a reliable estimator; we exclude models where this ratio differs from the expected value by more than $0.1$. Third, \textsc{harmonic} includes additional internal diagnostics to verify the quality of the evidence estimate, and we exclude models where these diagnostics fail. Full details of these diagnostics are provided in \citet{McEwen.etal2023_MachineLearningAssisted}.

\begin{figure}
 \includegraphics[width=\columnwidth]{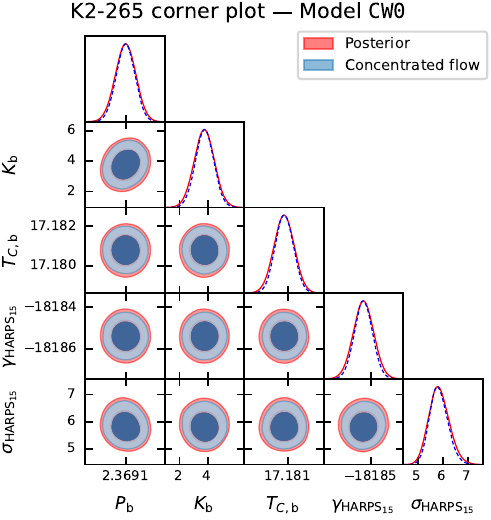}
 \caption{An example of a \textsc{harmonic} diagnostic corner plot, showing the MCMC posterior samples distributions (red) for each fitted parameter, and the LHME concentrated flow (blue). The flow successfully learns the shape of the posterior distribution while the tails remain contained within it, satisfying the requirement for reliable evidence estimation with the learned harmonic mean estimator (see Section~\ref{subsubsec:LHME}). The model shown here is the \texttt{CW0} model of K2-265 (TIC~146364192).}
 \label{fig:harmonic_corner}
\end{figure}

Finally, we also manually inspect posterior chains and corner plots, and exclude any models where the MCMC chains have clearly failed to converge, or where the LHME has visibly failed to learn the posterior samples distribution and shape, often due to multimodal posteriors. Fig.~\ref{fig:harmonic_corner} shows an example of the \textsc{harmonic} concentrated flow successfully approximating, and being contained within, the posterior samples of a model.
%

%%%%%%%%%%%%%%%% SECTION 5 RESULTS %%%%%%%%%%%%%%%%%
%
\section{Results}\label{sec:results}
We wished to investigate the performance of the LHME on a variety of different exoplanet systems by fitting our full set of model variants to the seven example systems using MCMC. Not every model produces well-behaved unimodal chains/posteriors suitable for Bayesian model comparison with the LHME: MCMC chains may fail to converge, or posteriors may exhibit multimodal or complex geometry that the normalising flows in \textsc{harmonic} cannot reliably approximate. In our analysis, this arises most commonly when uniform eccentricity prior models produce bimodal posteriors due to spurious high-eccentricity solutions fitting to outlier datapoints \citep{Hara.etal2019_BiasRobustnessEccentricity, Osborne.etal2025_HomogeneousPlanetMasses}, but also in poorly constrained GP hyperparameters, particularly timescales. Such models are identified and excluded from our analysis, using the diagnostic metrics and manual inspection as described in Section~\ref{subsec:fitting}.

Crucially, even in cases with unconverged parameters, traditional metrics such as $\chi^2$, log-likelihood, AIC, and BIC can still be computed and will return numerical values, even if the fitting failed and the model should be rejected. This highlights a key danger of relying on these statistics alone without thorough inspection of chains and posterior distributions, and the resulting RV fits and their residuals. 

\begin{figure*}
 \includegraphics[width=0.95\textwidth]{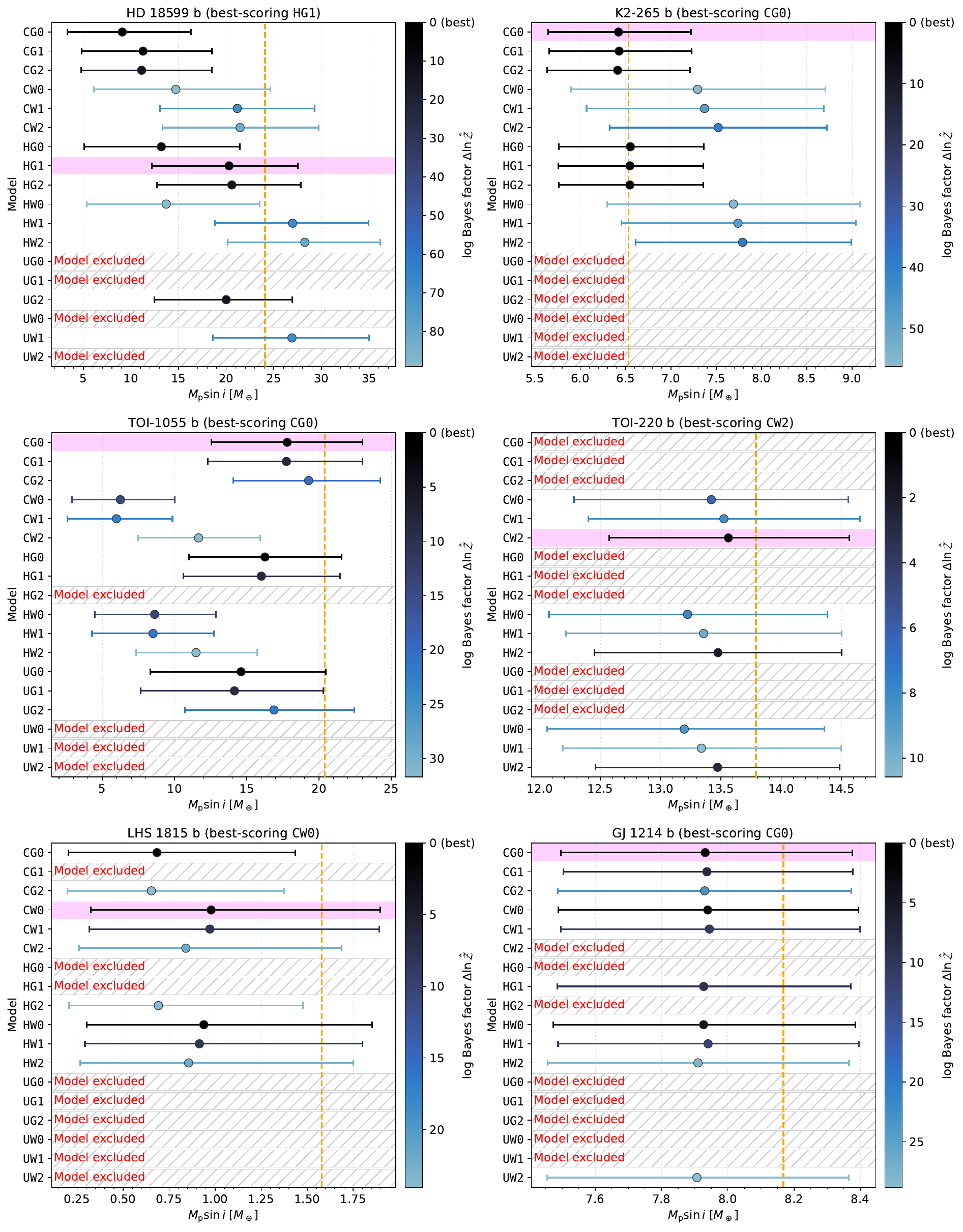}
 \caption{The derived minimum mass $\Mpsini$ values and uncertainties for all of the valid models in the six single-planet systems. The orange vertical lines show the derived minimum mass from the literature. Models that failed \textsc{harmonic} diagnostics (see Section~\ref{subsec:fitting}) are excluded, shown by the grey hatched areas. The row for the best-scoring model is highlighted in pink. The blue colourbar scale shows the log Bayes factor $\dlnZ$ relative to the best-scoring model for that system (lower is better; the best-scoring model has $\dlnZ=0$).}
 \label{fig:mpsini_grid}
\end{figure*}
\begin{figure*}
    \includegraphics[width=0.94\textwidth]{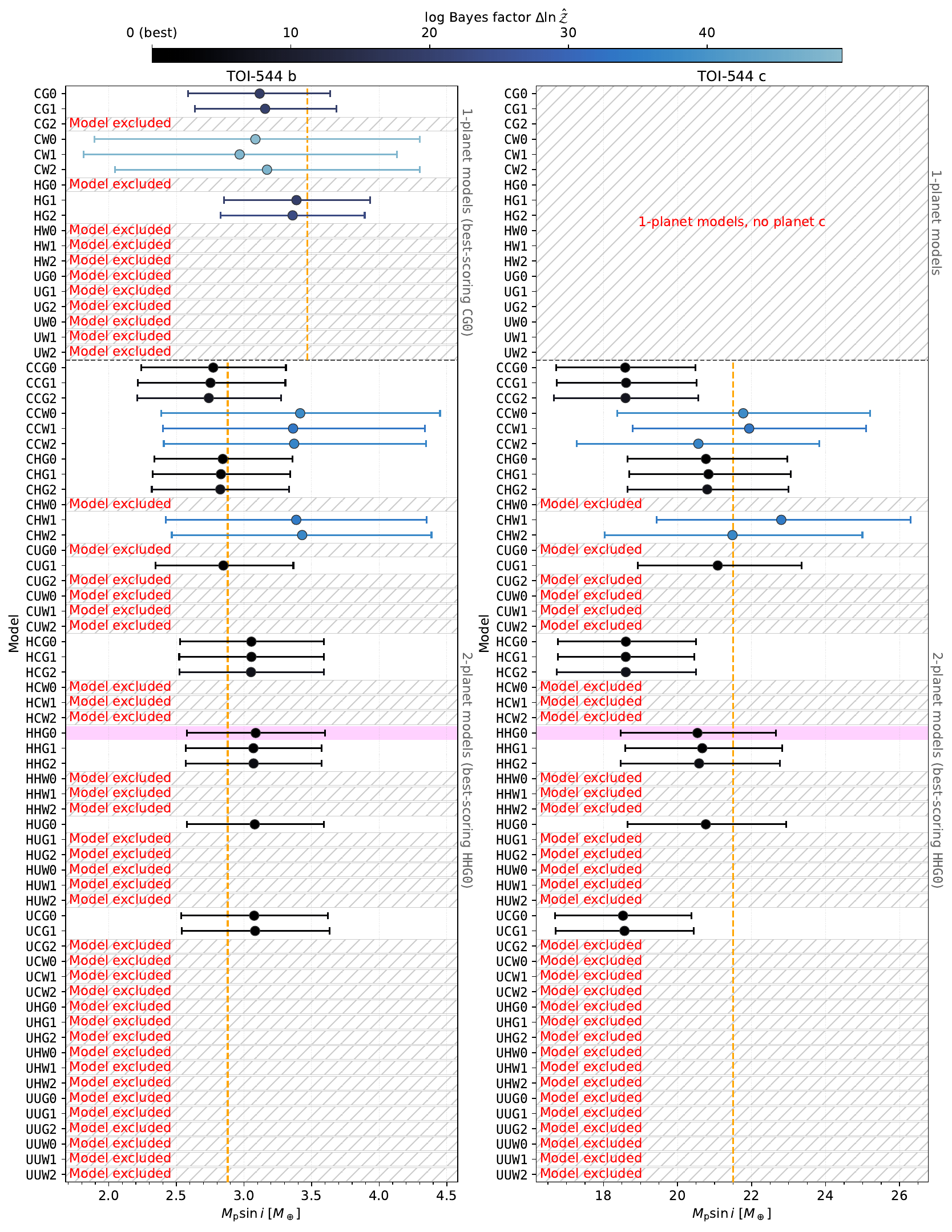}
    \caption{The derived minimum mass $\Mpsini$ values and uncertainties for all of the valid models for TOI-544, shown for planet~b (left panel) and planet~c (right panel). The orange vertical lines show the derived minimum masses from \citet{Osborne.etal2024_TOI544PotentialWaterworld}. Both the 18 one-planet and 54 two-planet models are shown. Models that failed \textsc{harmonic} diagnostics (see Section~\ref{subsec:fitting}) are excluded, shown by the grey hatched areas. For one-planet models, no planet~c signal is fitted; this is indicated by grey hatching. The row for the best-scoring model is highlighted in pink. The blue colourbar shows the log Bayes factor $\dlnZ$ relative to the best-scoring model \texttt{HHG0} (lower is better; \texttt{HHG0} has $\dlnZ=0$).}
    \label{fig:mpsini_grid_toi544}
\end{figure*}

We present results for each system in turn, identifying the preferred model by Bayesian evidence and comparing the model characteristics and masses to previous analyses in the literature. The derived minimum masses $\Mpsini$ for all valid models for the six single-planet systems are shown in Fig.~\ref{fig:mpsini_grid}, and for TOI-544 in Fig.~\ref{fig:mpsini_grid_toi544}, with the best-scoring model for each system highlighted, and the reference minimum mass $\Mpsini$ from the literature indicated for comparison. Tables showing the $\ln\Z$ estimates and log Bayes factor $\dlnZ$ between models for each system are shown in Appendix~\ref{app:extra_tables}.

%%%%%%%%%%%%%%%% SECTION 5.1 HD 18599 RESULTS %%%%%%%%%%%%%%%%%
%
\subsection{HD~18599 (TIC~207141131): an example of trend model comparison}\label{subsec:tic207_results}
\begin{figure*}
  \includegraphics[width=\textwidth]{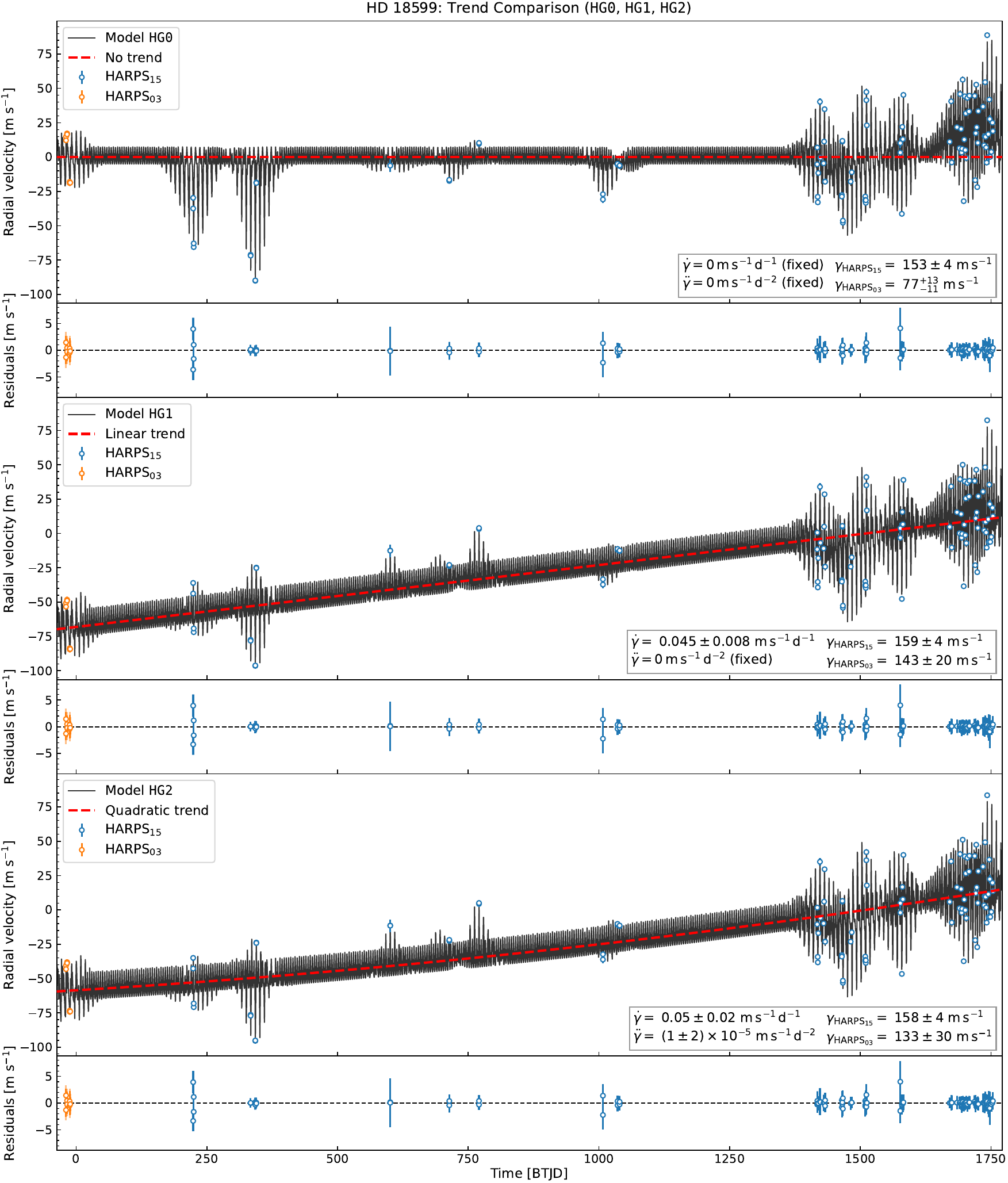}  
  \caption{
      Radial velocities for HD~18599 (TIC~207141131), with orange circles for $\mathrm{HARPS}_{03}$ and blue circles for $\mathrm{HARPS}_{15}$. Error bars show observed uncertainties, with lighter extensions indicating the contribution from the $\sjit$ jitter term added in quadrature. 
      The no-trend \texttt{HG0} model (upper panels), linear trend \texttt{HG1} model (middle panels) and quadratic trend \texttt{HG2} model (lower panels) are shown as the black curve and grey shaded region, representing the median and 68\% credible interval of the RV signal from the posterior samples. The trend model contribution to the RV is shown as the red dashed line.
      All models have a half-normal prior on eccentricity, and use a quasi-periodic GP, differing only in the long-term trend. 
      The RV offsets $\gamma$ and the linear and quadratic trend coefficients $\dot\gamma$ and $\ddot\gamma$ are labelled on the panels. 
      The lower panels show residuals relative to each RV model above.
      The adopted linear trend model \texttt{HG1} was strongly preferred over the no trend model \texttt{HG0} with log Bayes factor $\dlnZ=6.3$, and over the quadratic trend model \texttt{HG2} by $\dlnZ=12.7$.
      }
  \label{fig:207141131_trend_comparison}
\end{figure*}
\begin{table}
  \centering
  \setlength{\tabcolsep}{3pt}
  \caption{Fitted and derived parameters for the adopted model (\texttt{HG1}) for HD 18599 (TIC 207141131). Reported values and uncertainties are the median value and 68\% credible interval.}
  \label{tab:table2_tic207141131}
  \begin{tabular}{@{}lc@{}}
    \hline
    Parameter & Value \\
    \hline
    \multicolumn{2}{@{}l}{\textbf{Planet b parameters}} \\[2pt]
    Orbital period, $P$ [d] & $4.137437_{-0.000098}^{+0.000099}$ \\
    Transit time, $T_C$ [BTJD] & $2111.7395 \pm 0.0007$ \\
    $\secosw$ & $0.52_{-0.20}^{+0.08}$ \\
    $\sesinw$ & $-0.2_{-0.3}^{+0.4}$ \\
    Eccentricity, $e$ & $0.38_{-0.14}^{+0.15}$ \\
    Argument of periastron, $\wstar$ [$^\circ$] & $316_{-300}^{+30}$ \\
    RV semi-amplitude, $K$ [\ms] & $10 \pm 4$ \\
    Minimum mass, $\Mpsini$ [$M_\oplus$] & $20_{-8}^{+7}$ \\[4pt]
    \multicolumn{2}{@{}l}{\textbf{Instrument parameters}} \\[2pt]
    RV offset, $\gamma_{\mathrm{HARPS}_{03}}$ [\ms] & $143 \pm 20$ \\
    RV jitter, $\sigma_{\mathrm{HARPS}_{03}}$ [\ms] & $1.5_{-1.0}^{+2.0}$ \\
    RV offset, $\gamma_{\mathrm{HARPS}_{15}}$ [\ms] & $159 \pm 4$ \\
    RV jitter, $\sigma_{\mathrm{HARPS}_{15}}$ [\ms] & $0.3_{-0.2}^{+0.4}$ \\[4pt]
    \multicolumn{2}{@{}l}{\textbf{Trend parameters}} \\[2pt]
    Linear RV trend, $\dot{\gamma}$ [\msd] & $0.045 \pm 0.008$ \\[4pt]
    \multicolumn{2}{@{}l}{\textbf{GP hyperparameters}} \\[2pt]
    GP amplitude, $A_{\mathrm{GP}}$ [\ms] & $24 \pm 2$ \\
    GP period, $P_{\mathrm{GP}}$ [d] & $8.72 \pm 0.03$ \\
    Active region timescale, $\lambda_{\mathrm{e}}$ [d] & $24 \pm 4$ \\
    Inverse harmonic complexity, $\lambda_{\mathrm{p}}$ & $0.15 \pm 0.02$ \\[4pt]
    \hline
  \end{tabular}
\end{table}
\begin{figure*}
\centering
\begin{subfigure}{0.49\textwidth}
    \centering
    \includegraphics[width=\linewidth]{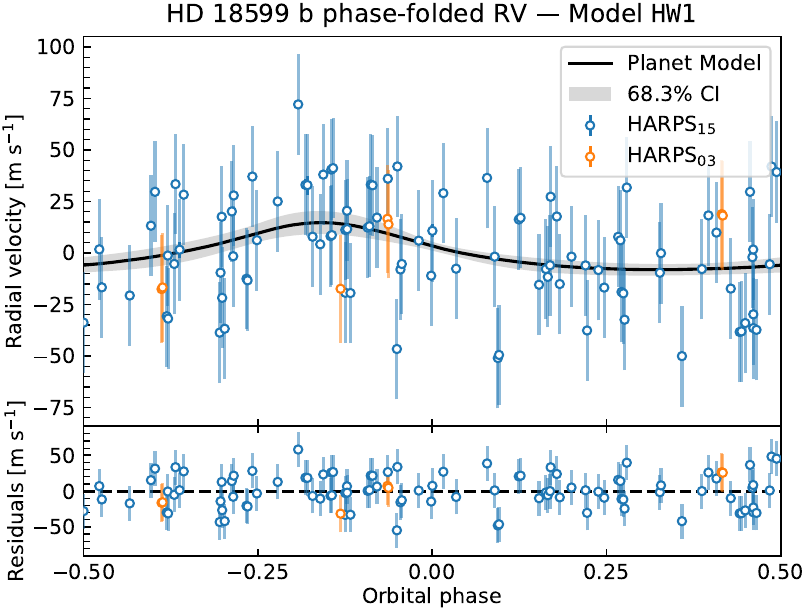}
    \caption{White-noise model \texttt{HW1}.}
    \label{fig:TIC207141131_posterior_phase_HW1}
\end{subfigure}
\hfill
\begin{subfigure}{0.49\textwidth}
    \centering
    \includegraphics[width=\linewidth]{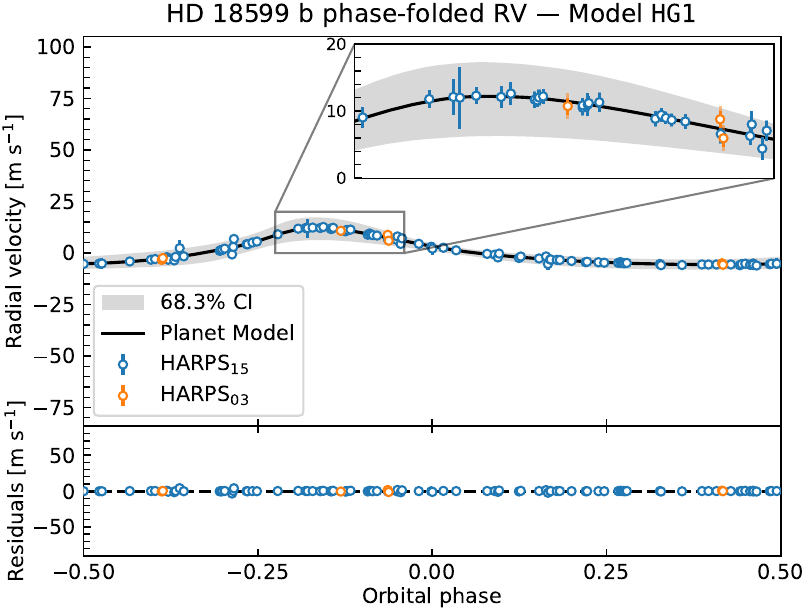}
    \caption{GP model \texttt{HG1}, on the same axis limits as (a); inset shows zoomed region.}
    \label{fig:TIC207141131_posterior_phase_HG1}
\end{subfigure}
\caption{
      Phase-folded radial velocities for HD~18599 (TIC~207141131), comparing the best-scoring white-noise model \texttt{HW1} (panel a) with the overall best-scoring, preferred model \texttt{HG1} (panel b), which uses a Gaussian process.
      In both panels, orange circles show $\mathrm{HARPS}_{03}$ and blue circles $\mathrm{HARPS}_{15}$; error bars show the observed uncertainties, with lighter extensions indicating the contribution from the $\sjit$ jitter term added in quadrature.
      The black curve and grey shaded region show the median and 68\% credible interval of the planetary RV signal from the posterior samples.
      All other fitted model components -- the instrument-specific systemic offsets $\gamma$, the long-term trend (linear or quadratic, in \texttt{1} and \texttt{2} models), and the Gaussian process (in \texttt{G} models) -- have been subtracted from both the data and the model, leaving the phase-folded planetary signal; the lower sub-panels show the residuals.
      Panel (b) is plotted on the same y-axis ranges as panel (a) for a visual comparison of the fits, with the inset showing a zoomed region of the \texttt{HG1} phase-folded RV curve.
      The \texttt{HW1} model exhibits substantially larger residuals and jitter $\sjit$ than \texttt{HG1}, and is strongly disfavoured by a log Bayes factor of $\dlnZ = 68.0$; the semi-amplitudes $K$ and derived minimum masses $\Mpsini$ from the two models are consistent within $1\sigma$ (see Section~\ref{sub:tic207_noise_comparison} and Table~\ref{tab:table1_tic207141131}).
      }
\label{fig:TIC207141131_posterior_phase_comparison}
\end{figure*}

Long-term velocity trends can have several origins, with the most significant being the gravitational influence of a long-period companion. One such example is HD~18599, a K2V star \citep{Gray.etal2006_ContributionsNearbyStars} which hosts a transiting sub-Neptune planet with orbital period $P = 4.137$\,d. The system was characterised in a joint HARPS RV and \textit{TESS} photometry fit by \citet{Desidera.etal2023_TOI179YoungSystem}, who used a quasi-periodic Gaussian process to model stellar activity alongside fitting for a linear velocity trend, which they attribute to the Keplerian orbit of a long-period outer companion HD~18599~B, likely a high-mass brown dwarf or low-mass star. They used nested sampling to compare two models with linear and quadratic long-term trends, finding a strong preference for the linear trend over the quadratic by $\dlnZ = 10.2$.

In our RV analysis, of the 18 models evaluated, 14 passed the \textsc{harmonic} diagnostic tests and had reliable evidence estimates; these are listed in  Table~\ref{tab:table1_tic207141131}. We find that the preferred model is \texttt{HG1} (half-normal eccentricity prior, GP noise model, linear trend), consistent with \citet{Desidera.etal2023_TOI179YoungSystem}. For the trend, we also find that linear models are strongly preferred over quadratic trend models ($\dlnZ = 12.7$ for \texttt{HG1} over \texttt{HG2}), while the linear trend is also strongly preferred over no trend ($\dlnZ = 6.3$ for \texttt{HG1} over \texttt{HG0}). The effects of each trend choice on the RV are illustrated in Fig.~\ref{fig:207141131_trend_comparison}. The preference for a GP over a white noise model is strong ($\dlnZ = 68.0$ for \texttt{HG1} over \texttt{HW1}). 
   
The eccentricity treatment is not as decisive: while we report $e = 0.38^{+0.15}_{-0.14}$, consistent with $e = 0.34^{+0.07}_{-0.09}$ from \citet{Desidera.etal2023_TOI179YoungSystem}, we note that the eccentric \texttt{HG1} model is only weakly preferred over the circular \texttt{CG1} model by $\dlnZ = 1.7$. HD~18599~b was also independently characterised by \citet{Vines.etal2023_DenseMiniNeptuneOrbiting} using HARPS and FEROS data, who report $e = 0.2^{+0.1}_{-0.2}$, which is consistent with a circular orbit within $1\sigma$. Together, the range of eccentricity values across two independent analyses, combined with the only weak Bayesian model preference for eccentric over circular models, suggests that further observations are needed to robustly constrain the orbital eccentricity of HD~18599~b.

The parameters for our preferred model \texttt{HG1} can be found in Table~\ref{tab:table2_tic207141131}. We measure a semi-amplitude $K = 10 \pm 4$\,\ms and derive minimum planetary mass $\Mpsini = 20^{+7}_{-8}\,M_{\earth}$. This is consistent with \citet{Desidera.etal2023_TOI179YoungSystem} who report $K = 11.3^{+3.3}_{-3.6}$\,\ms and $\Mp = 24.1^{+7.1}_{-7.7} \, M_{\earth}$, and with \citet{Vines.etal2023_DenseMiniNeptuneOrbiting} who report $K = 11 \pm 3$\,\ms and $\Mp = 25.5 \pm 4.6 \, M_{\earth}$. The linear trend coefficient $\dot{\gamma} = 0.045 \pm 0.008$\,\msd is in agreement with the values of $\dot{\gamma} = 0.048 \pm 0.008$\,\msd reported by \citet{Desidera.etal2023_TOI179YoungSystem}, and with $\dot{\gamma} = 0.047 \pm 0.008$\,\msd reported by \citet{Vines.etal2023_DenseMiniNeptuneOrbiting}. The RV data phase-folded on the orbital period for the \texttt{HG1} model is shown in Fig.~\ref{fig:TIC207141131_posterior_phase_HG1}.

%%%%%%%%%%%%%%%% SECTION 5.1.1 NOISE MODEL COMPARISON %%%%%%%%%%%%%%%%%
%
\subsubsection{Noise model comparison}\label{sub:tic207_noise_comparison}

In the phase-folded RV plot for the preferred GP model \texttt{HG1} (Fig.~\ref{fig:TIC207141131_posterior_phase_HG1}) the posterior RV 68\% credible interval on the planetary RV model is relatively wide compared to both the scatter of the residuals and the jitter values ($\sigma_{\mathrm{HARPS}_{03}} = 1.5^{+2.0}_{-1.0}$\,\ms, $\sigma_{\mathrm{HARPS}_{15}} = 0.3^{+0.4}_{-0.2}$\,\ms). To investigate if the GP is overfitting, we compare \texttt{HG1} to the best-scoring white noise \texttt{W} model, \texttt{HW1}. In the phase-folded RV plot for \texttt{HW1} (Fig.~\ref{fig:TIC207141131_posterior_phase_HW1}), the 68\% credible interval is slightly larger than for \texttt{HG1} (up to $1.5\times$ at orbital phase $0.0$--$0.2$), but the jitter values are substantially larger ($\sigma_{\mathrm{HARPS}_{03}} = 26^{+12}_{-7}$\,\ms, $\sigma_{\mathrm{HARPS}_{15}} = 24 \pm 2$\,\ms), far exceeding the typical observed HARPS formal uncertainties of $\sim$1--3\,\ms. The residuals are also significantly larger for \texttt{HW1} (where they often exceed the planetary Keplerian semi-amplitude $K$) than they are for \texttt{HG1}.

In the white noise model \texttt{HW1}, only the planetary Keplerian and the linear trend contribute to the modelled RVs -- it has no mechanism to account for correlated noise. Therefore, with structured stellar variability in the observations, the only mechanism available is to increase the jitter. The GP in the \texttt{HG1} model adds flexibility to the overall RV model, and in particular the quasi-periodic kernel function is well-suited to modelling rotationally modulated stellar activity \citep{Aigrain.Foreman-Mackey2023_GaussianProcessRegression}. The stellar variability is now being modelled by the GP component rather than as jitter, hence significantly lower jitter values.

\texttt{HW1} also recovers a higher semi-amplitude ($K = 13 \pm 4$\,\ms\ versus $K = 10 \pm 4$\,\ms\ for \texttt{HG1}) with a correspondingly higher minimum mass ($\Mpsini = 27 \pm 8\,M_{\earth}$ versus $20^{+7}_{-8}\,M_{\earth}$), both being consistent within 1$\sigma$. This suggests that the white noise model may be partly attributing stellar variability to the planetary Keplerian signal, albeit not at a significant level. HD~18599 is known to be an active star, and \citet{Desidera.etal2023_TOI179YoungSystem} consider their use of the QP GP justified by recovering similar stellar rotation period and decay timescales in both their \textit{TESS} photometry and RV analyses. Our values for the GP hyperparameters ($A_{\mathrm{GP}}$, $P_{\mathrm{GP}}$, $\lambda_{\mathrm{e}}$ and $\lambda_{\mathrm{p}}$) are consistent with those reported in \citet{Desidera.etal2023_TOI179YoungSystem} within 1$\sigma$. We also find the Bayesian evidence strongly favours the overall best-scoring model \texttt{HG1} over the best-scoring white noise model \texttt{HW1} ($\dlnZ = 68.0$). Together with the smaller residuals and jitter visible in Fig.~\ref{fig:TIC207141131_posterior_phase_HG1} compared to Fig.~\ref{fig:TIC207141131_posterior_phase_HW1}, we believe that the QP kernel function is capturing genuine rotationally modulated stellar variability present in the RV signal. Although we cannot entirely rule out some degree of overfitting, we consider the use of the QP GP model over the white noise model justified, and we therefore adopt \texttt{HG1} as our preferred RV model for HD~18599. We further discuss the possibility of GP overfitting more generally in Section~\ref{subsec:discussion_implications}.

%%%%%%%%%%%%%%%% SECTION 5.2 K2-265 RESULTS %%%%%%%%%%%%%%%%%
%
\subsection{K2-265 (TIC~146364192): an example of eccentricity model comparison}\label{subsec:tic146_results}
\begin{figure}  
  \includegraphics[width=\columnwidth]{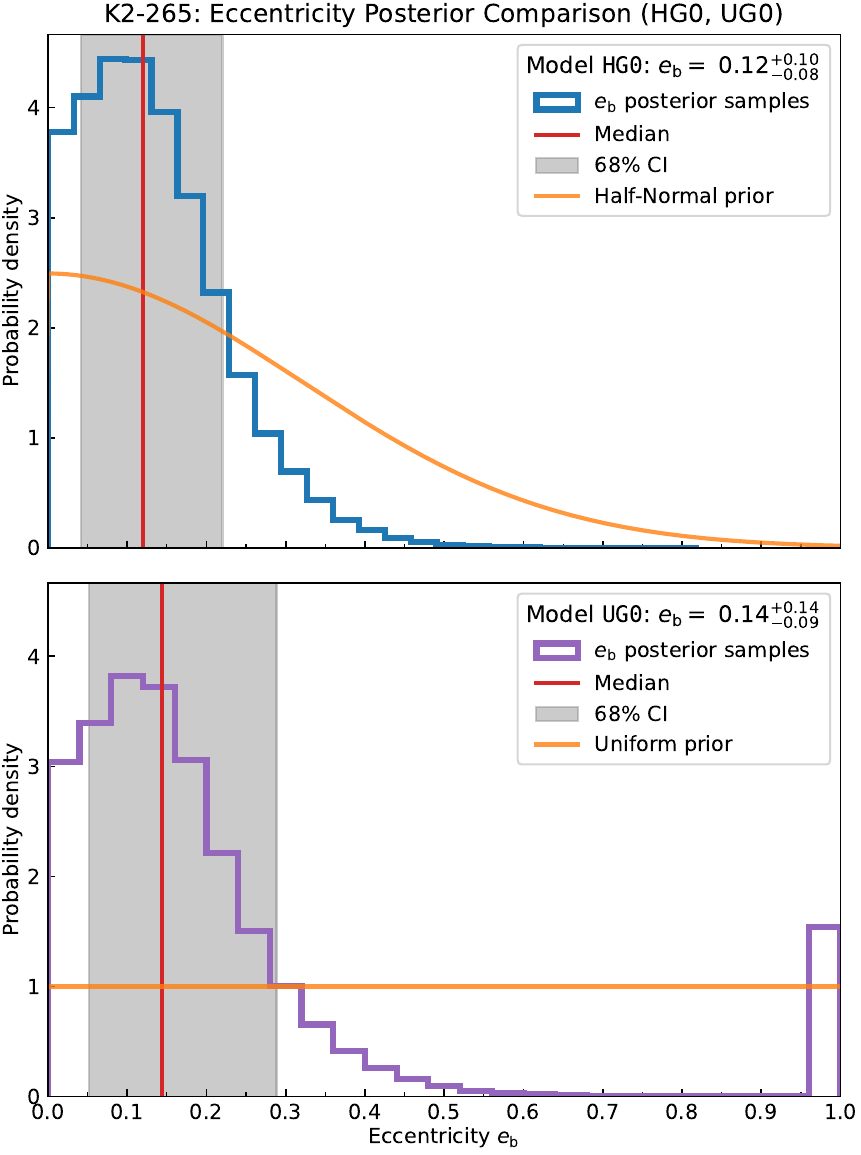}  
  \caption{
  Eccentricity posterior samples for the half-normal prior \texttt{HG0} (upper panel, blue) and uniform prior (lower panel, purple) \texttt{UG0} models, for K2-265 (TIC~146364192). 
  The prior distributions are shown with the orange lines. 
  The half-normal prior model \texttt{HG0} has a posterior peaked at $e \approx 0.12$, away from zero. The uniform prior model \texttt{UG0} (purple) is peaked at $e \approx 0.15$, but shows a significant large secondary peak near $e \approx 1$: these high-eccentricity solutions likely arise from the model spuriously fitting to outlier data \citep{Hara.etal2019_BiasRobustnessEccentricity, Osborne.etal2025_HomogeneousPlanetMasses}. 
  This model, along with the other five \texttt{U} models, failed the \textsc{harmonic} diagnostic tests, likely due to the multimodality of the posterior, and is excluded from the model comparison.}
  \label{fig:146_eccentricity_comparison} 
\end{figure}
\begin{table}
  \centering
  \setlength{\tabcolsep}{3pt}
  \caption{Fitted and derived parameters for the adopted model (\texttt{CG0}) for K2-265 (TIC 146364192). Reported values and uncertainties are the median value and 68\% credible interval.}
  \label{tab:table2_tic146364192}
  \begin{tabular}{@{}lc@{}}
    \hline
    Parameter & Value \\
    \hline
    \multicolumn{2}{@{}l}{\textbf{Planet b parameters}} \\[2pt]
    Orbital period, $P$ [d] & $2.36916 \pm 0.00008$ \\
    Transit time, $T_C$ [BTJD] & $17.1808 \pm 0.0005$ \\
    Eccentricity, $e$ & $0$ (fixed) \\
    Argument of periastron, $\wstar$ [$^\circ$] & $90$ (fixed) \\
    RV semi-amplitude, $K$ [\ms] & $3.3 \pm 0.4$ \\
    Minimum mass, $\Mpsini$ [$M_\oplus$] & $6.4 \pm 0.8$ \\[4pt]
    \multicolumn{2}{@{}l}{\textbf{Instrument parameters}} \\[2pt]
    RV offset, $\gamma_{\mathrm{HARPS}_{15}}$ [\ms] & $(-1.8187 \pm 0.0002) \times 10^{4}$ \\
    RV jitter, $\sigma_{\mathrm{HARPS}_{15}}$ [\ms] & $2.1 \pm 0.3$ \\[4pt]
    \multicolumn{2}{@{}l}{\textbf{GP hyperparameters}} \\[2pt]
    GP amplitude, $A_{\mathrm{GP}}$ [\ms] & $6.0_{-0.8}^{+0.9}$ \\
    GP period, $P_{\mathrm{GP}}$ [d] & $32.0 \pm 0.4$ \\
    Active region timescale, $\lambda_{\mathrm{e}}$ [d] & $36 \pm 7$ \\
    Inverse harmonic complexity, $\lambda_{\mathrm{p}}$ & $0.29 \pm 0.06$ \\[4pt]
    \hline
  \end{tabular}
\end{table}
\begin{figure}
    \centering
    \includegraphics[width=\columnwidth]{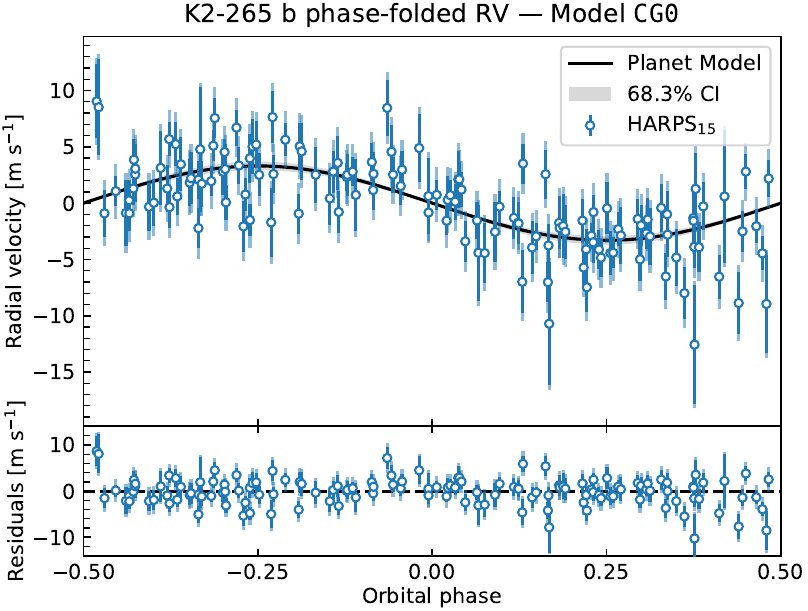}
    \caption{
        Phase-folded radial velocities for K2-265 (TIC 146364192). Same as Fig.~\ref{fig:TIC207141131_posterior_phase_comparison}, but with blue circles for $\mathrm{HARPS}_{15}$. The adopted model shown is \texttt{CG0}.
    }
    \label{fig:TIC146364192_posterior_phase}
\end{figure}
K2-265~b is a super-Earth, first discovered by \citet{Lam.etal2018_K2265TransitingRocky} reporting a mass $\Mp = 6.54 \pm 0.84\,M_{\earth}$, transiting the G8V host star on an orbital period of $P=2.37$\,d.
  
K2-265 provides a clear illustration of the pitfalls of uniform eccentricity priors: all six \texttt{U} model variants failed the \textsc{harmonic} diagnostic tests and are excluded from the model comparison, leaving 12 competitive models (listed in Table~\ref{tab:table1_tic146364192}). Fig.~\ref{fig:146_eccentricity_comparison} shows an example of why the \texttt{U} models failed: the \texttt{UG0} eccentricity posterior exhibits a large secondary peak near $e \approx 1$, a spurious high-eccentricity solution likely caused by the model fitting to outlier data \citep{Hara.etal2019_BiasRobustnessEccentricity, Osborne.etal2025_HomogeneousPlanetMasses}. The half-normal prior model \texttt{HG0}, by contrast, shows a unimodal posterior from which we could report $e = 0.12^{+0.10}_{-0.08}$, consistent with $e = 0.084 \pm 0.079$ reported by \citet{Lam.etal2018_K2265TransitingRocky}.
   
The Bayesian evidence is inconclusive between the circular model \texttt{CG0} and \texttt{HG0} ($\dlnZ = 0.4$), and we therefore adopt \texttt{CG0} (circular orbit, GP noise model, no velocity trend) as our preferred model, as it is the simpler model with fewer free parameters. This differs from \citet{Lam.etal2018_K2265TransitingRocky} in the treatment of eccentricity; however, their reported value of $e = 0.084 \pm 0.079$ is nearly consistent with a circular orbit, suggesting that the available observations are insufficient to conclusively support an eccentric solution. For the noise model and velocity trend though, we find the same conclusions: \citet{Lam.etal2018_K2265TransitingRocky} also modelled stellar activity using a quasi-periodic GP, which our analysis strongly corroborates ($\dlnZ = 39.1$ for \texttt{CG0} over the best white noise model \texttt{HW2}), and we also adopt a model with no velocity trend, which is both weakly preferred over the linear and quadratic trend models (\texttt{CG0} over \texttt{CG1}: $\dlnZ = 1.3$; over \texttt{CG2}: $\dlnZ = 1.1$) and is the simpler model with fewer free parameters.

The parameters for our preferred model \texttt{CG0} are listed in Table~\ref{tab:table2_tic146364192}. We measure a semi-amplitude $K = 3.3 \pm 0.4$\,\ms and a minimum mass $\Mpsini = 6.4 \pm 0.8\,M_{\earth}$, in agreement with \citet{Lam.etal2018_K2265TransitingRocky} who report $K = 3.34 \pm 0.43$\,\ms and $\Mp = 6.54 \pm 0.84\,M_{\earth}$. The RV data phase-folded on the orbital period is shown in Fig.~\ref{fig:TIC146364192_posterior_phase}.

%%%%%%%%%%%%%%%% SECTION 5.3 TOI-1055 RESULTS %%%%%%%%%%%%%%%%%
%
\subsection{TOI-1055 (TIC~320004517): an example of noise model comparison}\label{subsec:tic320_results}
\begin{figure*}  
  \includegraphics[width=\textwidth]{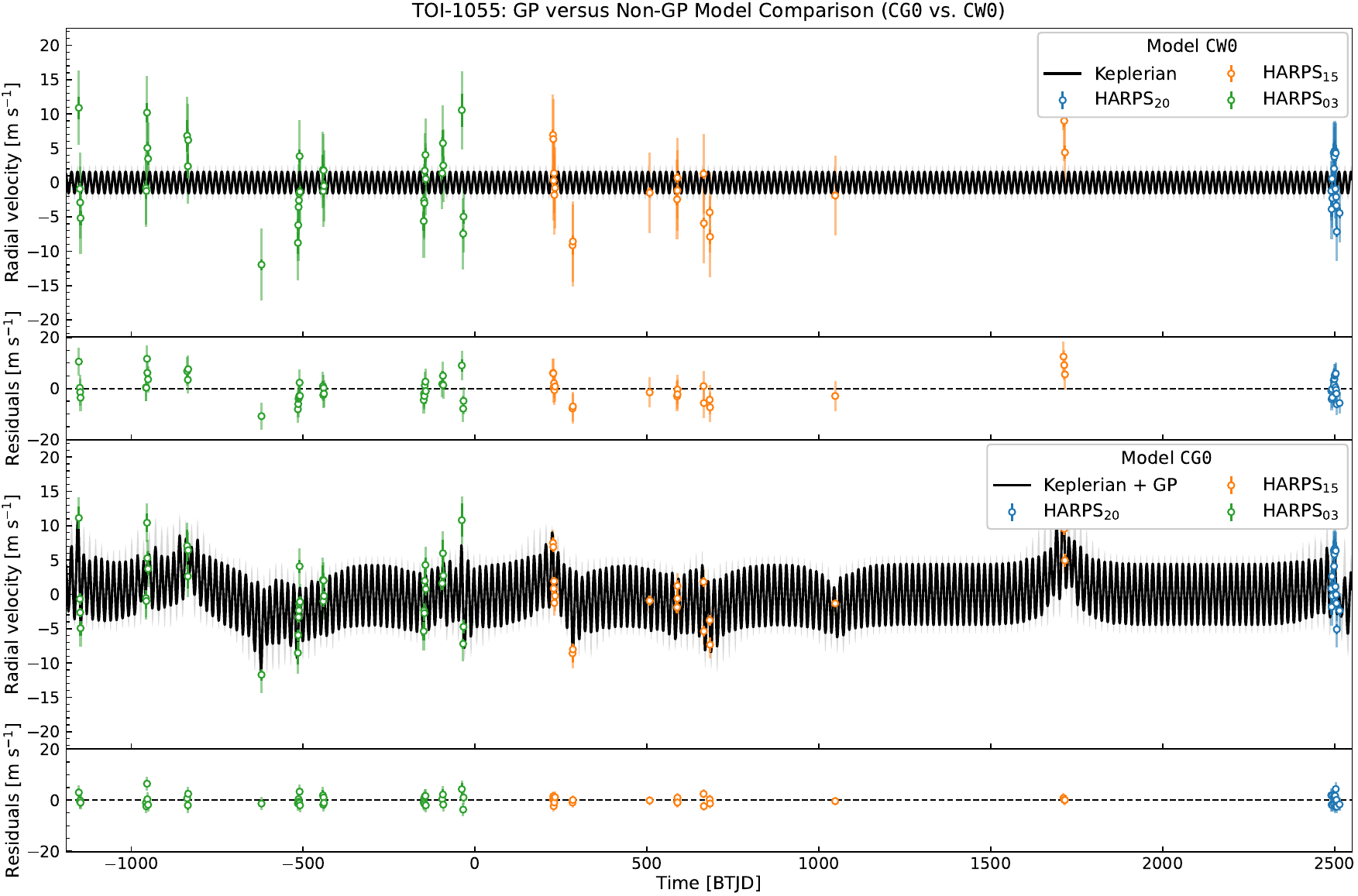}  
  \caption{
    Radial velocities for TOI-1055 (TIC~320004517), with green circles for $\mathrm{HARPS}_{03}$, orange circles for $\mathrm{HARPS}_{15}$ and blue circles for $\mathrm{HARPS}_{20}$. Error bars show observed uncertainties, with lighter extensions indicating the contribution from the $\sjit$ jitter terms added in quadrature. 
    The white noise \texttt{CW0} (upper panels) and GP \texttt{CG0} (lower panels) models are shown as the black curve and grey shaded region, representing the median and 68\% credible interval of the RV signal from the posterior samples. 
    Both models have circular orbit planets and no trend, differing only in the noise model.
    The lower panels show residuals relative to each RV model above.    
    The GP model is strongly preferred over the white noise model with log Bayes factor $\dlnZ = 14.4$, and has less scatter in the residuals (RMS of 2.05\,\ms versus 4.99\,\ms).
    }
  \label{fig:320_gp_nogp_comparison} 
\end{figure*}
\begin{table}
  \centering
  \setlength{\tabcolsep}{3pt}
  \caption{Fitted and derived parameters for the adopted model (\texttt{CG0}) for TOI-1055 (TIC 320004517). Reported values and uncertainties are the median value and 68\% credible interval.}
  \label{tab:table2_tic320004517}
  \begin{tabular}{@{}lc@{}}
    \hline
    Parameter & Value \\
    \hline
    \multicolumn{2}{@{}l}{\textbf{Planet b parameters}} \\[2pt]
    Orbital period, $P$ [d] & $17.471290_{-0.000099}^{+0.000100}$ \\
    Transit time, $T_C$ [BTJD] & $1661.06265 \pm 0.00097$ \\
    Eccentricity, $e$ & $0$ (fixed) \\
    Argument of periastron, $\wstar$ [$^\circ$] & $90$ (fixed) \\
    RV semi-amplitude, $K$ [\ms] & $4.4 \pm 1.3$ \\
    Minimum mass, $\Mpsini$ [$M_\oplus$] & $18 \pm 5$ \\[4pt]
    \multicolumn{2}{@{}l}{\textbf{Instrument parameters}} \\[2pt]
    RV offset, $\gamma_{\mathrm{HARPS}_{03}}$ [\ms] & $(-1.5544 \pm 0.0002) \times 10^{4}$ \\
    RV jitter, $\sigma_{\mathrm{HARPS}_{03}}$ [\ms] & $2.5_{-0.6}^{+0.7}$ \\
    RV offset, $\gamma_{\mathrm{HARPS}_{15}}$ [\ms] & $(-1.5524 \pm 0.0002) \times 10^{4}$ \\
    RV jitter, $\sigma_{\mathrm{HARPS}_{15}}$ [\ms] & $1.7_{-0.5}^{+0.7}$ \\
    RV offset, $\gamma_{\mathrm{HARPS}_{20}}$ [\ms] & $-1.5520_{-0.0005}^{+0.0004} \times 10^{4}$ \\
    RV jitter, $\sigma_{\mathrm{HARPS}_{20}}$ [\ms] & $2.4_{-0.7}^{+1.1}$ \\[4pt]
    \multicolumn{2}{@{}l}{\textbf{GP hyperparameters}} \\[2pt]
    GP amplitude, $A_{\mathrm{GP}}$ [\ms] & $6.0_{-1.1}^{+1.5}$ \\
    GP period, $P_{\mathrm{GP}}$ [d] & $24 \pm 5$ \\
    Active region timescale, $\lambda_{\mathrm{e}}$ [d] & $44 \pm 30$ \\
    Inverse harmonic complexity, $\lambda_{\mathrm{p}}$ & $0.8_{-0.3}^{+0.5}$ \\[4pt]
    \hline
  \end{tabular}
\end{table}
\begin{figure}
\centering
\includegraphics[width=\columnwidth]{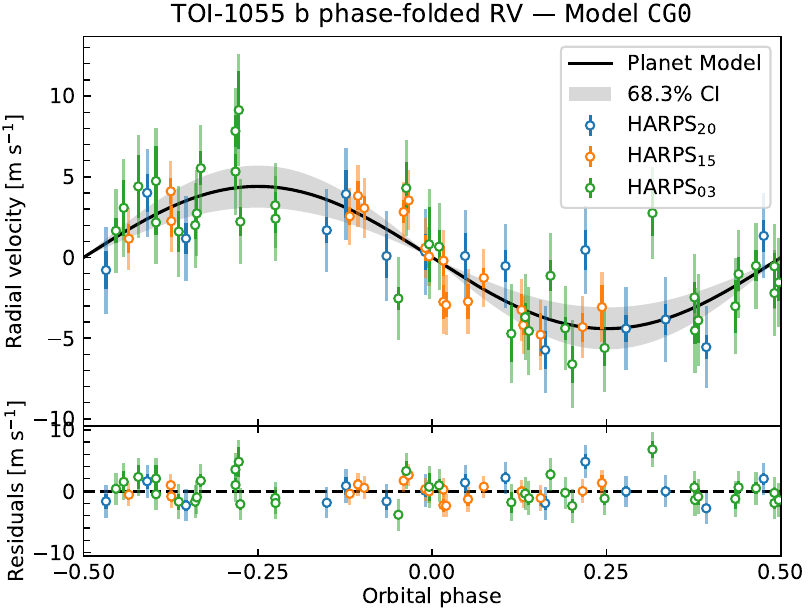}
\caption{
    Phase-folded radial velocities for TOI-1055 (TIC 320004517). Same as Fig.~\ref{fig:TIC207141131_posterior_phase_comparison}, but with green circles for $\mathrm{HARPS}_{03}$, orange circles for $\mathrm{HARPS}_{15}$ and blue circles for $\mathrm{HARPS}_{20}$. The adopted model shown is \texttt{CG0}.
}
\label{fig:TIC320004517_posterior_phase}
\end{figure}
TOI-1055~b is a warm Neptune transiting a G5V host star, with an orbital period of $P = 17.47$\,d, independently confirmed by \citet{Gan.etal2021_HD183579bWarm} and \citet{Palatnick.etal2021_ValidationHD183579b}, who reported discrepant semi-amplitudes of $K = 2.7 \pm 1.3$\,\ms\, and $K = 4.9^{+0.9}_{-1.0}$\,\ms\, respectively -- a tension of approximately $1.7\sigma$ \citep{Bonfanti.etal2023_TOI1055NeptunianPlanet}, attributed to differences in the number of RV observations used and how they were extracted, and the treatment of stellar variability \citep{Gan.etal2021_HD183579bWarm, Bonfanti.etal2023_TOI1055NeptunianPlanet}. \citet{Bonfanti.etal2023_TOI1055NeptunianPlanet} resolved the tension using the same 56 archival HARPS spectra plus 16 new observations, reporting $K = 5.03^{+0.60}_{-0.59}$\,\ms\, and $\Mp = 20.4^{+2.6}_{-2.5}\,M_{\earth}$, consistent with \citet{Palatnick.etal2021_ValidationHD183579b}. We use the same archival HARPS data as \citet{Bonfanti.etal2023_TOI1055NeptunianPlanet}, but reduced using the standard HARPS \textsc{drs} pipeline \citep{Barbieri2023_ESOHARPSRadial}, rather than their skew-normal CCF extraction and breakpoint detrending.

Of the 18 models evaluated, 14 passed the \textsc{harmonic} diagnostic tests and are listed in Table~\ref{tab:table1_tic320004517}. Our model comparison strongly favours a GP noise model: the circular GP model with no trend \texttt{CG0} is strongly preferred over the equivalent white noise model \texttt{CW0} with $\dlnZ = 14.4$; AIC and BIC also prefer \texttt{CG0}. This preference has a direct impact on the inferred semi-amplitude: \texttt{CG0} yields $K = 4.4 \pm 1.3$\,\ms, while \texttt{CW0} yields $K = 1.5^{+0.9}_{-0.8}$\,\ms\, -- a factor of three difference. \citet{Gan.etal2021_HD183579bWarm} tested a GP noise model, finding $\dlnZ = 8$ in its favour and recovering $K = 4.3 \pm 1.0$\,\ms, but ultimately rejected it, suspecting the stellar activity was not well sampled by sparse RV data which could risk the GP overfitting to the data. Looking at the resulting RV and residuals in Fig.~\ref{fig:320_gp_nogp_comparison}, the GP appears to be capturing genuine stellar activity -- particularly given our longer observation baseline from having more RV data points available than \citet{Gan.etal2021_HD183579bWarm} did, and the stellar rotation period $P_\mathrm{rot}=24.8^{+5.4}_{-4.4}$\,d reported by \citet{Bonfanti.etal2023_TOI1055NeptunianPlanet} from chromospheric $\log{R}'_\mathrm{HK}$ measurements, which we adopted as an informative prior on the GP period $P_\mathrm{GP}$ hyperparameter. Furthermore, our \texttt{CG0} result of $K = 4.4 \pm 1.3$\,\ms\, is consistent within $1\sigma$ with both the SN-fit-based breakpoint analysis ($K_\mathrm{b} = 5.03^{+0.60}_{-0.59}$\,\ms) and the HARPS \textsc{drs}-based comparison analysis ($K_{\mathrm{DRS},\mathrm{b}} = 5.05^{+0.71}_{-0.69}$\,\ms) reported by \citet{Bonfanti.etal2023_TOI1055NeptunianPlanet}.

There is strong evidence against a velocity trend ($\dlnZ = 7.8$ for \texttt{CG0} over \texttt{CG1}), consistent with \citet{Bonfanti.etal2023_TOI1055NeptunianPlanet} who do not report a velocity trend. The eccentricity treatment is inconclusive, with $\dlnZ = 0.5$ for \texttt{CG0} over \texttt{HG0}. While \citet{Bonfanti.etal2023_TOI1055NeptunianPlanet} reported a non-zero eccentricity of $e = 0.061^{+0.061}_{-0.042}$, this is nearly equally consistent both with a circular orbit and with the low eccentricity $e=0.13^{+0.20}_{-0.09}$ reported in our \texttt{HG0} model, which is in turn consistent with $e=0$ within $1\sigma$. This suggests that the data cannot robustly constrain eccentricity and that further observations are needed, therefore it is not surprising that the simpler model \texttt{CG0} (with two fewer free parameters) was slightly favoured in this case.

Our preferred model \texttt{CG0} parameters are listed in Table~\ref{tab:table2_tic320004517}: we measure $K = 4.4 \pm 1.3$\,\ms\, and a minimum mass $\Mpsini = 18 \pm 5\,M_{\earth}$, consistent with \citet{Bonfanti.etal2023_TOI1055NeptunianPlanet}. The RV data phase-folded on the orbital period is shown in Fig.~\ref{fig:TIC320004517_posterior_phase}.

%%%%%%%%%%%%%%%% SECTION 5.4 TOI-220 RESULTS %%%%%%%%%%%%%%%%%
%
\subsection{TOI-220 (TIC~150098860)}\label{subsec:tic150_results}
\begin{table}
  \centering
  \setlength{\tabcolsep}{3pt}
  \caption{Fitted and derived parameters for the adopted model (\texttt{CW2}) for TOI-220 (TIC 150098860). Reported values and uncertainties are the median value and 68\% credible interval.}
  \label{tab:table2_tic150098860}
  \begin{tabular}{@{}lc@{}}
    \hline
    Parameter & Value \\
    \hline
    \multicolumn{2}{@{}l}{\textbf{Planet b parameters}} \\[2pt]
    Orbital period, $P$ [d] & $10.695262_{-0.000099}^{+0.000100}$ \\
    Transit time, $T_C$ [BTJD] & $1335.9020 \pm 0.0014$ \\
    Eccentricity, $e$ & $0$ (fixed) \\
    Argument of periastron, $\wstar$ [$^\circ$] & $90$ (fixed) \\
    RV semi-amplitude, $K$ [\ms] & $4.5 \pm 0.3$ \\
    Minimum mass, $\Mpsini$ [$M_\oplus$] & $13.56_{-0.99}^{+1.00}$ \\[4pt]
    \multicolumn{2}{@{}l}{\textbf{Instrument parameters}} \\[2pt]
    RV offset, $\gamma_{\mathrm{HARPS}_{15}}$ [\ms] & $(2.61293 \pm 0.00002) \times 10^{4}$ \\
    RV jitter, $\sigma_{\mathrm{HARPS}_{15}}$ [\ms] & $1.6 \pm 0.2$ \\[4pt]
    \multicolumn{2}{@{}l}{\textbf{Trend parameters}} \\[2pt]
    Linear RV trend, $\dot{\gamma}$ [\msd] & $-0.025 \pm 0.005$ \\
    Quadratic RV trend, $\ddot{\gamma}$ [\msdd] & $(1.6 \pm 0.3) \times 10^{-4}$ \\[4pt]
    \hline
  \end{tabular}
\end{table}
\begin{figure}
\centering
\includegraphics[width=\columnwidth]{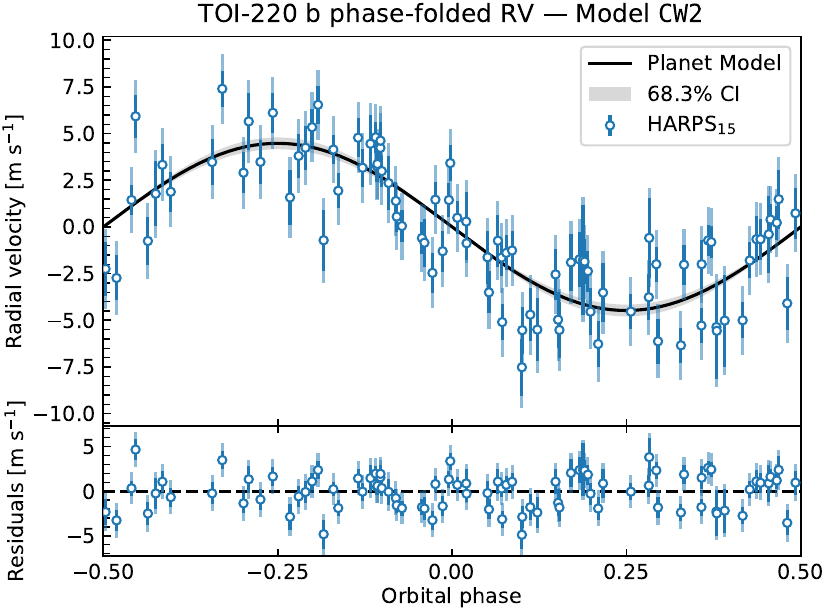}
\caption{
    Phase-folded radial velocities for TOI-220 (TIC 150098860). Same as Fig.~\ref{fig:TIC207141131_posterior_phase_comparison}, but with blue circles for $\mathrm{HARPS}_{15}$. The adopted model shown is \texttt{CW2}.
}
\label{fig:TIC150098860_posterior_phase}
\end{figure}
TOI-220~b is a sub-Neptune first discovered by \citet{Hoyer.etal2021_TOI220WarmSubNeptune} reporting a mass $\Mp = 13.8 \pm 1.0\,M_{\earth}$, transiting the K0V host star on an orbital period of $P=10.69$\,d.  
All nine GP model variants are excluded for this system. While some of the \texttt{G} models passed the automated \textsc{harmonic} diagnostics, visual inspection of the MCMC posteriors revealed that the GP hyperparameters (especially $P_{\mathrm{GP}}$ the GP period) are unconstrained and multimodal across all GP models, indicating that the data do not support a GP noise model. This is consistent with the low magnetic activity level of the host star, with $\log R'_\mathrm{HK} = -5.07 \pm 0.05$ \citep{Hoyer.etal2021_TOI220WarmSubNeptune} being below the minimum limit of $\approx-4.8$ given by \citet{Wright2005_RadialVelocityJitter} for a star of this $B - V$ colour index to be considered active. Furthermore, no coherent rotation period was detected in past photometric studies of the star \citep{Hoyer.etal2021_TOI220WarmSubNeptune}, suggesting that rotationally modulated stellar activity is not a significant source of quasi-periodic noise in the RV observations. We therefore manually exclude all GP models and restrict comparison to the nine white noise \texttt{W} models only.

Our preferred model is \texttt{CW2} -- a circular orbit, white noise model, and quadratic long-term trend -- as summarised in Table~\ref{tab:table1_tic150098860}. The quadratic trend is strongly preferred over no trend (\texttt{CW2} vs \texttt{CW0}: $\dlnZ = 6.4$), consistent with \citet{Hoyer.etal2021_TOI220WarmSubNeptune} who also preferred a quadratic trend model in their statistical comparisons by $\Delta\mathrm{AIC}=15$; we find $\Delta\mathrm{AIC}=22$. The circular orbit is weakly preferred over the half-normal eccentric model \texttt{HW2} by $\dlnZ = 1.8$, and moderately preferred over the uniform eccentric model \texttt{UW2} by $\dlnZ = 2.7$.

The parameters for our preferred model \texttt{CW2} can be found in Table~\ref{tab:table2_tic150098860}. We derive a RV semi-amplitude of $K = 4.5 \pm 0.3$\,\ms, in agreement with $K = 4.56 \pm 0.32$\,\ms\ reported by \citet{Hoyer.etal2021_TOI220WarmSubNeptune}. Our recovered trend coefficients of $\dot{\gamma} = -0.025 \pm 0.005$\,\msd\, and $\ddot{\gamma} = (1.6 \pm 0.3) \times 10^{-4}$\,\msdd\, are broadly consistent with those of \citet{Hoyer.etal2021_TOI220WarmSubNeptune} ($\dot{\gamma} = -0.0379 \pm 0.0080$\,\msd; $\ddot{\gamma} = 0.000149 \pm 0.000031$\,\msdd). This leads to a minimum mass of $\Mpsini = 13.56^{+1.00}_{-0.99}\,M_{\earth}$, in agreement with the mass of $13.8 \pm 1.0\,M_{\earth}$ reported by \citet{Hoyer.etal2021_TOI220WarmSubNeptune}. The RV data phase-folded on the orbital period is shown in Fig.~\ref{fig:TIC150098860_posterior_phase}.

%%%%%%%%%%%%%%%% SECTION 5.5 LHS 1815 RESULTS %%%%%%%%%%%%%%%%%
%
\subsection{LHS~1815 (TIC~260004324)}\label{subsec:tic260_results}
\begin{table}
  \centering
  \setlength{\tabcolsep}{3pt}
  \caption{Fitted and derived parameters for the adopted model (\texttt{CW0}) for LHS 1815 (TIC 260004324). Reported values and uncertainties are the median value and 68\% credible interval.}
  \label{tab:table2_tic260004324}
  \begin{tabular}{@{}lc@{}}
    \hline
    Parameter & Value \\
    \hline
    \multicolumn{2}{@{}l}{\textbf{Planet b parameters}} \\[2pt]
    Orbital period, $P$ [d] & $3.81433 \pm 0.00008$ \\
    Transit time, $T_C$ [BTJD] & $1327.000 \pm 0.004$ \\
    Eccentricity, $e$ & $0$ (fixed) \\
    Argument of periastron, $\wstar$ [$^\circ$] & $90$ (fixed) \\
    RV semi-amplitude, $K$ [\ms] & $0.6_{-0.4}^{+0.6}$ \\
    Minimum mass, $\Mpsini$ [$M_\oplus$] & $1.0_{-0.7}^{+0.9}$ \\[4pt]
    \multicolumn{2}{@{}l}{\textbf{Instrument parameters}} \\[2pt]
    RV offset, $\gamma_{\mathrm{HARPS}_{03}}$ [\ms] & $(4.2673 \pm 0.0001) \times 10^{4}$ \\
    RV jitter, $\sigma_{\mathrm{HARPS}_{03}}$ [\ms] & $2.2_{-1.3}^{+2.0}$ \\
    RV offset, $\gamma_{\mathrm{HARPS}_{15}}$ [\ms] & $(4.26728 \pm 0.00005) \times 10^{4}$ \\
    RV jitter, $\sigma_{\mathrm{HARPS}_{15}}$ [\ms] & $3.4_{-0.4}^{+0.5}$ \\[4pt]
    \hline
  \end{tabular}
\end{table}
\begin{figure}
  \centering
  \includegraphics[width=\columnwidth]{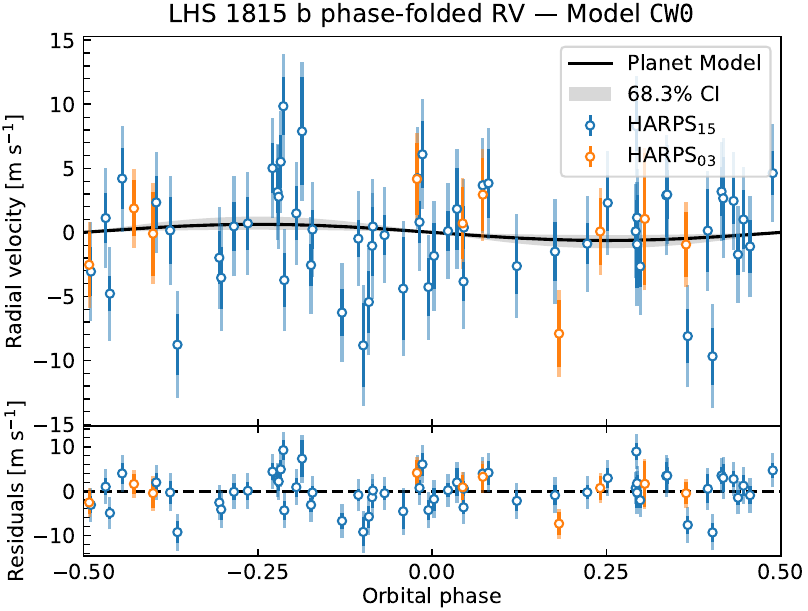}
  \caption{
    Phase-folded radial velocities for LHS 1815 (TIC 260004324). Same as Fig.~\ref{fig:TIC207141131_posterior_phase_comparison}, but with blue circles for $\mathrm{HARPS}_{15}$ and orange circles for $\mathrm{HARPS}_{03}$. The adopted model shown is \texttt{CW0}.
  }
  \label{fig:TIC260004324_posterior_phase}
\end{figure}
LHS~1815~b is a super-Earth on an orbital period of $P = 3.814$\,d around the M2.5V host star, first detected by \citet{Gan.etal2020_LHS1815bFirst}. They reported a $3\sigma$ upper mass limit of $\Mp < 8.7\,M_{\earth}$ from 22 HARPS observations. \citet{Luque.Palle2022_DensityNotRadius} subsequently added 64 further HARPS RVs, and used the 86 total RVs to constrain the mass to $\Mp = 1.58^{+0.64}_{-0.60}\,M_{\earth}$.

Of the 18 models evaluated, only 9 passed the \textsc{harmonic} diagnostic tests and are listed in Table~\ref{tab:table1_tic260004324}. The evidence is completely indistinguishable between \texttt{CW0} and \texttt{HW0} ($\dlnZ = 0.0$); we therefore adopt \texttt{CW0} as the preferred model as it is the simpler model with fewer free parameters. The \texttt{CW0} model is consistent with \citet{Luque.Palle2022_DensityNotRadius} in having circular orbits and no long-term velocity trend, and is weakly preferred over the next model \texttt{CG0} by $\dlnZ = 1.2$. Although \citet{Gan.etal2020_LHS1815bFirst} detected a stellar rotation period of $P_\mathrm{rot} = 47.8 \pm 0.7$\,d from \textit{TESS} photometry and WASP-South light curves, they found no significant correlation between the RVs and spectroscopic activity indicators (chromatic index and differential line width), suggesting that the main source of RV variation is the planetary Keplerian signal, not rotationally modulated quasi-periodic stellar activity. We find models with velocity trends are strongly disfavoured: over linear trend \texttt{1} models by $\dlnZ \geq 8.1$, and quadratic trend \texttt{2} models by $\dlnZ \geq 22.1$.

The parameters for our preferred model \texttt{CW0} can be found in Table~\ref{tab:table2_tic260004324}. The RV data phase-folded on the orbital period is shown in Fig.~\ref{fig:TIC260004324_posterior_phase}. We measure a semi-amplitude $K = 0.6^{+0.6}_{-0.4}$\,\ms, and a minimum mass $\Mpsini = 1.0^{+0.9}_{-0.7}\,M_{\earth}$. Our result is formally consistent within $1\sigma$ with \citet{Luque.Palle2022_DensityNotRadius} who report $K = 1.02^{+0.41}_{-0.39}$\,\ms and $\Mp = 1.58^{+0.64}_{-0.60}\,M_{\earth}$ but the difference in central values is notable. We attribute this to a combination of differing data preparation (see Section~\ref{sec:data_prep}; after cutting outliers we end up with 71 observations versus their 86), different RV extraction pipeline (\textsc{serval} versus the HARPS \textsc{drs} pipeline), and the use of spectral activity indicators by \citet{Luque.Palle2022_DensityNotRadius}, which are not used in our analysis.

%%%%%%%%%%%%%%%% SECTION 5.6 GJ 1214 RESULTS %%%%%%%%%%%%%%%%%
%
\subsection{GJ~1214 (TIC~467929202)}\label{subsec:tic467_results}  
\begin{table}
  \centering
  \setlength{\tabcolsep}{3pt}
  \caption{Fitted and derived parameters for the adopted model (\texttt{CG0}) for GJ 1214 (TIC 467929202). Reported values and uncertainties are the median value and 68\% credible interval.}
  \label{tab:table2_tic467929202}
  \begin{tabular}{@{}lc@{}}
    \hline
    Parameter & Value \\
    \hline
    \multicolumn{2}{@{}l}{\textbf{Planet b parameters}} \\[2pt]
    Orbital period, $P$ [d] & $1.580404 \pm 0.000006$ \\
    Transit time, $T_C$ [BTJD] & $2639.781262_{-0.000100}^{+0.000099}$ \\
    Eccentricity, $e$ & $0$ (fixed) \\
    Argument of periastron, $\wstar$ [$^\circ$] & $90$ (fixed) \\
    RV semi-amplitude, $K$ [\ms] & $13.6 \pm 0.7$ \\
    Minimum mass, $\Mpsini$ [$M_\oplus$] & $7.9 \pm 0.4$ \\[4pt]
    \multicolumn{2}{@{}l}{\textbf{Instrument parameters}} \\[2pt]
    RV offset, $\gamma_{\mathrm{HARPS}_{03}}$ [\ms] & $(2.1146 \pm 0.0002) \times 10^{4}$ \\
    RV jitter, $\sigma_{\mathrm{HARPS}_{03}}$ [\ms] & $3.02_{-1.10}^{+0.97}$ \\
    RV offset, $\gamma_{\mathrm{HARPS}_{15}}$ [\ms] & $(2.1143 \pm 0.0002) \times 10^{4}$ \\
    RV jitter, $\sigma_{\mathrm{HARPS}_{15}}$ [\ms] & $3.5_{-1.4}^{+1.2}$ \\[4pt]
    \multicolumn{2}{@{}l}{\textbf{GP hyperparameters}} \\[2pt]
    GP amplitude, $A_{\mathrm{GP}}$ [\ms] & $3.0_{-0.8}^{+0.9}$ \\
    GP period, $P_{\mathrm{GP}}$ [d] & $124 \pm 5$ \\
    Active region timescale, $\lambda_{\mathrm{e}}$ [d] & $235 \pm 60$ \\
    Inverse harmonic complexity, $\lambda_{\mathrm{p}}$ & $1.410 \pm 0.014$ \\[4pt]
    \hline
  \end{tabular}
\end{table}
\begin{figure}
\centering
\includegraphics[width=\columnwidth]{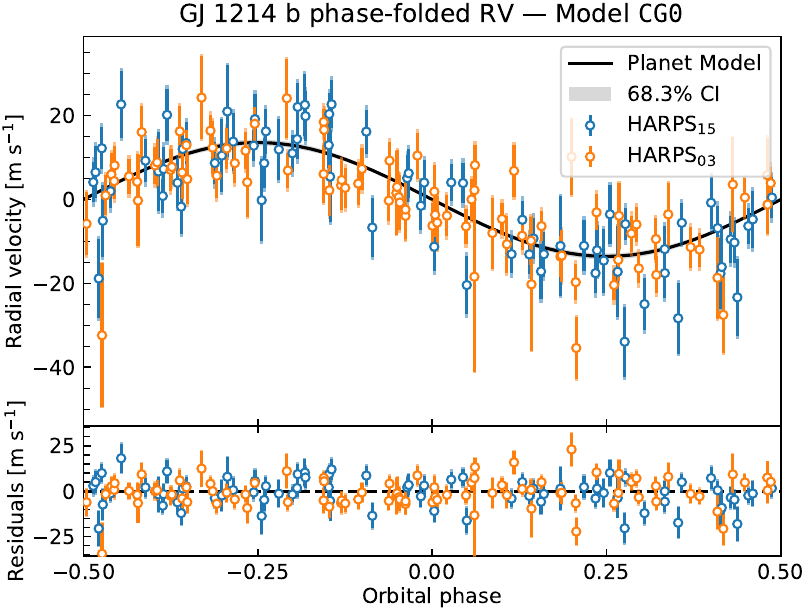}
\caption{
    Phase-folded radial velocities for GJ 1214 (TIC 467929202). Same as Fig.~\ref{fig:TIC207141131_posterior_phase_comparison}, but with orange circles for $\mathrm{HARPS}_{03}$ and blue circles for $\mathrm{HARPS}_{15}$. The adopted model shown is \texttt{CG0}.
}
\label{fig:TIC467929202_posterior_phase}
\end{figure}
GJ~1214~b is a sub-Neptune orbiting an M4.5V host star on an orbital period of $P = 1.580$\,d. The planet was discovered using transit photometry by \citet{Charbonneau.etal2009_SuperEarthTransitingNearby}, who identified it as a prime target for atmospheric characterisation via transmission spectroscopy, owing to the small size of the host star and the proximity to our solar system. Furthermore, as a sub-Neptune it is of particular interest because these planets are one of the most common types in the Galaxy but have no analogue within the Solar system \citep{Cassan.etal2012_OneMoreBound, Kreidberg.etal2014_CloudsAtmosphereSuperEarth}. For small planets in the Earth to Neptune range, the density obtained from planetary mass and radius alone is not enough to constrain the compositions due to degeneracies between different compositions with the same overall densities \citep{Adams.etal2008_OceanPlanetThick}, with atmospheric studies required to resolve the degeneracy and understand the composition \citep{Miller-Ricci.etal2009_AtmosphericSignaturesSuperEarths, Kreidberg.etal2014_CloudsAtmosphereSuperEarth}. This has led to extensive interest in atmospheric characterisation of GJ~1214~b. Atmospheric characterisation requires a precise planetary mass \citep{Batalha.etal2019_PrecisionMassMeasurements}, motivating interest in RV model comparison to ensure robust mass characterisation.

\citet{Charbonneau.etal2009_SuperEarthTransitingNearby} originally reported a planetary mass $\Mp=6.55\pm0.98\,M_{\earth}$ from 21 HARPS RV observations. Additional RVs were later obtained by \citet{Cloutier.etal2021_MorePreciseMass}, who performed a joint transit and RV analysis of 165 HARPS observations, incorporating a quasi-periodic GP to model stellar activity because of the strong photometric variability of the host star (amplitude 2--5\%; stellar rotation period $P_\mathrm{rot} = 124.7^{+5.0}_{-4.8}$\,d). Their joint fit constrains the eccentricity to $e < 0.063$ at 95\% confidence, and they report a semi-amplitude $K = 14.36 \pm 0.53$\,\ms and mass $\Mp=8.17\pm0.43\,M_{\earth}$. In our analysis, we have 164 observations available after the data preparation and sigma-clipping (see Section~\ref{sec:data_prep} and Table~\ref{tab:systems}).

Of the 18 models tested in our analysis, 10 passed the \textsc{harmonic} diagnostic tests and are included in Table~\ref{tab:table1_tic467929202}. Our preferred model is \texttt{CG0} (circular, GP, no trend). The Bayesian evidence moderately prefers the GP noise model over white noise ($\dlnZ = 2.5$ for \texttt{CG0} over \texttt{CW0}); however, AIC and BIC disagree with $\Delta\mathrm{AIC} = 6.4$ and $\Delta\mathrm{BIC} = 18.8$ in favour of \texttt{CW0}. The $\chi^2$ and $\lnL$ are marginally better for the GP model, confirming the inclusion of a QP GP does improve the fit; the AIC and BIC nonetheless favour \texttt{CW0} because their fixed penalty on parameter count (see equations~(\ref{eq:AIC}) and (\ref{eq:BIC})) outweighs the improvement in the residuals -- they treat all parameters equally regardless of whether the data actually constrains them. The Bayesian evidence, being the prior-weighted average of the likelihood, penalises only complexity that is not supported by the data, rather than a blanket penalty per parameter, so the moderate preference for \texttt{CG0} over \texttt{CW0} means the improved fit justifies the added flexibility from the GP. Furthermore, previous studies show independent photometric evidence for quasi-periodic rotationally modulated stellar activity \citep{Mallonn.etal2018_GJ1214Rotation, Cloutier.etal2021_MorePreciseMass}, supporting the use of a GP with the QP kernel. For these reasons, we adopt the \texttt{CG0} model.

Velocity trends are strongly disfavoured ($\dlnZ = 7.7$ for \texttt{CG0} over \texttt{CG1}). For eccentricity, the \texttt{HG0} did not pass the diagnostic tests and was excluded, preventing a direct comparison with \texttt{CG0}, however the best scoring eccentric model \texttt{HW0} is moderately disfavoured over \texttt{CG0} at $\dlnZ=3.8$, and \texttt{HG1} is strongly disfavoured at $\dlnZ=9.3$. This is consistent with the constraint $e < 0.063$ at 95\% confidence found by \citet{Cloutier.etal2021_MorePreciseMass}.

A notable feature of this system is the robustness of the planetary parameters to model choice: every model in Table~\ref{tab:table1_tic467929202} recovers $K = 13.6$\,\ms (uncertainties $\pm 0.7$--$0.8$\,\ms) and minimum mass estimate $\Mpsini = 7.9\,M_{\earth}$ (uncertainties $\pm 0.4$--$0.5\,M_{\earth}$), and the eccentric models find low eccentricities consistent with zero ($e = 0.04$--$0.05$, with 1$\sigma$ uncertainties of $0.03$--$0.05$).

The parameters for our preferred model \texttt{CG0} can be found in Table~\ref{tab:table2_tic467929202}. We measure $K = 13.6 \pm 0.7$\,\ms deriving a minimum mass $\Mpsini = 7.9 \pm 0.4\,M_{\earth}$, consistent with \citet{Cloutier.etal2021_MorePreciseMass} within $1\sigma$. The RV data phase-folded on the orbital period is shown in Fig.~\ref{fig:TIC467929202_posterior_phase}.

%%%%%%%%%%%%%% SECTION 5.7 TOI-544 RESULTS %%%%%%%%%%%%%%%

\subsection{TOI-544 (TIC~50618703): an example of comparing 1- and 2-planet models}\label{subsec:toi544_results}
\begin{table*}
  \centering
  \setlength{\tabcolsep}{3pt}
  \caption{Fitted and derived parameters for the best 1-planet and 2-planet (adopted) models for TOI-544 (TIC 50618703): two-planet (preferred, \texttt{HHG0}) and one-planet (\texttt{CG0}). Reported values and uncertainties are the median value and 68\% credible interval. $M_{\mathrm{p,b}}$ is the true planetary mass; $M_{\mathrm{p,c}}\sin i_{\mathrm{c}}$ is the minimum mass (2-planet models only).}
  \label{tab:table2_toi544_combined}
  \begin{tabular}{lcc}
    \hline
    Parameter & Two-planet model (\texttt{HHG0}) & One-planet model (\texttt{CG0}) \\
    \hline
    \multicolumn{3}{l}{\textbf{Planet b parameters}} \\[2pt]
    Orbital period, $P_{\mathrm{b}}$ [d] & $1.54836 \pm 0.00009$ & $1.54840 \pm 0.00009$ \\
    Transit time, $T_{C,\mathrm{b}}$ [BTJD] & $2199.0314 \pm 0.0007$ & $2199.0314 \pm 0.0007$ \\
    $\sqrt{e_{\mathrm{b}}}\cos{\omega_{\star,\mathrm{b}}}$ & $0.36_{-0.20}^{+0.12}$ & --- \\
    $\sqrt{e_{\mathrm{b}}}\sin{\omega_{\star,\mathrm{b}}}$ & $0.3_{-0.3}^{+0.2}$ & --- \\
    Eccentricity, $e_{\mathrm{b}}$ & $0.27_{-0.12}^{+0.14}$ & $0$ (fixed) \\
    Argument of periastron, $\omega_{\star,\mathrm{b}}$ [$^\circ$] & $50_{-30}^{+200}$ & $90$ (fixed) \\
    RV semi-amplitude, $K_{\mathrm{b}}$ [\ms] & $2.4 \pm 0.4$ & $2.3 \pm 0.4$ \\
    Mass, $M_{\mathrm{p,b}}$ [$M_\oplus$] & $3.1 \pm 0.5$ & $3.1 \pm 0.5$ \\[4pt]
    \multicolumn{3}{l}{\textbf{Planet c parameters}} \\[2pt]
    Orbital period, $P_{\mathrm{c}}$ [d] & $50.1_{-0.2}^{+0.3}$ & --- \\
    Transit time, $T_{C,\mathrm{c}}$ [BTJD] & $2213 \pm 2$ & --- \\
    $\sqrt{e_{\mathrm{c}}}\cos{\omega_{\star,\mathrm{c}}}$ & $0.46_{-0.15}^{+0.10}$ & --- \\
    $\sqrt{e_{\mathrm{c}}}\sin{\omega_{\star,\mathrm{c}}}$ & $0.1 \pm 0.2$ & --- \\
    Eccentricity, $e_{\mathrm{c}}$ & $0.26_{-0.10}^{+0.09}$ & --- \\
    Argument of periastron, $\omega_{\star,\mathrm{c}}$ [$^\circ$] & $39_{-30}^{+300}$ & --- \\
    RV semi-amplitude, $K_{\mathrm{c}}$ [\ms] & $5.0_{-0.5}^{+0.6}$ & --- \\
    Minimum mass, $M_{\mathrm{p,c}}\sin i_{\mathrm{c}}$ [$M_\oplus$] & $21 \pm 2$ & --- \\[4pt]
    \multicolumn{3}{l}{\textbf{Instrument parameters}} \\[2pt]
    RV offset, $\gamma_{\mathrm{HARPS}_{20}}$ [\ms] & $6 \pm 2$ & $6 \pm 2$ \\
    RV jitter, $\sigma_{\mathrm{HARPS}_{20}}$ [\ms] & $1.7 \pm 0.4$ & $1.7 \pm 0.5$ \\
    RV offset, $\gamma_{\mathrm{HARPS\text{-}N}}$ [\ms] & $-15_{-2}^{+3}$ & $-15_{-2}^{+3}$ \\
    RV jitter, $\sigma_{\mathrm{HARPS\text{-}N}}$ [\ms] & $2.6_{-1.3}^{+2.0}$ & $3.4_{-1.1}^{+2.0}$ \\[4pt]
    \multicolumn{3}{l}{\textbf{GP hyperparameters}} \\[2pt]
    GP amplitude, $A_{\mathrm{GP}}$ [\ms] & $7.7_{-1.4}^{+2.0}$ & $8.1_{-1.0}^{+1.3}$ \\
    Active region timescale, $\lambda_{\mathrm{e}}$ [d] & $95 \pm 30$ & $16 \pm 3$ \\
    Inverse harmonic complexity, $\lambda_{\mathrm{p}}$ & $0.29 \pm 0.07$ & $0.37 \pm 0.06$ \\
    GP characteristic period, $P_{\mathrm{GP}}$ [d] & $19.27 \pm 0.10$ & $19.5_{-0.5}^{+0.6}$ \\[4pt]
    \multicolumn{3}{l}{\textbf{Model comparison statistics}} \\[2pt]
    Number of free parameters & $18$ & $11$ \\
    $\chi^2$ & $108.8$ & $111.0$ \\
    Log-likelihood, $\lnL$ & $-323.9$ & $-356.2$ \\
    AIC & $683.8$ & $734.4$ \\
    BIC & $733.8$ & $765.0$ \\
    Log-evidence, $\ln\Z$ & $-366.4126_{-0.0095}^{+0.0096}$ & $-386.698 \pm 0.004$ \\
    $\dlnZ$ & $0.0$ & $20.3$ \\
    \hline
  \end{tabular}
\end{table*}
\begin{figure}
\centering

\begin{subfigure}{\columnwidth}
    \centering
    \includegraphics[width=0.98\columnwidth]{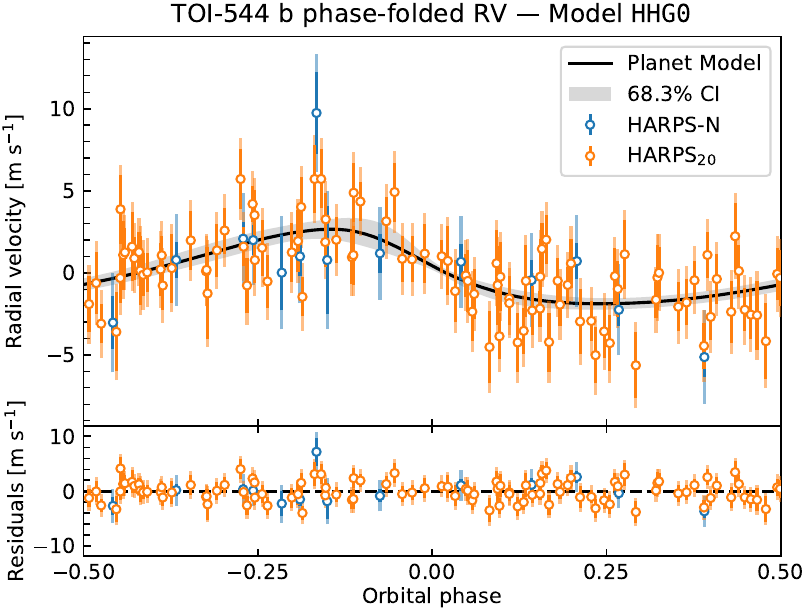}
\end{subfigure}

\begin{subfigure}{\columnwidth}
    \centering
    \includegraphics[width=0.98\columnwidth]{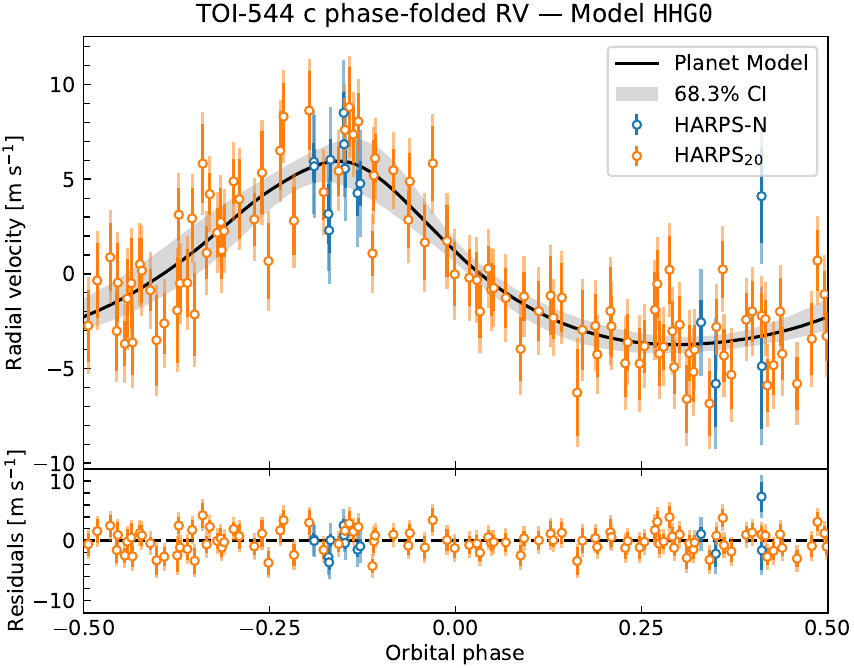}
\end{subfigure}

\caption{
    Phase-folded radial velocities for TOI-544, shown for planet b (upper panel) and planet c (lower panel) respectively. 
    Same as Fig.~\ref{fig:TIC207141131_posterior_phase_comparison}, but with orange circles for $\mathrm{HARPS}_{20}$ and blue circles for HARPS-N. The adopted model shown is the two-planet model \texttt{HHG0}.
}
\label{fig:toi544_2pl_bc_posterior_phase}
\end{figure}

TOI-544 is a K7V star \citep{Gore.etal2024_MetallicitiesRefinedStellar} hosting a transiting planet TOI-544~b first validated by \citet{Giacalone.etal2022_Validation13Hot}. A second non-transiting planet was later discovered by \citet{Osborne.etal2024_TOI544PotentialWaterworld}, who performed a joint photometric and RV analysis of the system using 108 HARPS and 14 HARPS-N radial velocities extracted using \textsc{serval}. For the RV modelling, they used the \textsc{pyaneti} code with a multidimensional quasi-periodic GP, incorporating the S-index activity indicator alongside the RV to mitigate stellar activity. \citet{Osborne.etal2024_TOI544PotentialWaterworld} reported that the inner planet TOI-544~b is on a circular orbit of period $P_{\mathrm{b}} = 1.55$\,d, with semi-amplitude $K_{\mathrm{b}} = 2.17 \pm 0.36$\,\ms and mass $M_{p,{\mathrm{b}}} = 2.89 \pm 0.48\,M_{\earth}$. They reported that the outer planet TOI-544~c is on an eccentric orbit ($e_{\mathrm{c}} = 0.32^{+0.08}_{-0.09}$) of period $P_{\mathrm{c}} = 50.09 \pm 0.24$\,d, with semi-amplitude $K_{\mathrm{c}} = 5.36 \pm 0.56$\,\ms\, and minimum mass $M_{p,{\mathrm{c}}}\sin i_{\mathrm{c}} = 21.5 \pm 2.0\,M_{\earth}$. As part of their analysis, they compared a one-planet and two-planet RV fit using AIC and BIC.

In our RV analysis, we use the same \textsc{serval}-extracted RV data, but we excluded the 2 HARPS observations falling in the HARPS$_{15}$ epoch (see Section~\ref{sec:data_prep}): with only 2 observations in that epoch, fitting a separate RV offset $\gamma_{\mathrm{HARPS}_{15}}$ is unreliable. (\citealt{Osborne.etal2024_TOI544PotentialWaterworld} modelled HARPS$_{15}$ and HARPS$_{20}$ together with a single $\gamma$ and $\sjit$.) After also applying the sigma-clip described in Section~\ref{sec:data_prep}, this leaves 119 RVs in total (105 HARPS$_{20}$ and 14 HARPS-N observations) with each instrument modelled with its own RV offset $\gamma$ and jitter $\sjit$.

Unlike the six transiting single-planet systems analysed above, for which we compared a fixed set of 18 model variants, here we compare both one-planet and two-planet Keplerian models across all eccentricity, noise and trend combinations, to demonstrate how the LHME can be used to help determine the strength of evidence for a suspected non-transiting planet. For the one-planet models, we use the same 18-model \texttt{XYZ} naming convention as before. For the two-planet models, we extend the naming to four characters \texttt{X$_{\mathrm{b}}$X$_{\mathrm{c}}$YZ}, with the first two characters indicating the eccentricity treatment for planets b and c respectively. This gives 54 two-planet model variants, which combined with the one-planet models means we evaluate 72 models in total.

Of the 18 one-planet models, 7 pass the \textsc{harmonic} diagnostic tests. Of the 54 two-planet models, 23 passed the automated diagnostics; however, visual inspection of the MCMC posteriors for two of these (\texttt{UHW2} and \texttt{HUW0}) revealed unphysical, strongly peaked high-eccentricity solutions at $e_{\mathrm{b}} \approx 1$ for both models. We suspect these are both caused by high-$e$ models fitting to outlier datapoints \citep{Hara.etal2019_BiasRobustnessEccentricity, Osborne.etal2025_HomogeneousPlanetMasses}, especially as both models have uniform priors on $e_{\mathrm{b}}$, so we therefore manually exclude these two models from the analysis, leaving 21 two-planet models for comparison. The 28 retained models (7 one-planet + 21 two-planet) are listed in Table~\ref{tab:table1_toi544_combined}.

The two-planet models are strongly preferred by the Bayesian evidence: the best-scoring two-planet model \texttt{HHG0} is preferred over the best-scoring one-planet model \texttt{HG1} by $\dlnZ = 20.1$, and every passing two-planet \texttt{G} model is preferred over all passing one-planet models. This is consistent with \citet{Osborne.etal2024_TOI544PotentialWaterworld} who found their two-planet model preferred over the one-planet model by $\Delta\mathrm{BIC}=1192.66$ and $\Delta\mathrm{AIC}=1199.56$, reinforcing the robustness of the detection of TOI-544~c.

Within both one- and two-planet models, the GP \texttt{G} models are strongly preferred over white noise \texttt{W} models: the best one-planet white noise model \texttt{CW2} is strongly disfavoured by $\dlnZ = 29.0$ relative to \texttt{HG1}, and the best two-planet white noise models (\texttt{CHW1} and \texttt{CCW1}, tied at $\dlnZ = 34.9$) are similarly strongly disfavoured relative to \texttt{HHG0}. 

The eccentricity treatment is less conclusive than the planet number and noise model comparisons. In the two-planet class, the best-scoring model \texttt{HHG0} with both planets eccentric is weakly preferred over \texttt{CHG0} by $\dlnZ = 1.6$ and \texttt{HCG0} by $\dlnZ=1.7$, and moderately preferred over \texttt{CCG0} by $\dlnZ=3.4$. The evidence therefore disfavours models with circular orbits on both planets, preferring models where one or both of the orbits is eccentric. In the one-planet class, the evidence is inconclusive: the best scoring model \texttt{HG1} is inconclusive over the next model \texttt{CG0} with $\dlnZ=0.2$, while the preference over the more closely comparable model \texttt{CG1} (differing from \texttt{HG1} only in eccentricity treatment) is weak, with $\dlnZ=1.0$.

However, we note that while \citet{Osborne.etal2024_TOI544PotentialWaterworld} did not compare the eccentricity treatment using information criteria or evidence, they did assess the significance of the eccentric models by generating synthetic RV data for circular orbits, and then fitting eccentric models. They reported a $\sim3.5$\% probability that fits with $\eb \geq 0.3$ could arise from noise in the synthetic circular data, but only $\sim0.2$\% for $\ec \geq 0.3$. By applying a 1\% significance limit, they therefore adopted a final model with TOI-544~b fixed on a circular orbit, but allowing TOI-544~c to remain eccentric. 

This configuration corresponds to our \texttt{CH} and \texttt{CU} type models. Of these, our most preferred variant \texttt{CHG0} is weakly disfavoured relative to \texttt{HHG0} by $\dlnZ = 1.6$, while the \texttt{CUG0} model did not pass the \textsc{harmonic} diagnostic tests and was excluded, the only surviving model of the \texttt{CU}-type being \texttt{CUG1}, which was moderately disfavoured by $\dlnZ=3.7$. Our \texttt{CHG0} model value of $\ec={0.270}^{+0.090}_{-0.097}$ is within $1\sigma$ of their final adopted value $\ec = 0.32^{+0.08}_{-0.09}$. When allowing both planets to be eccentric, \citet{Osborne.etal2024_TOI544PotentialWaterworld} reported $\eb=0.35^{+0.14}_{-0.12}$ and $\ec=0.30 \pm 0.09$, both consistent within $1\sigma$ of our \texttt{HHG0} values of $\eb=0.27^{+0.14}_{-0.12}$ and $\ec=0.26^{+0.09}_{-0.10}$. With a log Bayes factor of $\dlnZ=1.6$ for \texttt{HHG0} over \texttt{CHG0}, the evidence weakly favours fitting both planets with an eccentric orbit.

Models with no velocity trends are preferred: in the two-planet class, \texttt{HHG0} is weakly preferred over \texttt{HHG1} by $\dlnZ = 1.7$ and strongly preferred over \texttt{HHG2} by $\dlnZ = 5.3$. In the one-planet class, the no-trend model \texttt{CG0} is inconclusive ($\dlnZ=0.8$) over the linear-trend model \texttt{CG1}, but is the simpler model so is adopted; and \texttt{CG0} is moderately preferred ($\dlnZ=3.4$) over the best-scoring quadratic trend model \texttt{HG2}. \citet{Osborne.etal2024_TOI544PotentialWaterworld} did not fit for linear or quadratic trends in their models, but report no visible trends in their final model.

Of the two-planet models, \texttt{HHG0} (both planets eccentric with half-normal prior, GP, no trend) is the preferred model. Of the one-planet models, we adopt \texttt{CG0} (circular, GP, no trend): the evidence preference for \texttt{HG1} over \texttt{CG0} is inconclusive ($\dlnZ = 0.2$), and \texttt{CG0} has fewer free parameters and is the simpler model. Overall for TOI-544, our preferred model is the two-planet \texttt{HHG0}, as this is strongly favoured over the one-planet model \texttt{CG0} by $\dlnZ = 20.3$. This matches the findings of \citet{Osborne.etal2024_TOI544PotentialWaterworld} who also favoured two-planet models, reporting $\Delta\mathrm{BIC}=1192.66$ and $\Delta\mathrm{AIC}=1199.56$.

The fitted parameters and uncertainties for \texttt{HHG0} are given in Table~\ref{tab:table2_toi544_combined} and phase-folded RV plots for \texttt{HHG0} for TOI-544~b and TOI-544~c can be seen in Fig.~\ref{fig:toi544_2pl_bc_posterior_phase}. For comparison, we also provide the parameter values from the one-planet model \texttt{CG0}. Using the stellar mass $M_\star$ and the orbital inclination $i_{\mathrm{b}}$ of TOI-544~b from \citet{Osborne.etal2024_TOI544PotentialWaterworld}, we derive a mass for TOI-544~b of $M_{p,{\mathrm{b}}} = 3.1 \pm 0.5\,M_{\earth}$, and a minimum mass for the non-transiting TOI-544~c of $M_{p,{\mathrm{c}}}\sin i_{\mathrm{c}} = 21 \pm 2\,M_{\earth}$, within $1\sigma$ of the masses in \citet{Osborne.etal2024_TOI544PotentialWaterworld}, even despite the difference in eccentricity treatment for TOI-544~b.

Inspecting Table~\ref{tab:table1_toi544_combined}, the semi-amplitude $K_{\mathrm{b}}$ of the signal for planet~b is robust to model choice, with the values for all 28 models being self-consistent within $1\sigma$. The planet~c semi-amplitude, by contrast, appears more sensitive particularly to the eccentricity treatment, with models with circular planet~c having somewhat smaller values of $K_{\mathrm{c}}$ and $M_{p,{\mathrm{c}}}\sin i_{\mathrm{c}}$.

A notable difference between the two-planet and one-planet models is the behaviour of the active region evolution timescale $\lambda_{\mathrm{e}}$ GP hyperparameter. While the other hyperparameters are similar between the two models, $\lambda_{\mathrm{e}}$ changes from $95 \pm 30$\,d in the two-planet model \texttt{HHG0} to $16 \pm 3$\,d in the one-planet model \texttt{CG0}. Because $\lambda_{\mathrm{e}}$ has a broad uninformative prior, we interpret this shift as the GP compensating for the unmodelled periodic Keplerian signal of TOI-544~c in the one-planet fit.

With the preferred \texttt{HHG0} model, we find TOI-544~b on a $P=1.55$\,d orbit, with semi-amplitude $K_{\mathrm{b}} = 2.4 \pm 0.4$\,\ms and eccentricity $\eb = 0.27^{+0.14}_{-0.12}$. For TOI-544~c, we find $P=50.1^{+0.3}_{-0.2}$\,d, $K_{\mathrm{c}} = 5.0^{+0.6}_{-0.5}$\,\ms\, and $\ec = 0.26^{+0.09}_{-0.10}$.

%%%%%%%%%%%%%%%% SECTION 6 DISCUSSION %%%%%%%%%%%%%%%%%
%
\section{Discussion}\label{sec:discussion}
%
%%%%%%%%%%%%%%%% SECTION 6.1 BAYESIAN MODEL COMPARISON FOR RADIAL VELOCITIES %%%%%%%%%%%%%%%%%
%
\subsection{Bayesian model comparison for radial velocities}\label{subsec:discussion_bayesian}
We applied 18 model variants to six single-planet systems and 72 model variants (18 one-planet and 54 two-planet) to TOI-544, computing the Bayesian evidence for each using the learned harmonic mean estimator. To our knowledge, this is the first application of the LHME to RV model selection. We find that different systems prefer different models; no single combination of eccentricity treatment, noise model and velocity trend is universally preferred. This reinforces the need to fit multiple model variants and compare them quantitatively, rather than assuming and adopting a single model believed to be most suitable \emph{a priori}, if the resulting planetary masses are to be robust. The \textsc{ravest} and \textsc{harmonic} packages make this workflow straightforward: once MCMC sampling is complete, computing the Bayesian evidence requires only a few additional lines of code and negligible additional computational cost.

Some works use the chi-squared statistic $\chi^2$, or similarly the maximised log-likelihood $\lnL_\mathrm{max}$, to measure the goodness-of-fit between the model and the data. However, $\chi^2$ and $\lnL_\mathrm{max}$ do not incorporate any Occam's razor effect to penalise more complex models, making them unsuitable for robust model comparison. Reduced $\chi^2$ partially addresses this by penalising additional parameters via a degrees-of-freedom term \citep[e.g.][used to diagnose overfitting in GP noise models]{Cale.etal2021_DivingSeaStellar}, but, like AIC and BIC (equation~(\ref{eq:AIC}), equation~(\ref{eq:BIC})), it is a fixed penalty per parameter and does not encode any prior information on the parameters. AIC and BIC remain widely used for model comparison in RV analyses, including for planet detection \citep[e.g.][]{Cloutier.etal2021_MorePreciseMass, Barragan.etal2023_RevisitingK2233Spectroscopic, Osborne.etal2024_TOI544PotentialWaterworld}, trend model comparison \citep[e.g.][]{Palatnick.etal2021_ValidationHD183579b, Morgan.etal2025_ExploringWarmJupiter}, noise model comparison \citep[e.g.][]{Tang.etal2026_RVxTESSModelingAsteroseismic}, and in larger-scale homogeneous studies involving multiple modelling choices \citep[e.g.][]{Bonomo.etal2023_ColdJupitersImproved, Osborne.etal2025_HomogeneousPlanetMasses}. However, their derivations require assumptions that do not hold well for typical RV modelling with sparse datasets \citep{Nelson.etal2020_QuantifyingBayesianEvidence}: they treat all parameters equally regardless of the data's ability to constrain them and the model's sensitivity towards them, are unsuited for multimodal or non-Gaussian posteriors, and depend only on $\lnL_\mathrm{max}$ and parameter count $k$, incorporating none of the prior information on parameters \citep{Ford.Gregory2007_BayesianModelSelection, Trotta2008_BayesSkyBayesian}.

Our results highlight these limitations: for example, with GJ~1214 (Section~\ref{subsec:tic467_results}) the AIC and BIC both indicate preference for a white noise model over a GP, penalising the additional GP hyperparameters, yet the Bayesian evidence moderately favours the GP model, consistent with independent photometric evidence for quasi-periodic rotationally modulated stellar variability in this system \citep{Mallonn.etal2018_GJ1214Rotation, Cloutier.etal2021_MorePreciseMass}. Because the Bayesian evidence is the prior-weighted average of the likelihood (equation~(\ref{eq:evidence})), it naturally penalises GP hyperparameters only if they are not well constrained by the data (through the Occam's razor mechanism described in Section~\ref{subsec:model_comparison}), rather than applying a fixed dimensionality-based penalty which may not truly reflect the balance between model flexibility and overfitting.

Nested sampling \citep{Skilling2004_NestedSampling} is a commonly-used method for estimating Bayesian evidence in RV analysis \citep{Nelson.etal2020_QuantifyingBayesianEvidence}. Where direct comparison is possible, our results are consistent with nested sampling: \citet{Desidera.etal2023_TOI179YoungSystem} used nested sampling for HD~18599 and found a linear trend preferred over quadratic by $\dlnZ = 10.2$; our result of $\dlnZ = 12.7$ reaches the same conclusion. Similarly, \citet{Gan.etal2021_HD183579bWarm} used nested sampling for TOI-1055 and found a GP noise model preferred with $\dlnZ = 8$; our result of $\dlnZ = 14.4$ again reaches the same conclusion, with stronger evidence likely reflecting our longer observational baseline. The practical advantage of the LHME over nested sampling is not only the significantly reduced computational cost, but workflow integration: nested sampling requires running a dedicated algorithm during fitting, limiting the choice of sampling method and often increasing the total computational time. By contrast, the LHME easily slots into existing MCMC workflows without any modification to the existing fitting procedure, allowing the use of the most efficient or suitable sampling method for the task rather than being restricted to nested methods, and can even be applied post-hoc to archived chains. Furthermore, the computational cost scales favourably with the number of model parameters, remaining practical for the dimensionalities typical of RV models (up to $\sim$20 free parameters in this work) and having been successfully applied to cosmological problems with $\sim$150 parameters \citep{Piras.etal2024_FutureCosmologicalLikelihoodbased}.

%%%%%%%%%%%%%%%% SECTION 6.2 IMPLICATIONS FOR RV MODELLING %%%%%%%%%%%%%%%%%
%
\subsection{Implications for RV modelling}\label{subsec:discussion_implications}
Circular models are preferred by the Bayesian evidence in five of the six single-transiting planet systems. However, this does not necessarily imply that these orbits are truly circular: small eccentricities can be challenging to measure precisely from RV data \citep[e.g.][]{Hara.etal2019_BiasRobustnessEccentricity, VanEylen.etal2019_OrbitalEccentricitySmall}. Eccentric models introduce two additional free parameters per planet ($\secosw$ and $\sesinw$), and where the data are insufficient to constrain them, the Bayesian evidence naturally penalises this additional complexity through its built-in Occam's razor (Section~\ref{subsec:model_comparison}). For K2-265, our adoption of a circular model leads to a disagreement with \citet{Lam.etal2018_K2265TransitingRocky}, who adopted an eccentric model and reported $e = 0.084 \pm 0.079$, although their eccentricity is consistent with zero within $1.06\sigma$. Our model comparison yields a Bayes factor of only $\dlnZ = 0.4$ between the circular and eccentric models (and similarly $\dlnZ = 0.5$ for TOI-1055 with $e = 0.13^{+0.20}_{-0.09}$), indicating that the available data cannot constrain the eccentricity enough to justify the additional flexibility of an eccentric model in either case. The Bayesian evidence thus provides a quantitative framework for determining whether the data support reporting an eccentric orbit, particularly for low eccentricities typical of short-period planets (due to tidal circularisation, e.g.\ \citealt{VanEylen.etal2019_OrbitalEccentricitySmall}) that are inherently difficult to constrain with RV data alone. 

Of the six single-planet systems, HD~18599~b is the sole system where a non-zero eccentricity is preferred, where we find $e = 0.38^{+0.15}_{-0.14}$, consistent with $e = 0.34^{+0.07}_{-0.09}$ reported by \citet{Desidera.etal2023_TOI179YoungSystem}, though even here the eccentric model is only weakly preferred over the circular model ($\dlnZ = 1.7$). For TOI-544, models with at least one eccentric planet are preferred: the fully eccentric two-planet model \texttt{HHG0} is moderately preferred over the fully circular \texttt{CCG0} by $\dlnZ = 3.4$. In all systems, \texttt{U} models with uniform eccentricity priors consistently underperform: across the six single-planet systems, no \texttt{U} model is ever the preferred model; for LHS~1815, K2-265, and for the TOI-544 one-planet models, all six \texttt{U} models fail the \textsc{harmonic} diagnostic tests entirely. Examining the K2-265 \texttt{UG0} eccentricity posterior (see Section~\ref{subsec:tic146_results} and Fig.~\ref{fig:146_eccentricity_comparison}) reveals a spurious secondary peak near $e \approx 1$ -- an unphysical solution, not least because the resulting periastron distance would lie inside the host star -- arising when the model fits to outlier data with implausibly high eccentricity. This was often the cause of \texttt{U} models failing the \textsc{harmonic} diagnostic checks while the equivalent \texttt{H} model succeeded, as the secondary peak at $e\approx1$ in the \texttt{U} model meant that the posterior distribution was no longer unimodal, causing the LHME to fail. We have verified that across all planets in this work, every model passing the diagnostic checks recovers an eccentricity well below the physical limit imposed by considering the periastron distance. Overall, the results of our model comparisons reinforce the recommendation of \citet{Osborne.etal2025_HomogeneousPlanetMasses} against uniform eccentricity priors in RV analyses: in our modelling, the half-normal prior from \citet{VanEylen.etal2019_OrbitalEccentricitySmall} consistently produces better-behaved posterior distributions for $e$, and the \texttt{H} models were consistently preferred over \texttt{U} models, motivating the use of informative prior distributions (e.g.\ Rayleigh, beta or half-normal distributions) on eccentricity where they can be justified. Robust values of eccentricity are important not just for the effect on RV semi-amplitude $K$ and planetary masses, but also because eccentricity provides constraints on formation and dynamical history, and affects atmospheric compositions and habitability through variations in stellar insolation and tidal heating \citep{Sagear.Ballard2023_OrbitalEccentricityDistribution}, motivating comparison of models (including choice of priors) to robustly determine eccentricity.

Gaussian Process noise models are preferred by the Bayesian evidence in five of the seven systems: K2-265, HD~18599, TOI-1055, GJ~1214 and TOI-544 -- all of which have independently detected rotationally modulated stellar activity. For TOI-544 this holds across both one- and two-planet models, with white noise models strongly disfavoured in both. TOI-220 and LHS~1815 are the exceptions where white noise models are preferred; in both cases, independent evidence (\citealt{Gan.etal2020_LHS1815bFirst, Hoyer.etal2021_TOI220WarmSubNeptune}; see Section~\ref{subsec:tic150_results} and Section~\ref{subsec:tic260_results}) indicates that quasi-periodic stellar activity is not dominant in the RV observations for those systems. We investigated the possibility of GP overfitting directly for HD~18599 (Section~\ref{sub:tic207_noise_comparison}), where we believe that the preferred GP model is justified over a white noise model because of significantly smaller residuals and jitter, and by previous photometric and periodogram analyses confirming both that HD~18599 exhibits rotationally modulated stellar variability, and that stellar activity dominates the RV time series \citep{Desidera.etal2023_TOI179YoungSystem}, though future studies could compare the use of alternative GP kernels. Across our results, we do not find evidence that GP models are preferred due to overfitting: for systems with independent evidence of quasi-periodic rotationally modulated stellar variability, the \texttt{G} models are preferred, and for systems where it does not significantly affect the RVs, the simpler \texttt{W} models are preferred -- the Bayesian evidence penalises the flexibility of the QP GP when the data do not warrant it. We note however that in general when fitting RVs, a preference for white noise models over a QP GP model alone does not necessarily imply that a star is inactive: the RV observations may not adequately sample the stellar rotation or spot evolution timescales \citep{Nicholson.Aigrain2022_QuasiperiodicGaussianProcesses, Osborne.etal2025_HomogeneousPlanetMasses} -- in which case additional observations to improve the observing coverage of the stellar variability could help to determine if a GP noise model is justified -- or there may be more suitable alternative kernels (e.g.\ Mat\'ern-5/2, QP cosine, Simple Harmonic Oscillator; \citealt{Hara.Ford2023_StatisticalMethodsExoplanet, Gupta.Bedell2024_FishingPlanetsComparative}). We also note the contrasting case highlighted in \citealt{Blunt.etal2023_OverfittingAffectsReliability} where previous $\ln\Z$ comparison was shown to favour a GP model that was overfitting -- with the difficulty of modelling stellar variability, especially with sparse RV data, the decision on whether the use of a GP and/or a specific kernel is justified can be bolstered by other analyses, such as studies of photometric data and spectral activity indicators to further investigate the activity level of the star \citep[e.g.][]{Gan.etal2020_LHS1815bFirst, Hoyer.etal2021_TOI220WarmSubNeptune, Polanski.etal2024_TESSKeckSurveyXX}.

Comparing $N$ versus $N+1$ planet models is one of the most challenging decisions in RV analysis \citep{Hara.Ford2023_StatisticalMethodsExoplanet}, particularly for non-transiting companions where there is no transit signal to confirm the additional Keplerian. In previous studies, $N$ versus $N+1$ planet model selection has been performed using BIC \citep{Bonomo.etal2023_ColdJupitersImproved} as well as direct estimates of the Bayesian evidence $\Z$ via a variety of methods, including thermodynamic integration with parallel-tempered MCMC \citep{Gregory2007_BayesianPeriodogramFinds}, the Chib--Jeliazkov estimator \citep{Dumusque.etal2014_Kepler10PlanetarySystem}, nested sampling \citep{Feroz.Hobson2014_BayesianAnalysisRadial} and importance sampling \citep{Nelson.etal2016_EmpiricallyDerivedThreedimensional}; while \citet{Nelson.etal2020_QuantifyingBayesianEvidence} shows the results of a data challenge that compared a range of Bayesian evidence estimators using simulated RV observations. Our TOI-544 analysis (Section~\ref{subsec:toi544_results}) demonstrates the application of the LHME to $N$ versus $N+1$ planet model comparison, finding strong evidence ($\dlnZ = 20.1$) for the two-planet model over the one-planet model, consistent with the AIC/BIC analysis of \citet{Osborne.etal2024_TOI544PotentialWaterworld}.

We find that long-term velocity trends are preferred in only two of seven systems: HD~18599 (linear, attributed to the known long-period companion HD~18599~B; \citealt{Desidera.etal2023_TOI179YoungSystem}) and TOI-220 (quadratic). For TOI-220, \citet{Hoyer.etal2021_TOI220WarmSubNeptune} found structured RV residuals inconsistent with a single-planet model and attributed the quadratic drift to a possible long-period companion, though the timespan of the RV data ($\sim$327\,d) was too short to constrain its orbital period (suspected to be $\sim300 \pm 100$\,d from Lomb--Scargle periodogram analysis). Our strong preference for the model including the quadratic trend ($\dlnZ = 6.4$) is consistent with their finding. For the remaining five systems, trends are strongly disfavoured by the Bayesian evidence, in agreement with the previous studies for those systems.

Semi-amplitudes from our preferred models are in agreement with the literature values within $1\sigma$ for all seven systems. However, the sensitivity of derived masses to model choice varies considerably between systems (see Fig.~\ref{fig:mpsini_grid} and Fig.~\ref{fig:mpsini_grid_toi544}), and this is where model comparison shows its importance. For GJ~1214, $K = 13.6 \pm 0.7$\,\ms is invariant across all ten retained models (see Section~\ref{subsec:tic467_results}), meaning the mass is robust regardless of model choice. At the other extreme, TOI-1055 (see Fig.~\ref{fig:320_gp_nogp_comparison}) shows a factor of three difference in semi-amplitude between the preferred GP model \texttt{CG0} ($K = 4.4 \pm 1.3$\,\ms) and the white noise model \texttt{CW0} ($K = 1.5^{+0.9}_{-0.8}$\,\ms), with the Bayesian evidence strongly favouring the GP model by $\dlnZ = 14.4$.

Perhaps more intriguing are cases where the evidence difference is smaller but the mass difference is large: for HD~18599, the preferred eccentric model \texttt{HG1} yields $K = 10 \pm 4$\,\ms and $\Mpsini=20^{+7}_{-8}\,M_{\earth}$, while the circular model \texttt{CG1} (weakly disfavoured by $\dlnZ = 1.7$) yields $K = 5 \pm 3$\,\ms and $\Mpsini=11^{+7}_{-6}\,M_{\earth}$, a factor of two difference. The Bayesian evidence helps identify even subtle differences in the quality of the fits, giving a clearer picture of which model is more suitable -- this is especially important where the choice of model would have significant ramifications for the resulting masses. In this case, the Bayesian evidence preference for the eccentric model is also consistent with previous studies of this system \citep{Desidera.etal2023_TOI179YoungSystem, Vines.etal2023_DenseMiniNeptuneOrbiting}. 

For TOI-544, $K_{\mathrm{b}}$ is similar across all 28 retained models, while for the 21 two-planet models $K_{\mathrm{c}}$ of the non-transiting outer planet is more sensitive to model choice, particularly the eccentricity treatment for planet~c, with circular planet~c models yielding somewhat smaller values of $K_{\mathrm{c}}$ and $M_{p,{\mathrm{c}}}\sin i_{\mathrm{c}}$. The stability of $K_{\mathrm{b}}$ across both one- and two-planet models means a single-planet analysis would yield a seemingly reasonable mass for planet~b, masking the omission of planet~c, thereby illustrating the importance of comparing $N$ vs $N+1$ planet models. Overall, these examples demonstrate that model comparison is important to ensure robust masses: modelling choices can significantly affect derived planetary masses, with downstream consequences for bulk densities, compositions and atmospheric characterisation.

%%%%%%%%%%%%%%%% SECTION 6.3 LIMITATIONS AND CAVEATS %%%%%%%%%%%%%%%%%
%
\subsection{Limitations and caveats}\label{subsec:discussion_caveats}
The Bayesian evidence $\Z$ is conditioned on the observed data $\bm{y}$ (see equation~(\ref{eq:evidence})): all models being compared must use the same dataset. This means, for example, that models fitted to different subsets of the RV observations cannot be directly compared; nor can a 1D RV-only GP be compared with a multidimensional GP incorporating spectroscopic activity indicators such as FWHM or BIS, or two multidimensional GPs using different sets of activity indicators; nor can models that use simultaneous photometric analysis to inform RV GP hyperparameters \citep[e.g.\ the $FF'$ method;][]{Aigrain.etal2012_SimpleMethodEstimate, Haywood.etal2014_PlanetsStellarActivity} be compared with models that do not; nor can models fitted to RVs extracted by different reduction pipelines (e.g.\ \textsc{terra} versus \textsc{drs}). In all such cases, different data enter the likelihood.

This last point is relevant when comparing our results with the literature: for example, \citet{Cloutier.etal2021_MorePreciseMass}, \citet{Luque.Palle2022_DensityNotRadius}, \citet{Bonfanti.etal2023_TOI1055NeptunianPlanet} and \citet{Desidera.etal2023_TOI179YoungSystem} used \textsc{terra} or \textsc{serval} to reduce HARPS RVs for GJ~1214 LHS~1815, TOI-1055 and HD~18599 respectively, whereas our data from \citet{Barbieri2023_ESOHARPSRadial} use the standard HARPS \textsc{drs} pipeline. As a result, not only are our absolute $\ln\Z$ values not directly comparable with those in the literature, but our derived RV parameter values (e.g.\ $K$, $e$) may also differ slightly, since the underlying data are not identical. Where we compare our model comparison conclusions with the literature (Section~\ref{subsec:discussion_bayesian}), we therefore compare qualitatively -- which model is preferred over another, and the strength of the preference -- rather than comparing the exact $\dlnZ$ values.

The version of \textsc{harmonic} used in this work uses normalising flows \citep{Polanska.etal2025_LearnedHarmonicMean} to learn the importance sampling target distribution required to estimate the evidence, and is therefore restricted to unimodal posterior distributions. In our analysis, multimodal posteriors arise most commonly from models with uniform eccentricity priors fitting spurious high-$e$ solutions (Section~\ref{subsec:tic146_results}), but they can also appear in GP hyperparameters (e.g.\ harmonics of stellar rotation period showing as secondary peaks in $P_\mathrm{GP}$), or in cases where the data are insufficient (e.g.\ not enough observations, poor cadence/phase coverage, or too short a baseline) to constrain the period of a periodic pattern in the data \citep[e.g.\ the search for a long-period companion in][]{Hoyer.etal2021_TOI220WarmSubNeptune}. Crucially, \textsc{harmonic} has diagnostics to help identify when the evidence estimate is not reliable (such as due to multimodal posteriors), whereas $\chi^2$, AIC and BIC -- being calculations based just on the residuals, number of datapoints and number of parameters -- will return numerical values regardless of whether the chains have converged or the posterior is well-behaved. This reinforces the need to perform convergence checks and inspect the chains and posterior sample distributions of models. Modern nested sampling algorithms such as \textsc{MultiNest} \citep{Feroz.etal2009_MULTINESTEfficientRobust}, \textsc{polychord} \citep{Handley.etal2015_POLYCHORDNextgenerationNested} and \textsc{dynesty} \citep{Speagle2020_DYNESTYDynamicNested} can sample from complex multimodal posteriors, an advantage over the normalising-flow implementation of \textsc{harmonic} used in this work. However, a recent release of \textsc{harmonic} incorporating flow-matching methods \citep{Polanska.McEwen2026_LearnedHarmonicMean} can also handle multimodal posteriors; this was released after our analysis was conducted, but could be applied in future studies.

Unlike the other model comparison metrics discussed, the Bayesian evidence is also sensitive to the priors. Because the priors are part of any model specification alongside the likelihood, this is a source of strength when priors are well motivated \citep{Thorngren.etal2026_BayesianModelComparison}, but if priors are arbitrary, then the sensitivity of Bayesian evidence to the volume of the priors can be problematic \citep{Hu.McEwen2026_EfficientPriorSensitivity}. If the support of the prior is increased between two otherwise identical models, then the Bayesian evidence can drop drastically, especially under a highly informative likelihood that would be nearly identical between the two models \citep{Hu.McEwen2026_EfficientPriorSensitivity, Thorngren.etal2026_BayesianModelComparison}. This motivates the use of well-justified, normalised priors. To ensure priors are normalised, care must be taken where a change of parameterisation is used (e.g.\ sampling in $\secosw$ and $\sesinw$ rather than $e$ and $\wstar$), and the appropriate Jacobian term must be included where required. Furthermore, if priors are specified in a different parameterisation than is being sampled in (e.g.\ for the \texttt{H} models sampling in ($\secosw$, $\sesinw$) but with priors on ($e$, $\wstar$)) the support of the priors must be considered to ensure they remain normalised, else evidence estimates from posterior samples could be biased for those models (see Appendix~\ref{app:app1} for further explanation).

Without a transit signal to provide prior knowledge of the orbital period and transit timing, the detection of non-transiting planets in RV analysis can be challenging due to the broad parameter space that needs to be explored. We emphasise that the LHME requires well-behaved (converged, unimodal) posterior samples and so is not directly suited to a fully blind search across very broad uninformative priors; identifying candidate signals in RV data remains the domain of dedicated techniques such as periodogram analysis. Where a candidate signal has been identified, the LHME can quantify the strength of preference for modelling that signal with an additional Keplerian, as we demonstrate with TOI-544 (Section~\ref{subsec:toi544_results}, Section~\ref{subsec:discussion_implications}). If the RV baseline is too short to fit a Keplerian orbit, different trend models can be compared, where the presence of a trend can potentially indicate a long-period companion, as in TOI-220 and HD~18599 (\citealt{Hoyer.etal2021_TOI220WarmSubNeptune, Desidera.etal2023_TOI179YoungSystem}; see Section~\ref{subsec:tic150_results} and Section~\ref{subsec:tic207_results}). However, further investigation is needed to distinguish between planetary, stellar and instrumental causes of any detected trend.

%%%%%%%%%%%%%%%% SECTION 7 CONCLUSIONS %%%%%%%%%%%%%%%%%

\section{Conclusions}\label{sec:conclusions}
We have demonstrated the first application of the learned harmonic mean estimator \citep{McEwen.etal2023_MachineLearningAssisted, Polanska.etal2025_LearnedHarmonicMean} to radial velocity model selection, fitting 18 model variants -- spanning eccentricity treatment, noise model, and velocity trend -- to six single-planet systems, and 72 model variants to TOI-544 (additionally comparing one- and two-planet fits), estimating the Bayesian evidence for each. Our aim is to demonstrate the applicability of the method across common RV modelling choices, not to provide a comprehensive survey of the exoplanet population. The LHME estimates the evidence directly from MCMC posterior samples, requiring no modification to the RV fitting procedure and negligible additional computational cost. Where direct comparison with previous model comparisons in the literature that used nested sampling is possible -- HD~18599 \citep{Desidera.etal2023_TOI179YoungSystem} and TOI-1055 \citep{Gan.etal2021_HD183579bWarm} -- our results reach the same conclusions. By also applying the LHME to TOI-544, comparing 18 one-planet and 54 two-planet model variants, we find strong evidence for a two-planet model ($\dlnZ = 20.1$) in agreement with the findings of \citet{Osborne.etal2024_TOI544PotentialWaterworld}, and provide a demonstration of Bayesian model comparison for the $N$ versus $N+1$ planets problem.

Our results reinforce that no single model is universally appropriate for RV fitting: different systems prefer different combinations of eccentricity treatment, noise model and velocity trend. Circular models are preferred in five of six single-planet systems, reflecting the difficulty of constraining small eccentricities with sparse RV data. Uniform eccentricity priors consistently underperform, with no \texttt{U} model ever preferred across any system, supporting the use of informative priors such as half-normal, beta or Rayleigh distributions \citep{VanEylen.etal2019_OrbitalEccentricitySmall}. Models with quasi-periodic GPs are preferred for systems where independent evidence for rotationally modulated stellar variability exists, and white noise models preferred in systems where it does not. Critically, model choice directly affects derived planetary masses: semi-amplitudes vary by factors of two to three between competing models in some systems, with downstream consequences for bulk densities, compositions and atmospheric characterisation.

Based on our findings, we make the following recommendations for RV model comparison:
\begin{enumerate}
  \item Fit multiple model variants and compare them using the Bayesian evidence, rather than adopting a single preferred model \emph{a priori}.
  \item Use informative eccentricity priors, rather than uniform priors.
  \item For accurate evidence estimation, ensure all priors are properly normalised -- when changes of parameterisation are used, the Jacobian term from the change of variables and normalisation over the support of the prior need to be considered (see Appendix~\ref{app:app1}).
\end{enumerate}
The LHME is implemented in the open-source \textsc{harmonic} package, and our RV fitting framework is available in the open-source \textsc{ravest} package. The LHME integrates easily with any MCMC sampling technique, making rigorous Bayesian model comparison accessible to the RV community, but also applicable to any domain where MCMC posterior samples exist and model comparison is needed.

Several avenues for future work remain. Within RV modelling, the comparison of models that use multidimensional GPs incorporating different spectral activity indicator time series remains challenging, as these models have different data entering the likelihood and so cannot be compared directly through their Bayesian evidence $\Z$ (Section~\ref{subsec:discussion_caveats}). The recently-introduced Multi-GP Information Criterion $\mathrm{MGIC}_{\mathrm{rv}}$ \citep{Barragan2026_ModelSelectionCriterion} treats the multidimensional inference as a larger one-dimensional inference, and reanalysis of previous multidimensional GP RV studies demonstrates that it successfully quantifies the performance of different activity indicators and kernel functions; though as an information criterion it shares the limitations of AIC and BIC by not including the prior specification or the full likelihood distribution (Section~\ref{subsec:discussion_bayesian}). Future studies could investigate the feasibility of a fully Bayesian route, via a sequential pre-training approach: a multidimensional GP is first fitted to the spectral activity indicator time series only, producing posterior distributions for the GP kernel hyperparameters, which are then used as data-driven informative priors on the hyperparameters of a 1D (RV-only) GP; competing models can then all be evaluated on the same RV dataset, allowing the use of the LHME while still leveraging the additional information from ancillary indicators. Another future application of the LHME could be for model comparisons of different GP kernel functions alongside the quasi-periodic kernel used in this work -- building upon the work in previous studies such as \citet{Gupta.Bedell2024_FishingPlanetsComparative} -- in particular investigating whether the optimal kernel varies with stellar type. There is undoubtedly still more to learn about RV fitting from future model comparisons, particularly in regard to modelling stellar variability.

Beyond RV, the LHME is applicable to any problem using MCMC sampling where model comparison is needed -- natural extensions in exoplanet science include photometric transit fitting and transit-timing variation surveys \citep[e.g.][]{Naponiello2026_HomogeneousTransittimingvariationInvestigation}, and atmospheric retrievals -- and it has already been applied in other areas of astrophysics such as cosmology \citep[e.g.][]{Carrion.etal2025_HarmonicExample, Du.etal2025_HarmonicExample,  Stiskalek.etal2025_HarmonicExample, Stiskalek.etal2026_HarmonicExample}. We encourage its use wherever model selection decisions currently rely on AIC, BIC or similar approximations, and as a computationally inexpensive alternative to nested sampling.

%%%%%%%%%%%%%%%%%%%%%%%%%%%%%%%%%%%%%%%%%%%%%%%%%%
\section*{Acknowledgements}

We thank the anonymous referee for reviewing the paper and for their valuable suggestions which have improved the manuscript.

R.S.D. thanks Alicja A. Pola\'{n}ska for useful discussions regarding the learned harmonic mean estimator and the \textsc{harmonic} software package.

R.S.D. has been supported by the UCL Centre for Doctoral Training in Data Intensive Science (STFC grant number ST/P006736/1) including departmental and industry contributions.

V.V.E. has been supported by the UK's Science \& Technology Facilities Council through the STFC grants ST/W001136/1 and ST/S000216/1, and UKRI1180.

K.S. has been supported by the UK's Science \& Technology Facilities Council through the STFC grant UKRI1180.

This research has made use of the NASA Exoplanet Archive, which is operated by the California Institute of Technology, under contract with the National Aeronautics and Space Administration under the Exoplanet Exploration Program.

This research has made use of the SIMBAD database, operated at CDS, Strasbourg, France \citep{Wenger.etal2000_SIMBADAstronomicalDatabase}.

This research has made use of the VizieR catalogue access tool, CDS, Strasbourg, France \citep{10.26093/cds/vizier}. The original description of the VizieR service was published in \citet{vizier2000}.

This work makes use of the \textsc{NumPy} \citep{harris2020array}, \textsc{Matplotlib} \citep{Hunter:2007}, \textsc{pandas} \citep{mckinney-proc-scipy-2010}, \textsc{Astropy} \citep{astropy:2013, astropy:2018, astropy:2022}, \textsc{SciPy} \citep{2020SciPy-NMeth}, \textsc{emcee} \citep{Foreman-Mackey.etal2013_EmceeMCMCHammer}, \textsc{tinygp} \citep{foreman_mackey_2024_10463641}, \textsc{GetDist} \citep{Lewis2025_GetDistPythonPackage}, and \textsc{harmonic} \citep{McEwen.etal2023_MachineLearningAssisted, Polanska.etal2025_LearnedHarmonicMean} software packages. This work makes use of the colormaps available in the \textsc{CMasher} software package \citep{2020JOSS....5.2004V}.

%%%%%%%%%%%%%%%%%%%%%%%%%%%%%%%%%%%%%%%%%%%%%%%%%%
\section*{Data Availability}
The HARPS radial velocity data for the K2-265, TOI-220, HD~18599, LHS~1815, TOI-1055 and GJ~1214 systems are from \citet{Osborne.etal2025_HomogeneousPlanetMasses}, available via the CDS VizieR service entry \href{https://doi.org/10.26093/cds/vizier.36930004}{J/A+A/693/A4}, and are also publicly available from the ESO archive as described by \citet{Barbieri2023_ESOHARPSRadial}. The HARPS and HARPS-N \textsc{drs} radial velocity data for the TOI-544 system were obtained from the online tables in appendix~A of \citet{Osborne.etal2024_TOI544PotentialWaterworld}.



\bibliographystyle{mnras}
\bibliography{references}

%%%%%%%%%%%%%%%%%%%%%%%%%%%%%%%%%%%%%%%%%%%%%%%%%%

%%%%%%%%%%%%%%%%% APPENDICES %%%%%%%%%%%%%%%%%%%%%

\appendix

%%%%%%%%%%%%%%%%%%%% appendix 1: prior normalisation %%%%%%%%%%%%%%%%%%%%%%%
\section{Prior normalisation}\label{app:app1}
\subsection{Importance for Bayesian model comparison}
\label{app:priors_why}

When performing Bayesian parameter inference with MCMC, under a fixed model $M$, the evidence $\Z$ acts only as a normalisation constant and does not affect the relative probability of any parameter values. For each proposed parameter vector $\bm{\theta}$, the posterior is therefore sampled up to a constant (the evidence $\Z$), i.e.\
\begin{equation}
  p(\bm{\theta}|\bm{y},M) \propto 
  \Like(\bm{\theta})\,\pi(\bm{\theta}),
\end{equation}
or more commonly (for numerical stability) the logarithm of the above equation:
\begin{equation}
  \ln p(\bm{\theta}|\bm{y},M) \propto 
  \ln\Like(\bm{\theta}) + \ln \pi(\bm{\theta}),
\end{equation}
thus sampling the log-posterior $\ln p(\bm{\theta}|\bm{y},M)$ up to a normalising constant, the log-evidence $-\ln\Z$ (see equation~(\ref{eq:bayes}) and Section~\ref{sec:bayesian_comparison}). Here $\Like(\bm{\theta})$ denotes the likelihood and $\pi(\bm{\theta})$ the prior; for brevity we drop the explicit conditioning on $\bm{y}$ and $M$ throughout this appendix. 

In the Metropolis--Hastings MCMC algorithm, the acceptance ratio depends only on the ratio of posterior values between proposed and current states:
\begin{equation}
  \label{eq:metropolis_hastings}
  \alpha = \min\left(1,\,
  \frac{\Like(\bm{\theta}')\,\pi(\bm{\theta}')}{\Like(\bm{\theta})\,\pi(\bm{\theta})}
  \times
  \frac{q(\bm{\theta} | \bm{\theta}')}{q(\bm{\theta}' | \bm{\theta})}
  \right),
\end{equation}
where $\bm{\theta}$ is the current parameter vector, $\bm{\theta}'$ is the proposed parameter vector, and $q(\bm{\theta}'|\bm{\theta})$ is the proposal distribution \citep{Ford2006_ImprovingEfficiencyMarkov, Foreman-Mackey.etal2013_EmceeMCMCHammer}. Therefore it can be seen that any constant term $C$ in the ratio (i.e. any term that does not change with the parameters) would simply cancel:
\begin{equation}
  \frac{\Like(\bm{\theta}')\cdot C\pi(\bm{\theta}')}{\Like(\bm{\theta})\cdot C\pi(\bm{\theta})} =
  \frac{\Like(\bm{\theta}')\,\pi(\bm{\theta}')}{\Like(\bm{\theta})\,\pi(\bm{\theta})}.
\end{equation}
In this work, we use the affine-invariant ensemble sampler from the \textsc{emcee} package \citep{Foreman-Mackey.etal2013_EmceeMCMCHammer}, which uses a stretch move with acceptance criterion
\begin{equation}\label{eq:affine_invariant}
  \alpha = \min\left(1,\, a^{N-1} \cdot
  \frac{\Like(\bm{\theta}')\,\pi(\bm{\theta}')}{\Like(\bm{\theta})\,\pi(\bm{\theta})}
  \right),
\end{equation}
where $a$ is the stretch scale factor and $N$ is the number of dimensions. It can be seen that any constant normalisation term in $\pi(\bm{\theta})$ cancels identically in this ratio. Therefore, MCMC parameter inference does not require the prior $\pi(\bm{\theta})$ to be correctly normalised (nor the posterior as a whole; \citealt{Ford2006_ImprovingEfficiencyMarkov}).

However, for Bayesian model comparison this is not the case. If the prior is not correctly normalised, then the estimated evidence $\hat{\Z}$ is affected because it is the integral of the likelihood with respect to the prior \citep{Gelman.etal2014_BayesianDataAnalysis}:
\begin{equation}
\Z = \int \Like(\bm{\theta})\,\pi(\bm{\theta})\,\mathrm{d}\bm{\theta}.
\end{equation}

In this work, the LHME from \textsc{harmonic} uses log-posterior samples from MCMC to estimate the log evidence. Therefore if the implemented prior density, which we can denote $\tilde{\pi}(\bm{\theta})$, is not correctly normalised and differs from the properly normalised density $\pi(\bm{\theta})$ by a constant factor $C$, such that $\tilde{\pi}(\bm{\theta}) = C\cdot\pi(\bm{\theta})$, then
\begin{equation}
  \label{eq:lnZ_shift}
  \ln\Like + \ln\pi  = \ln\Like + \ln\tilde{\pi} - \ln C.
\end{equation}
meaning the estimated log evidence value $\ln{\hat{\Z}}$ will be systematically shifted by the missing normalisation $\ln C$ across models that share the same prior construction. Therefore, if the models being compared via evidences do not share the same priors on parameters, and if for one of those models the prior is not correctly normalised, then the improperly normalised prior will systematically bias the evidence estimates for that model, meaning the model comparison will not be fair. 

We highlight this issue because in RV fitting we are often sampling in transformed parameterisations. A common reparameterisation arises because sampling directly in $(e, \wstar)$ is not recommended, partly due to the Lucy--Sweeney bias \citep{Lucy.Sweeney1971_SpectroscopicBinariesCircular, Eastman.etal2013_EXOFASTFastExoplanetary} for low eccentricities caused by the boundary at $e=0$, and also due to slow convergence speeds especially when $e$ is low and $\wstar$ is poorly constrained \citep{Ford2005_QuantifyingUncertaintyOrbits}. Instead, alternative parameterisations are commonly adopted, such as sampling for $(\secosw, \sesinw)$ in place of $(e, \wstar)$ \citep[e.g.][]{Anderson.etal2010_WASP30b61MJup, Eastman.etal2013_EXOFASTFastExoplanetary}. However, when sampling in transformed parameterisations, care must be taken to ensure all priors in all models are correctly normalised.

The principle to enforce is that the prior $\pi(\bm{\theta})$ used must be a proper probability density with respect to the sampling parameterisation -- so that it integrates to unity over the supported region -- in order for the evidence estimated by \textsc{harmonic} to be unbiased. Two common ways this can fail for the eccentric orbit models used in this paper (where we sample in $(\secosw, \sesinw)$) are omission of the Jacobian determinant under a change of variables (where the sampling and prior parameterisations differ), and inconsistency between the support of a prior distribution and the physically-allowed region for a parameter. In this work, this affects the half-normal eccentricity prior \texttt{H} models, and also the uniform eccentricity prior \texttt{U} models, for two different reasons which we explain below. In both cases, an additional term must be added to $\ln\pi(\bm{\theta})$ to correctly normalise the priors in the log-posterior equation used in the MCMC sampling.

\subsection{Notation}
\label{app:priors_notation}
In our work, the only parameters affected are eccentricity $e$ and argument of periastron $\wstar$, and their alternative parameterisations $\secosw$ and $\sesinw$. For brevity in this appendix, we adopt the notation $u \equiv \secosw$ and $v \equiv \sesinw$. Furthermore, we omit the other model parameters and focus only on $e, \wstar, u, v$, as all other parameters are unchanged in parameterisation, form, and priors between the models we compare.

The mapping between the MCMC sampling parameterisation $(u,v)$ and the physical parameters $(e,\wstar)$ used in the Keplerian RV equation in the likelihood functions is
\begin{equation}\label{eq:ew_to_uv}
u = \sqrt{e}\cos\wstar, \qquad v = \sqrt{e}\sin\wstar,
\end{equation}
with inverse mapping
\begin{equation}\label{eq:uv_to_ew}
e = u^2 + v^2, \qquad \wstar = \operatorname{atan2}(v,u),
\end{equation}
where $\operatorname{atan2}$ denotes the two-argument arctangent that returns the correct quadrant. The physical requirement $e \in [0,1)$ corresponds to the open unit disk formed by $u^2 + v^2 < 1$.

When the sampling parameterisation differs from the parameterisation in which the prior is defined, the prior density must be transformed using the change-of-variables formula \citep{Ford2006_ImprovingEfficiencyMarkov, Hogg.Foreman-Mackey2018_DataAnalysisRecipes}:
\begin{equation}\label{eq:change_of_variables}
  \pi(u,v) = \pi(e,\wstar)\,\left|\det \left(\frac{\partial(e,\wstar)}{\partial(u,v)}\right)\right|,
\end{equation}
where $J = \partial(e,\wstar)/\partial(u,v)$ is the Jacobian matrix of the transformation from $(u,v)$ to $(e,\wstar)$, and $|\det J|$ is its absolute determinant.

\subsection{Case 1: sampling and priors both in \texorpdfstring{$(e,\wstar)$}{(e, ω⋆)}} \label{app:case1}
In the default parameterisation, the sampler proposes values of $e$ and $\wstar$ directly, and the priors are defined on the same pair of variables. As there is no change of variables between the sampling space and the space in which the prior is defined, no Jacobian factor arises. A prior on $e$ has support on $[0,1)$ (i.e. it is defined to be zero for $e < 0$ and $e \geq 1$), and is therefore properly normalised over the physically allowed range for $e$. Assuming independent priors on $e$ and $\wstar$, such that the joint prior factorises as $\pi(e,\wstar) = \pi(e)\,\pi(\wstar)$, the log-posterior is simply
\begin{equation}
      \ln\Like(e,\wstar) + \ln\pi(e) + \ln\pi(\wstar),
\end{equation}
with no corrective terms needed. We can verify this directly -- for $\pi(e) = \mathcal{U}[0,1)$ and $\pi(\wstar) = \mathcal{U}[-\pi,\pi]$, the joint prior integrates to unity over its support:
\begin{equation}
\begin{split}
  &\int_{0}^{1}\int_{-\pi}^{\pi} \pi(e)\,\pi(\wstar)\,\mathrm{d}\wstar\,\mathrm{d}e \\
  = &\int_{0}^{1}\int_{-\pi}^{\pi}\, 1\cdot\frac{1}{2\pi}\, \mathrm{d}\wstar\,\mathrm{d}e \\
  = &\;1,
\end{split}
\end{equation}
confirming that no additive correction to $\ln\pi(e,\wstar)$ is required for the prior to be correctly normalised.

In this work, this case applies trivially to our \texttt{C} models, because the parameters are fixed at $e = 0$ and $\wstar = \pi/2$ and neither parameter contributes to the prior function at all; we simply mention this case as a reference configuration against which the other two cases can be validated.

\subsection{Case 2: sampling and priors both in \texorpdfstring{$(u,v)$}{(u,v)}}
\label{app:priors_case2}

In this case, the sampler proposes values of $(u,v)$, and the priors are also defined directly on $(u,v)$. Since the sampling and prior parameterisations are the same, no change of variables is required and no Jacobian correction arises. This is the case in our \texttt{U} models, where independent uniform priors $\mathcal{U}[-1,1]$ are placed on $u$ and $v$. For the specific transformation $u=\sqrt{e}\cos\wstar$, $v=\sqrt{e}\sin\wstar$, the Jacobian of the $(u,v)$ to $(e,\wstar)$ mapping is constant, and a uniform prior over the unit disk in $(u,v)$ corresponds to a uniform prior in both $e$ and $\wstar$ for parameter inference \citep{Anderson.etal2010_WASP30b61MJup, Eastman.etal2013_EXOFASTFastExoplanetary}. However, for evidence estimation the prior must be correctly normalised over its physical support, which, as we show below, requires an explicit correction. 

The uniform prior $\mathcal{U}[-1,1]$ has density $1/2$ within its support and $0$ otherwise, therefore for valid $(u,v)$ the resulting joint prior density is
\begin{equation}
    \pi(u,v) = \pi(u)\pi(v) = \frac{1}{2}\cdot\frac{1}{2}=\frac{1}{4},
\end{equation}
uniform over the square formed by $[-1,1]^2$ which has area 4. 

\begin{figure}
 \includegraphics[width=\columnwidth]{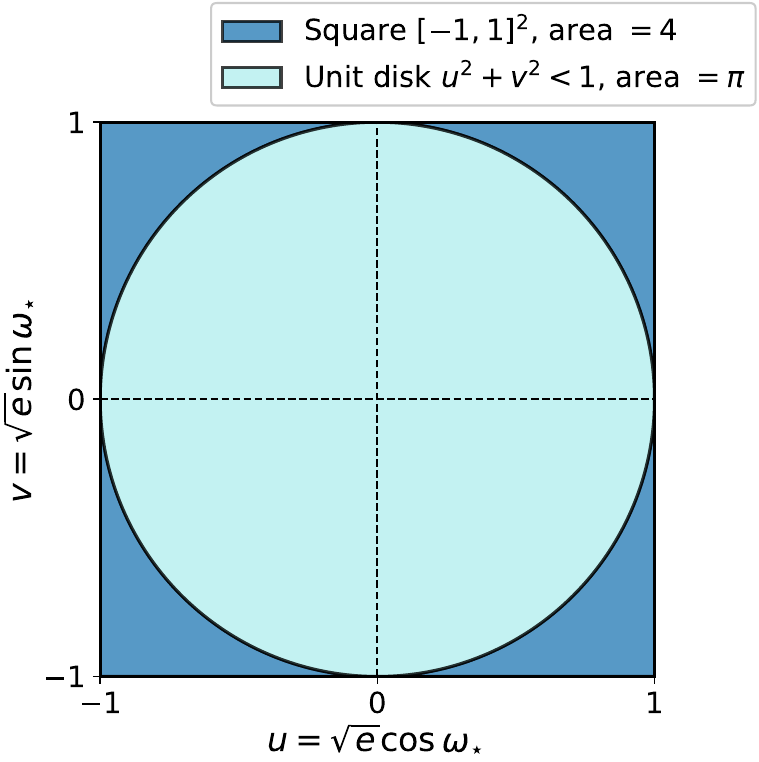}
 \caption{Visualisation of the prior space on the parameters $(u,v)=(\secosw,\sesinw)$. The dark blue square of area 4 is formed by two uniform priors $\mathcal{U}[-1,1]$ on $(u,v)$. The light blue unit disk of area $\pi$ is the truncated support formed by the condition $u^2+v^2<1$, corresponding to the physical requirement that eccentricity $0\leq e <1$.}
 \label{fig:case2_square_disk}
\end{figure}

However, consider that the physical requirement for $e < 1$ imposes the additional constraint $u^2 + v^2 < 1$, restricting the valid region to the unit disk, not the square of area $4$, as illustrated in Fig.~\ref{fig:case2_square_disk}. If this constraint is enforced separately in the MCMC code from the prior function evaluation -- such as by rejecting proposals of $(u,v)$ if $u^2 + v^2 \geq 1$ -- then the effective joint prior becomes
\begin{equation}\label{eq:pi_u_pi_v_indicator}
  \pi(u,v) = \pi(u)\pi(v)\,\mathds{1}_{\{u^2+v^2<1\}},
\end{equation}
where the indicator function $\mathds{1}_{\{u^2+v^2<1\}}$ takes value $1$ within the unit disk and $0$ outside, thereby truncating the support to the unit disk. The resulting joint prior $\pi(u,v)$ is no longer separable and is not correctly normalised: the constant density $1/4$ from the two uniform priors is now supported on a region of area $\pi$, rather than of area $4$, meaning the prior will not integrate to $1$, unlike in the $(e,\wstar)$ parameterisation (see Appendix~\ref{app:case1}). It instead integrates to $\pi/4$:
\begin{equation}
\begin{split}
&\iint_{u^2+v^2<1}\pi(u,v)\,\mathrm{d}u\,\mathrm{d}v\\
= &\iint_{u^2+v^2<1}\pi(u)\pi(v)\cdot\mathds{1}_{\{u^2+v^2<1\}}\,\mathrm{d}u\,\mathrm{d}v\\
=&\iint_{u^2+v^2<1} \frac{1}{2}\cdot\frac{1}{2}\cdot 1\,\mathrm{d}u\,\mathrm{d}v \\
= \,&\frac{\pi}{4}.
\end{split}
\end{equation}

If the code implementation defines a correctly normalised joint prior function $\pi(u,v)$ with uniform density $1/\pi$ within the unit disk and $0$ outside, then no correction is needed. However, if the two uniform functions and the validity check $u^2+v^2<1$ are treated separately, then a multiplicative correction term is required:
\begin{equation}
    \pi_{\mathrm{corrected}}(u,v) = \pi(u)\pi(v)\cdot\mathds{1}_{\{u^2+v^2<1\}} \cdot \frac{4}{\pi}.
\end{equation}
In practice, since the MCMC evaluates the log-posterior, this corresponds to an additive term that must be included in the log-prior, such that for points $(u,v)$ inside the unit disk:
\begin{equation}
    \ln\pi_{\mathrm{corrected}}(u,v) = \ln\pi(u)+\ln\pi(v) + \ln\left(\frac{4}{\pi}\right).
\end{equation}

As discussed earlier in Appendix~\ref{app:priors_why}, this additive constant on the log-prior has no effect on MCMC parameter inference, as it cancels in the MCMC acceptance ratio (equations~(\ref{eq:metropolis_hastings}) and (\ref{eq:affine_invariant})) -- therefore, if only performing parameter inference in the $(u,v)$ parameterisation with priors on $u$ and $v$, an incorrect normalisation may go unnoticed. However, it directly affects the evidence estimate: when using MCMC samples with the LHME, all priors must be correctly normalised over their support, or the evidence for that model will be systematically biased.

The $(\secosw,\sesinw)$ parameterisation is often used with $\mathcal{U}[-1,1]$ priors on $u$ and $v$ \citep[e.g.][]{Ahrer.etal2021_HARPSSearchSouthern, Cloutier.etal2021_MorePreciseMass, Almenara.etal2022_GJ3090One, Barragan.etal2022_PyanetiIIMultidimensional, Desidera.etal2023_TOI179YoungSystem, Stock.etal2023_GaussianProcessesRadial, Osborne.etal2024_TOI544PotentialWaterworld, Polanski.etal2024_TESSKeckSurveyXX, Osborne.etal2025_HomogeneousPlanetMasses}. However, the normalisation issue described above is not intrinsic to the parameterisation itself -- it arises specifically from separately handling the independent uniform $\mathcal{U}[-1,1]$ priors and the physical validity check enforcing $u^2+v^2<1$; therefore it depends on how these are implemented in the MCMC code. We believe that this could be an easy oversight to make, since an incorrect normalisation has no effect on MCMC parameter inference and may go undetected if the LHME is used with an MCMC sampler in an existing RV workflow. We therefore highlight that regardless of which RV software is being used, the prior implementation should be inspected carefully to verify that the prior is correctly normalised over the physically allowed region.

In this work, the $(\secosw,\sesinw)$ parameterisation with independent $\mathcal{U}[-1,1]$ priors is used in \textsc{ravest} for the \texttt{U} models, and the correction term $+\ln(4/\pi)$ is included in the log-posterior used in the MCMC sampling, ensuring that \texttt{U} model evidence estimates are not systematically biased relative to \texttt{C} or \texttt{H} models when the samples are used with the LHME.

We note that the additive correction term $\ln(4/\pi)$ is specific to the use of independent $\mathcal{U}[-1,1]$ priors on $u$ and $v$. If uniform priors with different bounds are used, or if a different informative prior is placed on $(u,v)$, then the correction term to ensure proper normalisation will differ and must be derived separately for that specific prior construction.

\subsection{Case 3: sampling in \texorpdfstring{$(u,v)$}{(u,v)}, priors on \texorpdfstring{$(e,\wstar)$}{(e, ω⋆)}}
\label{app:priors_case3}

Like the \texttt{U} models, the \texttt{H} models sample in the $(u,v)$ parameterisation. However, the prior on $e$ is now the half-normal prior for single-planet systems from \citet{VanEylen.etal2019_OrbitalEccentricitySmall} with $\sigma = 0.32$, truncated over the range $0 \leq e < 1$, and the prior on $\wstar$ is $\mathcal{U}[-\pi,\pi]$. The sampling parameterisation and the prior parameterisation are therefore different: the sampler must evaluate $\ln\pi(u,v)$ at each proposed point, but only $\ln\pi(e,\wstar)$ is defined. The prior must therefore be transformed into the sampling parameterisation, and the change-of-variables formula requires the Jacobian determinant.

Applying equation~(\ref{eq:change_of_variables}) with the inverse mapping $e = u^2+v^2$ and $\wstar = \operatorname{atan2}(v,u)$, the Jacobian matrix is
\begin{equation}
\begin{split}
  J &= \frac{\partial(e,\wstar)}{\partial(u,v)}
  =
  \begin{pmatrix}
    \partial e/\partial u & \partial e/\partial v \\[2pt]
    \partial \wstar/\partial u & \partial \wstar/\partial v
  \end{pmatrix}\\
  &=
  \begin{pmatrix}
    2u & 2v \\[2pt]
    -v/(u^2+v^2) & u/(u^2+v^2)
  \end{pmatrix}
  =
  \begin{pmatrix}
    2u & 2v \\[2pt]
    -v/e & u/e
  \end{pmatrix},
\end{split}
\end{equation}
with determinant $|\det J| = 2$, which is constant and therefore independent of $(u,v)$. Substituting into equation~(\ref{eq:change_of_variables}) gives
\begin{equation}
  \pi(u,v) = 2\,\pi(e,\wstar) = 2\,\pi(e(u,v))\,\pi(\wstar(u,v)),
\end{equation}
so the log-posterior in the sampling parameterisation is
\begin{equation}
  \ln\Like(u,v) + \ln\pi(e(u,v)) + \ln\pi(\wstar(u,v)) + \ln 2.
\end{equation}

Since $\ln 2$ is a constant, it cancels in the MCMC acceptance ratio (equations~(\ref{eq:metropolis_hastings}) and (\ref{eq:affine_invariant})) and has no effect on MCMC parameter inference. However, it directly affects the evidence estimate: omitting it would bias the estimate of $\ln\Z$ for the \texttt{H} models by $-\ln 2 \approx -0.6931$ for each planet with free eccentricity sampled in $(u,v)$.

No further additive term to the log-posterior is required. Unlike in Case 2, there is no mismatch between the support of the prior and the physically allowed region: the half-normal prior on $e$ used in the \texttt{H} models is truncated to $[0,1)$, assigning zero density for $e \geq 1$ -- so if the sampler proposes a point $(u,v)$ with $u^2+v^2 \geq 1$ (i.e. $e \geq 1$), the move is rejected by the prior with no separate check required.

An important property of this correction is that it is independent of the functional form of the prior on $e$. The Jacobian determinant $|\det J| = 2$ is a geometric property of the transformation from $(u,v)$ space to $(e,\wstar)$ space alone, and does not depend on the form of the prior. The same $+\ln 2$ term therefore applies whether the prior on $e$ is a truncated half-normal distribution (as in our \texttt{H} models), or a uniform on $[0,1)$, Beta, Rayleigh, or any other distribution on $e$ provided it is properly normalised and truncated to $[0,1)$.

In this work, the $+\ln 2$ correction per planet with free eccentricity is included in the log-posterior used in the MCMC sampling in \textsc{ravest} for the \texttt{H} models, ensuring that their evidence estimates are not systematically biased relative to \texttt{C} or \texttt{U} models.

\subsection{Summary for the models in this work}
\label{app:priors_summary}

\begin{table}
\centering
\caption{Summary of prior normalisation corrections applied to the log-posterior in \textsc{ravest} for each eccentricity model family. The additive term is applied once per planet with free eccentricity. \texttt{C} models fix $e=0$ and require no correction. For \texttt{U} models (Case~2), which sample in $(u,v) = (\secosw,\sesinw)$ with independent $\mathcal{U}[-1,1]$ priors, the term corrects for the mismatch between the prior support (the unit square) and the physically allowed region (the unit disk formed by $u^2+v^2<1$, corresponding to $e<1$). For \texttt{H} models (Case~3), which also sample in $(u,v)$ but with the prior defined on $(e,\wstar)$, the term is the Jacobian factor arising from the change of variables. In both cases, the term does not affect parameter inference via MCMC but must be included for unbiased evidence estimation with the LHME.}
\label{tab:prior_corrections}
\setlength{\tabcolsep}{4pt}
\begin{tabular}{clcl}
  \hline
  Model & Eccentricity prior & Case & Additive term (per planet) \\
  \hline
  \texttt{C} & Fixed ($e=0$) & --- & $0$ \\
  
  \texttt{U} & $\mathcal{U}[-1,1]$ on $(u, v)$ & 2 & $+\ln(4/\pi) \approx +0.242$ \\
  
  \texttt{H} & $\mathcal{N_H}[0.32]$ on $e \in [0,1)$ & 3 & $+\ln 2 \approx +0.693$ \\
  \hline
\end{tabular}
\end{table}

Cases 2 and 3 therefore represent two different mechanisms leading to the same practical consequence: a constant offset in the log-posterior when the prior is not correctly specified in the sampling parameterisation. In Case 2 the offset arises from a support mismatch between the uniform priors and the physically allowed region; in Case 3 it arises from the Jacobian of the coordinate transformation.

Each model in this paper falls into one of the three cases above according to its eccentricity letter: \texttt{C}, \texttt{U}, or \texttt{H}. Table~\ref{tab:prior_corrections} records the case and the additive term applied to the log-posterior for each planet with free eccentricity parameters.

Where needed, the correction terms are included in the log-posterior function used during MCMC sampling in \textsc{ravest}. Since these corrections are constant, they cancel in the MCMC acceptance ratio and do not affect parameter inference. They are required when the posterior chains are used to estimate the log evidence $\ln\hat\Z$ via \textsc{harmonic}, or for any other evidence estimation technique that depends on log-posterior samples.

For systems with multiple planets, such as the two-planet models for TOI-544, the corrections apply independently to each planet with free eccentricity sampled in $(u,v)$, and the total correction is the sum of the individual per-planet corrections. The contribution for a planet depends on its own eccentricity case: $+\ln(4/\pi)$ for Case~2 (\texttt{U} uniform prior) and $+\ln 2$ for Case~3 (\texttt{H} half-normal prior). For example, a two-planet model \texttt{HH} in which both planets are Case~3 requires a total correction of $2\ln 2 \approx 1.386$, while a mixed Case~2 and Case~3 model (\texttt{UH} and \texttt{HU} models) requires $\ln(4/\pi) + \ln 2 \approx 0.935$. If one planet is modelled with a circular orbit (Case~1; \texttt{C} models), then the first planet does not require a correction and the total correction is simply that for the other eccentric planet alone. 

The \texttt{G} model family uses a different likelihood from the \texttt{W} models, but the same priors and sampling parameterisation for the orbital elements. Since the corrections depend only on the prior construction -- not on the likelihood -- the corrections for \texttt{G} models are identical to those for \texttt{W} models within the same eccentricity class.

The corrections were validated empirically on a two-planet system. Three sampling configurations were tested: Case~1, in which the sampler and priors were both defined directly on $(e,\wstar)$, using $\mathcal{U}[0,1)$ on $e$ and $\mathcal{U}[-\pi,\pi]$ on $\wstar$ for each planet; Case~2, in which the sampler and priors were both in $(u,v)$, using independent $\mathcal{U}[-1,1]$ priors on $u$ and $v$ with a separate $u^2+v^2<1$ validity check; and Case~3, in which the sampler was in $(u,v)$ with priors defined on $(e,\wstar)$, using $\mathcal{U}[0,1)$ on $e$ and $\mathcal{U}[-\pi,\pi]$ on $\wstar$. Case~1 is not recommended for RV fitting due to the Lucy--Sweeney bias and slow convergence, but served here as a reference configuration requiring no correction. 

All combinations for a two-planet system were tested, and in each configuration, $\ln\hat{\Z}$ was estimated with \textsc{harmonic}. Without corrections, the $\ln\hat{\Z}$ values were offset from the Case~1 reference by the expected amounts: $+\ln(4/\pi)$ per Case~2 planet, $+\ln 2$ per Case~3 planet, summing independently for mixed-case models. With corrections applied, all configurations agreed to within the stochastic uncertainty of the MCMC and \textsc{harmonic} estimates. The Case~3 configuration was then repeated replacing the uniform prior on $e$ with a truncated half-normal ($\mathcal{N_H}[0.32]$, $e \in [0,1)$); the correction remained $+\ln 2$ per planet, confirming that it is the Jacobian determinant of the transformation from sampling space $(u,v)$ to prior space $(e,\wstar)$ -- not the form of the prior on $e$ itself -- that determines the Case~3 term.

Finally, we note that the additive corrections in Cases~2 and~3 both rely on a specific property of the $(\secosw,\sesinw)$ parameterisation: the Jacobian of the $(u,v)$ to $(e,\wstar)$ mapping is a constant ($|\det J|=2$), independent of the parameter values. For alternative parameterisations such as $(e\cos\wstar,\,e\sin\wstar)$, the Jacobian has been shown to be proportional to $e$ \citep{Ford2006_ImprovingEfficiencyMarkov, Anderson.etal2010_WASP30b61MJup}, which is sample-dependent, therefore not cancelling in the MCMC acceptance ratio, and it cannot be absorbed into a fixed additive offset; it would instead require per-step evaluation during MCMC sampling. The proportionality to $e$ also implicitly imposes a non-uniform prior on $e$, which contributes to why the $(\secosw,\sesinw)$ parameterisation is more commonly used \citep{Anderson.etal2010_WASP30b61MJup, Eastman.etal2013_EXOFASTFastExoplanetary}.
%

%%%%%%%%%%%%%%%%%%%% appendix 2: extra tables %%%%%%%%%%%%%%%%%%%%%%%%%%%
\newpage
\section{Model comparison tables}\label{app:extra_tables}

\begin{table*}
  \centering
  \setlength{\tabcolsep}{3pt}
  \caption{Model comparison for HD 18599 (TIC 207141131). Only models passing diagnostic checks with successfully estimated log evidence $\ln\Z$ are shown, sorted by $\dlnZ$ (log Bayes factor relative to the log evidence of the best model).}
  \label{tab:table1_tic207141131}
  \begin{tabular}{@{}lccccccccccc@{}}
    \hline
    Model & $P$ & $T_C$ & $K$ & $e$ & $\Mpsini$ & $\chi^2$ & $\lnL$ & AIC & BIC & $\ln\Z$ & $\dlnZ$ \\
          & [d] & [BTJD] & [\ms] & & [$M_\oplus$] & & & & & & \\
    \hline
    \texttt{HG1} & $4.137437_{-0.000098}^{+0.000099}$ & $2111.7395 \pm 0.0007$ & $10 \pm 4$ & $0.38_{-0.14}^{+0.15}$ & $20_{-8}^{+7}$ & $95.8$ & $-412.8$ & $853.6$ & $890.9$ & $-440.546 \pm 0.007$ & $0.0$ \\
    \texttt{CG1} & $4.137426 \pm 0.000099$ & $2111.7395 \pm 0.0007$ & $5 \pm 3$ & $0$ (fixed) & $11_{-6}^{+7}$ & $97.7$ & $-416.3$ & $856.6$ & $888.5$ & $-442.277 \pm 0.004$ & $1.7$ \\
    \texttt{HG0} & $4.13744 \pm 0.00010$ & $2111.7395 \pm 0.0007$ & $6 \pm 4$ & $0.3 \pm 0.2$ & $13 \pm 8$ & $111.7$ & $-427.0$ & $880.0$ & $914.6$ & $-446.821 \pm 0.008$ & $6.3$ \\
    \texttt{CG0} & $4.137433_{-0.000100}^{+0.000098}$ & $2111.7395 \pm 0.0007$ & $4 \pm 3$ & $0$ (fixed) & $9_{-6}^{+7}$ & $110.7$ & $-427.7$ & $877.3$ & $906.6$ & $-447.431 \pm 0.003$ & $6.9$ \\
    \texttt{UG2} & $4.13744 \pm 0.00010$ & $2111.7395 \pm 0.0007$ & $11 \pm 4$ & $0.5 \pm 0.2$ & $20_{-8}^{+7}$ & $95.4$ & $-412.1$ & $854.2$ & $894.1$ & $-453.100 \pm 0.009$ & $12.6$ \\
    \texttt{HG2} & $4.137440_{-0.000099}^{+0.000100}$ & $2111.7394 \pm 0.0007$ & $10 \pm 4$ & $0.39 \pm 0.14$ & $21_{-8}^{+7}$ & $95.5$ & $-412.6$ & $855.2$ & $895.2$ & $-453.263 \pm 0.007$ & $12.7$ \\
    \texttt{CG2} & $4.137426_{-0.000099}^{+0.000100}$ & $2111.7395 \pm 0.0007$ & $5 \pm 3$ & $0$ (fixed) & $11_{-6}^{+7}$ & $97.2$ & $-416.2$ & $858.4$ & $893.1$ & $-455.015 \pm 0.005$ & $14.5$ \\
    \texttt{HW1} & $4.137434_{-0.000098}^{+0.000099}$ & $2111.7395 \pm 0.0007$ & $13 \pm 4$ & $0.33_{-0.14}^{+0.13}$ & $27 \pm 8$ & $98.2$ & $-485.8$ & $991.7$ & $1018.3$ & $-508.508 \pm 0.003$ & $68.0$ \\
    \texttt{UW1} & $4.137438 \pm 0.000099$ & $2111.7395 \pm 0.0007$ & $13 \pm 4$ & $0.40_{-0.14}^{+0.20}$ & $27 \pm 8$ & $97.7$ & $-485.7$ & $991.3$ & $1018.0$ & $-508.655 \pm 0.005$ & $68.1$ \\
    \texttt{CW1} & $4.137414 \pm 0.000099$ & $2111.7395 \pm 0.0007$ & $9 \pm 4$ & $0$ (fixed) & $21 \pm 8$ & $98.6$ & $-489.1$ & $994.2$ & $1015.5$ & $-510.133 \pm 0.002$ & $69.6$ \\
    \texttt{HW2} & $4.137434_{-0.000099}^{+0.000098}$ & $2111.7395 \pm 0.0007$ & $13 \pm 4$ & $0.35_{-0.13}^{+0.12}$ & $28 \pm 8$ & $96.8$ & $-484.9$ & $991.8$ & $1021.1$ & $-521.063 \pm 0.003$ & $80.5$ \\
    \texttt{CW2} & $4.137414_{-0.000100}^{+0.000099}$ & $2111.7395 \pm 0.0007$ & $9 \pm 4$ & $0$ (fixed) & $21 \pm 8$ & $97.8$ & $-488.7$ & $995.5$ & $1019.4$ & $-522.969 \pm 0.003$ & $82.4$ \\
    \texttt{CW0} & $4.13743 \pm 0.00010$ & $2111.7395 \pm 0.0007$ & $6 \pm 4$ & $0$ (fixed) & $14.7_{-9.0}^{+9.9}$ & $100.1$ & $-516.3$ & $1046.5$ & $1065.2$ & $-529.488 \pm 0.002$ & $88.9$ \\
    \texttt{HW0} & $4.13744 \pm 0.00010$ & $2111.7395 \pm 0.0007$ & $6_{-4}^{+5}$ & $0.19_{-0.14}^{+0.20}$ & $13.7_{-8.0}^{+9.8}$ & $99.8$ & $-516.2$ & $1050.4$ & $1074.3$ & $-529.589 \pm 0.002$ & $89.0$ \\
    \hline
  \end{tabular}
\end{table*}

\begin{table*}
  \centering
  \setlength{\tabcolsep}{3pt}
  \caption{Model comparison for K2-265 (TIC 146364192). Only models passing diagnostic checks with successfully estimated log evidence $\ln\Z$ are shown, sorted by $\dlnZ$ (log Bayes factor relative to the log evidence of the best model).}
  \label{tab:table1_tic146364192}
  \begin{tabular}{@{}lccccccccccc@{}}
    \hline
    Model & $P$ & $T_C$ & $K$ & $e$ & $\Mpsini$ & $\chi^2$ & $\lnL$ & AIC & BIC & $\ln\Z$ & $\dlnZ$ \\
          & [d] & [BTJD] & [\ms] & & [$M_\oplus$] & & & & & & \\
    \hline
    \texttt{CG0} & $2.36916 \pm 0.00008$ & $17.1808 \pm 0.0005$ & $3.3 \pm 0.4$ & $0$ (fixed) & $6.4 \pm 0.8$ & $141.9$ & $-393.2$ & $804.4$ & $830.9$ & $-409.756 \pm 0.002$ & $0.0$ \\
    \texttt{HG0} & $2.36911 \pm 0.00009$ & $17.1808_{-0.0006}^{+0.0005}$ & $3.4 \pm 0.4$ & $0.12_{-0.08}^{+0.10}$ & $6.6 \pm 0.8$ & $142.0$ & $-393.1$ & $808.1$ & $840.5$ & $-410.143 \pm 0.003$ & $0.4$ \\
    \texttt{CG2} & $2.36916 \pm 0.00008$ & $17.1808_{-0.0005}^{+0.0006}$ & $3.3 \pm 0.4$ & $0$ (fixed) & $6.4 \pm 0.8$ & $139.5$ & $-389.8$ & $801.7$ & $834.0$ & $-410.842 \pm 0.003$ & $1.1$ \\
    \texttt{CG1} & $2.36916 \pm 0.00008$ & $17.1808 \pm 0.0005$ & $3.3 \pm 0.4$ & $0$ (fixed) & $6.4 \pm 0.8$ & $141.1$ & $-392.6$ & $805.2$ & $834.6$ & $-411.052_{-0.002}^{+0.003}$ & $1.3$ \\
    \texttt{HG2} & $2.369106_{-0.000096}^{+0.000090}$ & $17.1808_{-0.0006}^{+0.0005}$ & $3.4 \pm 0.4$ & $0.125_{-0.080}^{+0.099}$ & $6.5 \pm 0.8$ & $139.3$ & $-389.5$ & $805.0$ & $843.2$ & $-411.160 \pm 0.004$ & $1.4$ \\
    \texttt{HG1} & $2.36911 \pm 0.00009$ & $17.1808_{-0.0006}^{+0.0005}$ & $3.4 \pm 0.4$ & $0.12_{-0.08}^{+0.10}$ & $6.5 \pm 0.8$ & $141.0$ & $-392.4$ & $808.8$ & $844.1$ & $-411.429 \pm 0.003$ & $1.7$ \\
    \texttt{HW2} & $2.369059 \pm 0.000098$ & $17.1808 \pm 0.0005$ & $4.2 \pm 0.6$ & $0.21_{-0.12}^{+0.14}$ & $7.8 \pm 1.2$ & $134.4$ & $-427.3$ & $872.5$ & $899.0$ & $-448.833 \pm 0.002$ & $39.1$ \\
    \texttt{CW2} & $2.36913 \pm 0.00009$ & $17.1808 \pm 0.0005$ & $3.9 \pm 0.6$ & $0$ (fixed) & $7.5 \pm 1.2$ & $135.1$ & $-429.5$ & $873.0$ & $893.6$ & $-449.357 \pm 0.002$ & $39.6$ \\
    \texttt{HW1} & $2.369053_{-0.000099}^{+0.000098}$ & $17.1808 \pm 0.0005$ & $4.2 \pm 0.7$ & $0.23_{-0.14}^{+0.20}$ & $7.7 \pm 1.3$ & $134.8$ & $-439.4$ & $894.8$ & $918.3$ & $-457.651 \pm 0.003$ & $47.9$ \\
    \texttt{CW1} & $2.36911 \pm 0.00009$ & $17.1808_{-0.0006}^{+0.0005}$ & $3.8 \pm 0.7$ & $0$ (fixed) & $7.4 \pm 1.3$ & $135.8$ & $-441.8$ & $895.6$ & $913.3$ & $-458.1785_{-0.0011}^{+0.0012}$ & $48.4$ \\
    \texttt{HW0} & $2.369047 \pm 0.000098$ & $17.1808_{-0.0005}^{+0.0006}$ & $4.1 \pm 0.8$ & $0.23_{-0.14}^{+0.20}$ & $7.7 \pm 1.4$ & $135.3$ & $-450.7$ & $915.5$ & $936.1$ & $-465.533 \pm 0.002$ & $55.8$ \\
    \texttt{CW0} & $2.36909 \pm 0.00009$ & $17.1808 \pm 0.0005$ & $3.8 \pm 0.7$ & $0$ (fixed) & $7.3 \pm 1.4$ & $136.2$ & $-453.0$ & $916.1$ & $930.8$ & $-465.9227 \pm 0.0009$ & $56.2$ \\
    \hline
  \end{tabular}
\end{table*}

\begin{table*}
  \centering
  \setlength{\tabcolsep}{3pt}
  \caption{Model comparison for TOI-1055 (TIC 320004517). Only models passing diagnostic checks with successfully estimated log evidence $\ln\Z$ are shown, sorted by $\dlnZ$ (log Bayes factor relative to the log evidence of the best model).}
  \label{tab:table1_tic320004517}
  \begin{tabular}{@{}lccccccccccc@{}}
    \hline
    Model & $P$ & $T_C$ & $K$ & $e$ & $\Mpsini$ & $\chi^2$ & $\lnL$ & AIC & BIC & $\ln\Z$ & $\dlnZ$ \\
          & [d] & [BTJD] & [\ms] & & [$M_\oplus$] & & & & & & \\
    \hline
    \texttt{CG0} & $17.471290_{-0.000099}^{+0.000100}$ & $1661.06265 \pm 0.00097$ & $4.4 \pm 1.3$ & $0$ (fixed) & $18 \pm 5$ & $66.1$ & $-197.8$ & $421.6$ & $451.0$ & $-225.309 \pm 0.005$ & $0.0$ \\
    \texttt{HG0} & $17.471292_{-0.000100}^{+0.000099}$ & $1661.06264_{-0.00097}^{+0.00096}$ & $4.1 \pm 1.3$ & $0.13_{-0.09}^{+0.20}$ & $16 \pm 5$ & $65.5$ & $-197.5$ & $425.1$ & $459.0$ & $-225.838 \pm 0.008$ & $0.5$ \\
    \texttt{UG0} & $17.47129 \pm 0.00010$ & $1661.06260 \pm 0.00097$ & $3.8_{-1.3}^{+1.4}$ & $0.21_{-0.15}^{+0.30}$ & $15 \pm 6$ & $65.5$ & $-197.4$ & $424.7$ & $458.6$ & $-226.360 \pm 0.012$ & $1.1$ \\
    \texttt{CG1} & $17.471291_{-0.000099}^{+0.000100}$ & $1661.06263 \pm 0.00097$ & $4.4 \pm 1.3$ & $0$ (fixed) & $18 \pm 5$ & $65.2$ & $-198.1$ & $424.1$ & $455.8$ & $-233.126 \pm 0.009$ & $7.8$ \\
    \texttt{HG1} & $17.47129 \pm 0.00010$ & $1661.06263_{-0.00097}^{+0.00096}$ & $4.1 \pm 1.3$ & $0.13_{-0.09}^{+0.20}$ & $16 \pm 5$ & $65.4$ & $-197.8$ & $427.7$ & $463.9$ & $-233.606 \pm 0.012$ & $8.3$ \\
    \texttt{UG1} & $17.471289_{-0.000099}^{+0.000100}$ & $1661.06260_{-0.00090}^{+0.00098}$ & $3.7 \pm 1.4$ & $0.2_{-0.2}^{+0.3}$ & $14_{-7}^{+6}$ & $64.6$ & $-197.6$ & $427.2$ & $463.4$ & $-234.12 \pm 0.02$ & $8.8$ \\
    \texttt{HW0} & $17.471292 \pm 0.000099$ & $1661.06263_{-0.00096}^{+0.00090}$ & $2.3_{-1.1}^{+1.2}$ & $0.3 \pm 0.2$ & $9 \pm 4$ & $62.3$ & $-212.3$ & $446.6$ & $471.5$ & $-238.987 \pm 0.003$ & $13.7$ \\
    \texttt{CW0} & $17.471292_{-0.000100}^{+0.000098}$ & $1661.06263 \pm 0.00097$ & $1.5_{-0.8}^{+0.9}$ & $0$ (fixed) & $6_{-3}^{+4}$ & $62.9$ & $-213.8$ & $445.7$ & $466.0$ & $-239.741 \pm 0.002$ & $14.4$ \\
    \texttt{CG2} & $17.471294_{-0.000100}^{+0.000099}$ & $1661.06261_{-0.00096}^{+0.00098}$ & $4.8_{-1.3}^{+1.2}$ & $0$ (fixed) & $19 \pm 5$ & $64.3$ & $-194.1$ & $418.3$ & $452.2$ & $-245.5735_{-0.0099}^{+0.0100}$ & $20.3$ \\
    \texttt{UG2} & $17.471294_{-0.000098}^{+0.000100}$ & $1661.06266_{-0.00097}^{+0.00090}$ & $4.3_{-1.4}^{+1.3}$ & $0.15_{-0.11}^{+0.20}$ & $17 \pm 6$ & $63.6$ & $-194.0$ & $422.0$ & $460.4$ & $-246.956 \pm 0.013$ & $21.6$ \\
    \texttt{HW1} & $17.471290_{-0.000099}^{+0.000100}$ & $1661.06261_{-0.00096}^{+0.00097}$ & $2.3 \pm 1.2$ & $0.3 \pm 0.2$ & $9 \pm 4$ & $61.6$ & $-212.2$ & $448.4$ & $475.6$ & $-247.161 \pm 0.004$ & $21.9$ \\
    \texttt{CW1} & $17.471290_{-0.000099}^{+0.000100}$ & $1661.06262_{-0.00096}^{+0.00097}$ & $1.48_{-0.80}^{+0.96}$ & $0$ (fixed) & $6_{-3}^{+4}$ & $61.8$ & $-213.8$ & $447.7$ & $470.3$ & $-247.979 \pm 0.003$ & $22.7$ \\
    \texttt{CW2} & $17.471292_{-0.000100}^{+0.000099}$ & $1661.06262 \pm 0.00097$ & $2.9 \pm 1.0$ & $0$ (fixed) & $12 \pm 4$ & $60.6$ & $-206.1$ & $434.2$ & $459.1$ & $-256.851 \pm 0.003$ & $31.5$ \\
    \texttt{HW2} & $17.471289_{-0.000098}^{+0.000099}$ & $1661.06264_{-0.00097}^{+0.00090}$ & $2.9_{-1.0}^{+1.1}$ & $0.17_{-0.11}^{+0.20}$ & $11 \pm 4$ & $60.2$ & $-205.6$ & $437.3$ & $466.7$ & $-257.006 \pm 0.004$ & $31.7$ \\
    \hline
  \end{tabular}
\end{table*}

\begin{table*}
  \centering
  \setlength{\tabcolsep}{3pt}
  \caption{Model comparison for TOI-220 (TIC 150098860). Only models passing diagnostic checks with successfully estimated log evidence $\ln\Z$ are shown, sorted by $\dlnZ$ (log Bayes factor relative to the log evidence of the best model).}
  \label{tab:table1_tic150098860}
  \begin{tabular}{@{}lccccccccccc@{}}
    \hline
    Model & $P$ & $T_C$ & $K$ & $e$ & $\Mpsini$ & $\chi^2$ & $\lnL$ & AIC & BIC & $\ln\Z$ & $\dlnZ$ \\
          & [d] & [BTJD] & [\ms] & & [$M_\oplus$] & & & & & & \\
    \hline
    \texttt{CW2} & $10.695262_{-0.000099}^{+0.000100}$ & $1335.9020 \pm 0.0014$ & $4.5 \pm 0.3$ & $0$ (fixed) & $13.56_{-0.99}^{+1.00}$ & $86.1$ & $-190.1$ & $394.2$ & $411.7$ & $-208.365 \pm 0.002$ & $0.0$ \\
    \texttt{HW2} & $10.695265_{-0.000099}^{+0.000098}$ & $1335.9020 \pm 0.0014$ & $4.5 \pm 0.3$ & $0.04_{-0.02}^{+0.05}$ & $13.5 \pm 1.0$ & $85.2$ & $-190.0$ & $398.1$ & $420.7$ & $-210.214 \pm 0.002$ & $1.8$ \\
    \texttt{UW2} & $10.695263_{-0.000100}^{+0.000099}$ & $1335.9020 \pm 0.0014$ & $4.5 \pm 0.3$ & $0.04_{-0.03}^{+0.05}$ & $13.5 \pm 1.0$ & $85.2$ & $-190.0$ & $398.1$ & $420.7$ & $-211.109 \pm 0.002$ & $2.7$ \\
    \texttt{CW0} & $10.695262 \pm 0.000099$ & $1335.9020 \pm 0.0014$ & $4.4 \pm 0.4$ & $0$ (fixed) & $13.4 \pm 1.1$ & $86.2$ & $-203.1$ & $416.2$ & $428.8$ & $-214.7634 \pm 0.0010$ & $6.4$ \\
    \texttt{HW0} & $10.695261_{-0.000099}^{+0.000100}$ & $1335.9020 \pm 0.0014$ & $4.4 \pm 0.4$ & $0.05_{-0.04}^{+0.07}$ & $13.2_{-1.1}^{+1.2}$ & $85.3$ & $-202.7$ & $419.5$ & $437.1$ & $-216.221 \pm 0.002$ & $7.9$ \\
    \texttt{CW1} & $10.69526 \pm 0.00010$ & $1335.9020 \pm 0.0014$ & $4.5 \pm 0.4$ & $0$ (fixed) & $13.5 \pm 1.1$ & $85.7$ & $-202.0$ & $416.0$ & $431.1$ & $-216.5914 \pm 0.0012$ & $8.2$ \\
    \texttt{UW0} & $10.69526 \pm 0.00010$ & $1335.9020 \pm 0.0014$ & $4.4 \pm 0.4$ & $0.06_{-0.04}^{+0.08}$ & $13.2_{-1.1}^{+1.2}$ & $85.2$ & $-202.7$ & $419.4$ & $437.0$ & $-217.0934 \pm 0.0014$ & $8.7$ \\
    \texttt{HW1} & $10.695262 \pm 0.000099$ & $1335.9020 \pm 0.0014$ & $4.4 \pm 0.4$ & $0.05_{-0.04}^{+0.07}$ & $13.4 \pm 1.1$ & $84.6$ & $-201.7$ & $419.4$ & $439.4$ & $-218.069 \pm 0.002$ & $9.7$ \\
    \texttt{UW1} & $10.69526 \pm 0.00010$ & $1335.9020 \pm 0.0014$ & $4.4 \pm 0.4$ & $0.05_{-0.04}^{+0.07}$ & $13.3_{-1.1}^{+1.2}$ & $84.8$ & $-201.6$ & $419.3$ & $439.3$ & $-218.9442 \pm 0.0014$ & $10.6$ \\
    \hline
  \end{tabular}
\end{table*}

\begin{table*}
  \centering
  \setlength{\tabcolsep}{3pt}
  \caption{Model comparison for LHS 1815 (TIC 260004324). Only models passing diagnostic checks with successfully estimated log evidence $\ln\Z$ are shown, sorted by $\dlnZ$ (log Bayes factor relative to the log evidence of the best model).}
  \label{tab:table1_tic260004324}
  \begin{tabular}{@{}lccccccccccc@{}}
    \hline
    Model & $P$ & $T_C$ & $K$ & $e$ & $\Mpsini$ & $\chi^2$ & $\lnL$ & AIC & BIC & $\ln\Z$ & $\dlnZ$ \\
          & [d] & [BTJD] & [\ms] & & [$M_\oplus$] & & & & & & \\
    \hline
    \texttt{CW0} & $3.81433 \pm 0.00008$ & $1327.000 \pm 0.004$ & $0.6_{-0.4}^{+0.6}$ & $0$ (fixed) & $1.0_{-0.7}^{+0.9}$ & $67.7$ & $-197.3$ & $408.6$ & $424.5$ & $-215.4901 \pm 0.0015$ & $0.0$ \\
    \texttt{HW0} & $3.81433 \pm 0.00008$ & $1327.000 \pm 0.004$ & $0.6_{-0.4}^{+0.6}$ & $0.21_{-0.15}^{+0.20}$ & $0.9_{-0.6}^{+0.9}$ & $67.5$ & $-197.3$ & $412.7$ & $433.1$ & $-215.514 \pm 0.002$ & $0.0$ \\
    \texttt{CG0} & $3.81433 \pm 0.00008$ & $1327.000 \pm 0.004$ & $0.4_{-0.3}^{+0.5}$ & $0$ (fixed) & $0.7_{-0.5}^{+0.8}$ & $63.4$ & $-194.0$ & $410.1$ & $434.9$ & $-216.707 \pm 0.009$ & $1.2$ \\
    \texttt{CW1} & $3.81433 \pm 0.00008$ & $1327.000 \pm 0.004$ & $0.6_{-0.4}^{+0.6}$ & $0$ (fixed) & $1.0_{-0.7}^{+0.9}$ & $67.3$ & $-196.9$ & $409.8$ & $427.9$ & $-223.540 \pm 0.002$ & $8.1$ \\
    \texttt{HW1} & $3.81433 \pm 0.00008$ & $1327.000 \pm 0.004$ & $0.6_{-0.4}^{+0.6}$ & $0.21_{-0.15}^{+0.20}$ & $0.9_{-0.6}^{+0.9}$ & $67.1$ & $-196.9$ & $413.8$ & $436.4$ & $-223.575 \pm 0.003$ & $8.1$ \\
    \texttt{HW2} & $3.81433 \pm 0.00008$ & $1327.000 \pm 0.004$ & $0.6_{-0.4}^{+0.6}$ & $0.2 \pm 0.2$ & $0.9_{-0.6}^{+0.9}$ & $65.5$ & $-194.8$ & $411.6$ & $436.5$ & $-237.564 \pm 0.005$ & $22.1$ \\
    \texttt{CW2} & $3.81433 \pm 0.00008$ & $1327.000 \pm 0.004$ & $0.5_{-0.4}^{+0.5}$ & $0$ (fixed) & $0.8_{-0.6}^{+0.8}$ & $65.8$ & $-194.8$ & $407.6$ & $428.0$ & $-237.605 \pm 0.004$ & $22.1$ \\
    \texttt{HG2} & $3.81433 \pm 0.00008$ & $1327.000 \pm 0.004$ & $0.5_{-0.3}^{+0.6}$ & $0.2_{-0.2}^{+0.3}$ & $0.7_{-0.5}^{+0.8}$ & $62.8$ & $-193.2$ & $416.4$ & $450.3$ & $-239.40 \pm 0.02$ & $23.9$ \\
    \texttt{CG2} & $3.81433 \pm 0.00008$ & $1327.000 \pm 0.004$ & $0.4_{-0.3}^{+0.5}$ & $0$ (fixed) & $0.7_{-0.5}^{+0.7}$ & $62.5$ & $-193.1$ & $412.3$ & $441.7$ & $-239.514_{-0.013}^{+0.014}$ & $24.0$ \\
    \hline
  \end{tabular}
\end{table*}

\begin{table*}
  \centering
  \setlength{\tabcolsep}{3pt}
  \caption{Model comparison for GJ 1214 (TIC 467929202). Only models passing diagnostic checks with successfully estimated log evidence $\ln\Z$ are shown, sorted by $\dlnZ$ (log Bayes factor relative to the log evidence of the best model).}
  \label{tab:table1_tic467929202}
  \begin{tabular}{@{}lccccccccccc@{}}
    \hline
    Model & $P$ & $T_C$ & $K$ & $e$ & $\Mpsini$ & $\chi^2$ & $\lnL$ & AIC & BIC & $\ln\Z$ & $\dlnZ$ \\
          & [d] & [BTJD] & [\ms] & & [$M_\oplus$] & & & & & & \\
    \hline
    \texttt{CG0} & $1.580404 \pm 0.000006$ & $2639.781262_{-0.000100}^{+0.000099}$ & $13.6 \pm 0.7$ & $0$ (fixed) & $7.9 \pm 0.4$ & $176.1$ & $-560.3$ & $1142.7$ & $1176.8$ & $-581.141_{-0.002}^{+0.003}$ & $0.0$ \\
    \texttt{CW0} & $1.580403 \pm 0.000006$ & $2639.78126 \pm 0.00010$ & $13.6 \pm 0.7$ & $0$ (fixed) & $7.9 \pm 0.5$ & $178.7$ & $-561.2$ & $1136.3$ & $1158.0$ & $-583.6088 \pm 0.0015$ & $2.5$ \\
    \texttt{HW0} & $1.5804080_{-0.0000097}^{+0.0000110}$ & $2639.781264 \pm 0.000099$ & $13.6 \pm 0.8$ & $0.05_{-0.04}^{+0.05}$ & $7.9 \pm 0.5$ & $176.7$ & $-560.5$ & $1139.0$ & $1166.9$ & $-584.922 \pm 0.003$ & $3.8$ \\
    \texttt{CG1} & $1.580404 \pm 0.000006$ & $2639.78126 \pm 0.00010$ & $13.6 \pm 0.7$ & $0$ (fixed) & $7.9 \pm 0.4$ & $175.4$ & $-560.4$ & $1144.7$ & $1181.9$ & $-588.843 \pm 0.003$ & $7.7$ \\
    \texttt{HG1} & $1.580407_{-0.000009}^{+0.000010}$ & $2639.781262_{-0.000099}^{+0.000100}$ & $13.6 \pm 0.7$ & $0.04_{-0.03}^{+0.04}$ & $7.9 \pm 0.4$ & $173.9$ & $-560.1$ & $1148.1$ & $1191.5$ & $-590.414 \pm 0.004$ & $9.3$ \\
    \texttt{CW1} & $1.580404 \pm 0.000007$ & $2639.781262_{-0.000100}^{+0.000099}$ & $13.6 \pm 0.7$ & $0$ (fixed) & $7.9_{-0.4}^{+0.5}$ & $177.7$ & $-561.1$ & $1138.2$ & $1163.0$ & $-592.158 \pm 0.002$ & $11.0$ \\
    \texttt{HW1} & $1.5804080_{-0.0000097}^{+0.0000110}$ & $2639.781262_{-0.000098}^{+0.000099}$ & $13.6_{-0.8}^{+0.7}$ & $0.05_{-0.04}^{+0.05}$ & $7.9 \pm 0.5$ & $175.8$ & $-560.5$ & $1141.0$ & $1172.0$ & $-593.504 \pm 0.003$ & $12.4$ \\
    \texttt{CG2} & $1.580404 \pm 0.000006$ & $2639.78126 \pm 0.00010$ & $13.6 \pm 0.7$ & $0$ (fixed) & $7.9 \pm 0.4$ & $174.8$ & $-560.4$ & $1146.7$ & $1187.0$ & $-604.254 \pm 0.003$ & $23.1$ \\
    \texttt{HW2} & $1.5804091_{-0.0000098}^{+0.0000110}$ & $2639.781261_{-0.000099}^{+0.000100}$ & $13.6_{-0.8}^{+0.7}$ & $0.05_{-0.04}^{+0.05}$ & $7.9 \pm 0.5$ & $174.2$ & $-559.9$ & $1141.7$ & $1175.8$ & $-609.051 \pm 0.003$ & $27.9$ \\
    \texttt{UW2} & $1.5804092_{-0.0000097}^{+0.0000110}$ & $2639.781262_{-0.000099}^{+0.000100}$ & $13.6_{-0.8}^{+0.7}$ & $0.05_{-0.04}^{+0.05}$ & $7.9 \pm 0.5$ & $173.9$ & $-559.9$ & $1141.7$ & $1175.8$ & $-609.944 \pm 0.003$ & $28.8$ \\
    \hline
  \end{tabular}
\end{table*}

\begin{landscape}
 \begin{table}
  \centering
  \setlength{\tabcolsep}{3pt}
  \caption{Model comparison for TOI-544 (TIC 50618703). Only models passing diagnostic checks with successfully estimated log evidence $\ln\Z$ are shown, sorted by $\dlnZ$ (log Bayes factor relative to the log evidence of the best model). $M_{\mathrm{p,b}}$ is the derived mass for planet b; $M_{\mathrm{p,c}}\sin i_{\mathrm{c}}$ is the derived minimum mass for planet c (2-planet models only).}
  \label{tab:table1_toi544_combined}
  \begin{tabular}{@{}lcccccccccccccc@{}}
    \hline
    Model & $P_{\mathrm{b}}$ & $K_{\mathrm{b}}$ & $e_{\mathrm{b}}$ & $M_{\mathrm{p,b}}$ & $P_{\mathrm{c}}$ & $K_{\mathrm{c}}$ & $e_{\mathrm{c}}$ & $M_{\mathrm{p,c}}\sin i_{\mathrm{c}}$ & $\chi^2$ & $\lnL$ & AIC & BIC & $\ln\Z$ & $\dlnZ$ \\
          & [d] & [\ms] & & [$M_\oplus$] & [d] & [\ms] & & [$M_\oplus$] & & & & & & \\
    \hline
    \texttt{HHG0} & $1.54836 \pm 0.00009$ & $2.4 \pm 0.4$ & $0.27_{-0.12}^{+0.14}$ & $3.1 \pm 0.5$ & $50.1_{-0.2}^{+0.3}$ & $5.0_{-0.5}^{+0.6}$ & $0.26_{-0.10}^{+0.09}$ & $21 \pm 2$ & $108.8$ & $-323.9$ & $683.8$ & $733.8$ & $-366.4126_{-0.0095}^{+0.0096}$ & $0.0$ \\
    \texttt{HUG0} & $1.54836 \pm 0.00009$ & $2.4 \pm 0.4$ & $0.26_{-0.12}^{+0.14}$ & $3.1 \pm 0.5$ & $50.1_{-0.2}^{+0.3}$ & $5.1 \pm 0.6$ & $0.280_{-0.099}^{+0.090}$ & $21 \pm 2$ & $108.7$ & $-323.6$ & $683.2$ & $733.2$ & $-366.9128_{-0.0097}^{+0.0098}$ & $0.5$ \\
    \texttt{CHG0} & $1.54841 \pm 0.00009$ & $2.1 \pm 0.4$ & $0$ (fixed) & $2.9 \pm 0.5$ & $50.1 \pm 0.3$ & $5.1 \pm 0.6$ & $0.270_{-0.097}^{+0.090}$ & $21 \pm 2$ & $110.1$ & $-327.4$ & $686.8$ & $731.3$ & $-367.999 \pm 0.008$ & $1.6$ \\
    \texttt{HHG1} & $1.548363_{-0.000095}^{+0.000090}$ & $2.4 \pm 0.4$ & $0.27_{-0.12}^{+0.14}$ & $3.1 \pm 0.5$ & $50.1 \pm 0.3$ & $5.1 \pm 0.6$ & $0.264_{-0.097}^{+0.090}$ & $21 \pm 2$ & $107.9$ & $-323.0$ & $684.0$ & $736.8$ & $-368.067 \pm 0.014$ & $1.7$ \\
    \texttt{HCG0} & $1.54836 \pm 0.00009$ & $2.4_{-0.4}^{+0.5}$ & $0.28_{-0.12}^{+0.14}$ & $3.1 \pm 0.5$ & $50.3 \pm 0.3$ & $4.4 \pm 0.4$ & $0$ (fixed) & $19 \pm 2$ & $111.3$ & $-328.8$ & $689.6$ & $734.1$ & $-368.100 \pm 0.006$ & $1.7$ \\
    \texttt{UCG0} & $1.54836 \pm 0.00009$ & $2.5 \pm 0.5$ & $0.35_{-0.15}^{+0.20}$ & $3.1 \pm 0.5$ & $50.3 \pm 0.3$ & $4.4 \pm 0.4$ & $0$ (fixed) & $19 \pm 2$ & $110.6$ & $-328.4$ & $688.8$ & $733.2$ & $-368.420 \pm 0.007$ & $2.0$ \\
    \texttt{CHG1} & $1.54841 \pm 0.00009$ & $2.1 \pm 0.4$ & $0$ (fixed) & $2.8 \pm 0.5$ & $50.1 \pm 0.3$ & $5.1 \pm 0.6$ & $0.274_{-0.099}^{+0.090}$ & $21 \pm 2$ & $108.9$ & $-326.6$ & $687.2$ & $734.5$ & $-369.685 \pm 0.009$ & $3.3$ \\
    \texttt{CCG0} & $1.548402_{-0.000095}^{+0.000090}$ & $2.1 \pm 0.4$ & $0$ (fixed) & $2.8 \pm 0.5$ & $50.3 \pm 0.3$ & $4.4 \pm 0.4$ & $0$ (fixed) & $19 \pm 2$ & $113.2$ & $-332.7$ & $693.3$ & $732.2$ & $-369.836 \pm 0.006$ & $3.4$ \\
    \texttt{HCG1} & $1.54836 \pm 0.00009$ & $2.4_{-0.4}^{+0.5}$ & $0.28_{-0.12}^{+0.14}$ & $3.1 \pm 0.5$ & $50.3 \pm 0.3$ & $4.4 \pm 0.4$ & $0$ (fixed) & $19 \pm 2$ & $110.6$ & $-328.0$ & $689.9$ & $737.2$ & $-369.869 \pm 0.007$ & $3.5$ \\
    \texttt{CUG1} & $1.548412_{-0.000095}^{+0.000090}$ & $2.1 \pm 0.4$ & $0$ (fixed) & $2.9 \pm 0.5$ & $50.1 \pm 0.3$ & $5.2_{-0.6}^{+0.7}$ & $0.300_{-0.098}^{+0.090}$ & $21 \pm 2$ & $108.6$ & $-326.3$ & $686.7$ & $733.9$ & $-370.151 \pm 0.008$ & $3.7$ \\
    \texttt{UCG1} & $1.54836 \pm 0.00009$ & $2.5 \pm 0.5$ & $0.35_{-0.15}^{+0.20}$ & $3.1_{-0.5}^{+0.6}$ & $50.3 \pm 0.3$ & $4.4 \pm 0.4$ & $0$ (fixed) & $19 \pm 2$ & $109.5$ & $-327.6$ & $689.1$ & $736.3$ & $-370.166 \pm 0.008$ & $3.8$ \\
    \texttt{CCG1} & $1.548406 \pm 0.000095$ & $2.1 \pm 0.4$ & $0$ (fixed) & $2.8_{-0.5}^{+0.6}$ & $50.3 \pm 0.3$ & $4.4 \pm 0.4$ & $0$ (fixed) & $19 \pm 2$ & $112.2$ & $-331.8$ & $693.5$ & $735.2$ & $-371.628 \pm 0.005$ & $5.2$ \\
    \texttt{HHG2} & $1.54836 \pm 0.00009$ & $2.4 \pm 0.4$ & $0.26_{-0.12}^{+0.14}$ & $3.1 \pm 0.5$ & $50.1 \pm 0.3$ & $5.1 \pm 0.6$ & $0.27_{-0.10}^{+0.09}$ & $21 \pm 2$ & $106.8$ & $-323.5$ & $687.0$ & $742.6$ & $-371.711 \pm 0.014$ & $5.3$ \\
    \texttt{CHG2} & $1.54841 \pm 0.00009$ & $2.1 \pm 0.4$ & $0$ (fixed) & $2.8 \pm 0.5$ & $50.1 \pm 0.3$ & $5.1 \pm 0.6$ & $0.278_{-0.098}^{+0.090}$ & $21 \pm 2$ & $108.2$ & $-327.0$ & $690.1$ & $740.1$ & $-373.352 \pm 0.010$ & $6.9$ \\
    \texttt{HCG2} & $1.54837 \pm 0.00009$ & $2.4_{-0.4}^{+0.5}$ & $0.28_{-0.12}^{+0.14}$ & $3.1 \pm 0.5$ & $50.3 \pm 0.3$ & $4.4 \pm 0.4$ & $0$ (fixed) & $19 \pm 2$ & $109.5$ & $-328.5$ & $693.0$ & $743.0$ & $-373.5376_{-0.0096}^{+0.0097}$ & $7.1$ \\
    \texttt{CCG2} & $1.548405_{-0.000096}^{+0.000090}$ & $2.1 \pm 0.4$ & $0$ (fixed) & $2.7 \pm 0.5$ & $50.3 \pm 0.3$ & $4.4 \pm 0.5$ & $0$ (fixed) & $19 \pm 2$ & $111.5$ & $-332.3$ & $696.6$ & $741.0$ & $-375.319 \pm 0.007$ & $8.9$ \\
    \texttt{\phantom{X}HG1} & $1.54836 \pm 0.00009$ & $2.6_{-0.4}^{+0.5}$ & $0.21_{-0.11}^{+0.20}$ & $3.4 \pm 0.5$ & -- & -- & -- & -- & $109.5$ & $-351.9$ & $731.8$ & $770.7$ & $-386.555 \pm 0.006$ & $20.1$ \\
    \texttt{\phantom{X}CG0} & $1.54840 \pm 0.00009$ & $2.3 \pm 0.4$ & $0$ (fixed) & $3.1 \pm 0.5$ & -- & -- & -- & -- & $111.0$ & $-356.2$ & $734.4$ & $765.0$ & $-386.698 \pm 0.004$ & $20.3$ \\
    \texttt{\phantom{X}CG1} & $1.54840 \pm 0.00009$ & $2.4 \pm 0.4$ & $0$ (fixed) & $3.2 \pm 0.5$ & -- & -- & -- & -- & $110.6$ & $-354.4$ & $732.8$ & $766.2$ & $-387.540 \pm 0.004$ & $21.1$ \\
    \texttt{\phantom{X}HG2} & $1.54836 \pm 0.00009$ & $2.6_{-0.4}^{+0.5}$ & $0.21_{-0.10}^{+0.14}$ & $3.4 \pm 0.5$ & -- & -- & -- & -- & $108.9$ & $-351.8$ & $733.7$ & $775.4$ & $-390.116 \pm 0.006$ & $23.7$ \\
    \texttt{CHW1} & $1.54839 \pm 0.00010$ & $2.5 \pm 0.7$ & $0$ (fixed) & $3.40 \pm 0.97$ & $50.4_{-0.6}^{+0.7}$ & $5.6_{-0.9}^{+1.2}$ & $0.2_{-0.2}^{+0.3}$ & $23 \pm 3$ & $110.6$ & $-372.1$ & $770.2$ & $806.3$ & $-401.309 \pm 0.007$ & $34.9$ \\
    \texttt{CCW1} & $1.548382_{-0.000098}^{+0.000099}$ & $2.5 \pm 0.7$ & $0$ (fixed) & $3.38_{-0.97}^{+0.98}$ & $50.2 \pm 0.4$ & $5.2 \pm 0.7$ & $0$ (fixed) & $22 \pm 3$ & $109.6$ & $-372.5$ & $767.1$ & $797.6$ & $-401.311 \pm 0.002$ & $34.9$ \\
    \texttt{CHW2} & $1.548383_{-0.000099}^{+0.000097}$ & $2.6 \pm 0.7$ & $0$ (fixed) & $3.44_{-0.97}^{+0.96}$ & $50.6_{-0.6}^{+0.4}$ & $5.4_{-0.9}^{+1.1}$ & $0.3 \pm 0.2$ & $21_{-3}^{+4}$ & $109.0$ & $-369.7$ & $767.4$ & $806.3$ & $-404.030 \pm 0.006$ & $37.6$ \\
    \texttt{CCW2} & $1.548378_{-0.000097}^{+0.000098}$ & $2.5 \pm 0.7$ & $0$ (fixed) & $3.38_{-0.97}^{+0.98}$ & $50.3_{-0.5}^{+0.4}$ & $4.9 \pm 0.8$ & $0$ (fixed) & $21 \pm 3$ & $108.8$ & $-371.0$ & $766.0$ & $799.3$ & $-404.218 \pm 0.003$ & $37.8$ \\
    \texttt{CCW0} & $1.548374_{-0.000097}^{+0.000098}$ & $2.6 \pm 0.8$ & $0$ (fixed) & $3.4 \pm 1.0$ & $50.5_{-0.4}^{+0.3}$ & $5.1 \pm 0.8$ & $0$ (fixed) & $22 \pm 3$ & $110.6$ & $-379.8$ & $779.7$ & $807.5$ & $-405.375 \pm 0.003$ & $39.0$ \\
    \texttt{\phantom{X}CW2} & $1.548370_{-0.000098}^{+0.000099}$ & $2.4 \pm 0.8$ & $0$ (fixed) & $3.2 \pm 1.1$ & -- & -- & -- & -- & $111.4$ & $-389.5$ & $797.0$ & $822.0$ & $-415.547 \pm 0.002$ & $49.1$ \\
    \texttt{\phantom{X}CW1} & $1.548374_{-0.000099}^{+0.000098}$ & $2.2 \pm 0.9$ & $0$ (fixed) & $3.0 \pm 1.2$ & -- & -- & -- & -- & $112.1$ & $-395.3$ & $806.6$ & $828.8$ & $-415.974 \pm 0.002$ & $49.6$ \\
    \texttt{\phantom{X}CW0} & $1.54837 \pm 0.00010$ & $2.3 \pm 0.9$ & $0$ (fixed) & $3.1 \pm 1.2$ & -- & -- & -- & -- & $113.0$ & $-399.5$ & $813.0$ & $832.5$ & $-417.508 \pm 0.002$ & $51.1$ \\
    \hline
  \end{tabular}
 \end{table}
\end{landscape}

%%%%%%%%%%%%%%%%%%%%%%%%%%%%%%%%%%%%%%%%%%%%%%%%%%

% Don't change these lines
\bsp	% typesetting comment
\label{lastpage}
\end{document}